\pdfoutput=1
\documentclass[12pt,a4paper]{article}

\usepackage{ifthen} 
\newboolean{pdflatex}
\setboolean{pdflatex}{true} 

\newboolean{articletitles}
\setboolean{articletitles}{true} 

\newboolean{uprightparticles}
\setboolean{uprightparticles}{false} 

\def\paperauthors{LHCb collaboration} 
\def\paperasciititle{Measurement of the branching fraction of the Lb -> J/psi Lambda decay and isospin asymmetry of B -> J/psi K decays} 
\def\papertitle{Measurement of the branching fraction of the 
$\Lb\to\jpsi\Lz$ decay and isospin asymmetry of $\B\to\jpsi\kaon$ decays} 
\def\paperkeywords{{High Energy Physics}, {LHCb}} 
\def\papercopyright{\the\year\ CERN for the benefit of the LHCb collaboration} 
\def\paperlicence{CC BY 4.0 licence}
\def\paperlicenceurl{https://creativecommons.org/licenses/by/4.0/}

\newif\ifEnableSectionTOCLinks
\EnableSectionTOCLinksfalse 

\usepackage[top=1in, bottom=1.25in, left=1in, right=1in]{geometry}

\usepackage{microtype}
\usepackage{lineno}  
\usepackage{xspace} 
\usepackage{caption} 

\usepackage{graphicx}  
\usepackage{color}
\usepackage{colortbl}
\graphicspath{{./figs/}} 

\usepackage{amsmath} 
\usepackage{amssymb}
\usepackage{amsfonts}
\usepackage{upgreek} 

\newcommand*\patchAmsMathEnvironmentForLineno[1]{%
\expandafter\let\csname old#1\expandafter\endcsname\csname #1\endcsname
\expandafter\let\csname oldend#1\expandafter\endcsname\csname
end#1\endcsname
 \renewenvironment{#1}%
   {\linenomath\csname old#1\endcsname}%
   {\csname oldend#1\endcsname\endlinenomath}%
}
\newcommand*\patchBothAmsMathEnvironmentsForLineno[1]{%
  \patchAmsMathEnvironmentForLineno{#1}%
  \patchAmsMathEnvironmentForLineno{#1*}%
}
\AtBeginDocument{%
\patchBothAmsMathEnvironmentsForLineno{equation}%
\patchBothAmsMathEnvironmentsForLineno{align}%
\patchBothAmsMathEnvironmentsForLineno{flalign}%
\patchBothAmsMathEnvironmentsForLineno{alignat}%
\patchBothAmsMathEnvironmentsForLineno{gather}%
\patchBothAmsMathEnvironmentsForLineno{multline}%
\patchBothAmsMathEnvironmentsForLineno{eqnarray}%
}

\usepackage[pdftex,
            pdfauthor={\paperauthors},
            pdftitle={\paperasciititle},
            pdfkeywords={\paperkeywords}]{hyperref}
\usepackage{hyperxmp}
\hypersetup{
    pdfcopyright={Copyright (C) \papercopyright},
    pdflicenseurl={\paperlicenceurl}
}

\usepackage[colorinlistoftodos,textsize=scriptsize]{todonotes}

\usepackage[bottom,flushmargin,hang,multiple]{footmisc}

\usepackage[all]{hypcap} 

\usepackage{xspace}
\usepackage{upgreek}

\def\lhcb   {\mbox{LHCb}\xspace}

\def\babar  {\mbox{BaBar}\xspace}
\def\belle  {\mbox{Belle}\xspace}

\def\cleo   {\mbox{CLEO}\xspace}
\def\cdf    {\mbox{CDF}\xspace}
\def\dzero  {\mbox{D0}\xspace}

\def\MagUp {\mbox{\em Mag\kern -0.05em Up}\xspace}

\ifthenelse{\boolean{uprightparticles}}%
{

 \def\Pmu         {\ensuremath{\upmu}\xspace}

 \def\Ppi         {\ensuremath{\uppi}\xspace}

 \def\Ppsi        {\ensuremath{\uppsi}\xspace}

 \def\PDelta      {\ensuremath{\Delta}\xspace}
 \def\PXi         {\ensuremath{\Xi}\xspace}
 \def\PLambda     {\ensuremath{\Lambda}\xspace}
 \def\PSigma      {\ensuremath{\Sigma}\xspace}
 \def\POmega      {\ensuremath{\Omega}\xspace}
 \def\PUpsilon    {\ensuremath{\Upsilon}\xspace}
 \let\oldPi\Pi
 \def\PPi         {\ensuremath{\oldPi}\xspace}

 \def\PB      {\ensuremath{\mathrm{B}}\xspace}
 \def\PD      {\ensuremath{\mathrm{D}}\xspace}
 \def\PJ      {\ensuremath{\mathrm{J}}\xspace}
 \def\PK      {\ensuremath{\mathrm{K}}\xspace}
 \def\Pb      {\ensuremath{\mathrm{b}}\xspace}
 \def\Pc      {\ensuremath{\mathrm{c}}\xspace}

 \def\Pp      {\ensuremath{\mathrm{p}}\xspace}

 \def\Ps      {\ensuremath{\mathrm{s}}\xspace}

 \def\thebaroffset{0.0em}
}
{

 \def\Pmu         {\ensuremath{\mu}\xspace}

 \def\Ppi         {\ensuremath{\pi}\xspace}

 \def\Ppsi        {\ensuremath{\psi}\xspace}
 
 \mathchardef\PDelta="7101
 \mathchardef\PXi="7104
 \mathchardef\PLambda="7103
 \mathchardef\PSigma="7106
 \mathchardef\POmega="710A
 \mathchardef\PUpsilon="7107
 \mathchardef\PPi="7105
 \def\PB      {\ensuremath{B}\xspace}
 \def\PD      {\ensuremath{D}\xspace}
 \def\PJ      {\ensuremath{J}\xspace}
 \def\PK      {\ensuremath{K}\xspace}
 \def\Pb      {\ensuremath{b}\xspace}
 \def\Pc      {\ensuremath{c}\xspace}

 \def\Pp      {\ensuremath{p}\xspace}

 \def\Ps      {\ensuremath{s}\xspace}

 \def\thebaroffset{0.18em}
}
\newcommand{\offsetoverline}[2][\thebaroffset]{\kern #1\overline{\kern -#1 #2}}%

\makeatletter
\ifcase \@ptsize \relax
  \newcommand{\miniscule}{\@setfontsize\miniscule{4}{5}}
\or
  \newcommand{\miniscule}{\@setfontsize\miniscule{5}{6}}
\or
  \newcommand{\miniscule}{\@setfontsize\miniscule{5}{6}}
\fi
\makeatother

\DeclareRobustCommand{\optbar}[1]{\shortstack{{\miniscule (\rule[.5ex]{1.25em}{.18mm})}
  \\ [-.7ex] $#1$}}

\def\mumu       {{\ensuremath{\Pmu^+\Pmu^-}}\xspace}

\def\squark    {{\ensuremath{\Ps}}\xspace}

\def\cquark    {{\ensuremath{\Pc}}\xspace}

\def\bquark    {{\ensuremath{\Pb}}\xspace}

\def\pion   {{\ensuremath{\Ppi}}\xspace}

\def\pip    {{\ensuremath{\pion^+}}\xspace}
\def\pim    {{\ensuremath{\pion^-}}\xspace}

\def\kaon    {{\ensuremath{\PK}}\xspace}
\def\Kbar    {{\ensuremath{\offsetoverline{\PK}}}\xspace}

\def\KorKbar {\kern \thebaroffset\optbar{\kern -\thebaroffset \PK}{}\xspace}
\def\Kz      {{\ensuremath{\kaon^0}}\xspace}
\def\Kzb     {{\ensuremath{\Kbar{}^0}}\xspace}
\def\Kp      {{\ensuremath{\kaon^+}}\xspace}
\def\Km      {{\ensuremath{\kaon^-}}\xspace}
\def\Kpm     {{\ensuremath{\kaon^\pm}}\xspace}

\def\KS      {{\ensuremath{\kaon^0_{\mathrm{S}}}}\xspace}

\def\D       {{\ensuremath{\PD}}\xspace}

\def\DorDbar {\kern \thebaroffset\optbar{\kern -\thebaroffset \PD}\xspace}
\def\Dz      {{\ensuremath{\D^0}}\xspace}

\def\Dp      {{\ensuremath{\D^+}}\xspace}
\def\Dm      {{\ensuremath{\D^-}}\xspace}

\def\DpDm    {\ensuremath{\Dp {\kern -0.16em \Dm}}\xspace}

\def\Dstarp  {{\ensuremath{\D^{*+}}}\xspace}

\def\B       {{\ensuremath{\PB}}\xspace}
\def\Bbar    {{\ensuremath{\offsetoverline{\PB}}}\xspace}

\def\BorBbar {\kern \thebaroffset\optbar{\kern -\thebaroffset \PB}\xspace}
\def\Bz      {{\ensuremath{\B^0}}\xspace}
\def\Bzb     {{\ensuremath{\Bbar{}^0}}\xspace}
\def\Bd      {{\ensuremath{\B^0}}\xspace}

\def\BdorBdbar {\kern \thebaroffset\optbar{\kern -\thebaroffset \Bd}\xspace}
\def\Bu      {{\ensuremath{\B^+}}\xspace}
\def\Bub     {{\ensuremath{\B^-}}\xspace}
\def\Bp      {{\ensuremath{\Bu}}\xspace}
\def\Bm      {{\ensuremath{\Bub}}\xspace}

\def\Bs      {{\ensuremath{\B^0_\squark}}\xspace}

\def\BsorBsbar {\kern \thebaroffset\optbar{\kern -\thebaroffset \Bs}\xspace}

\def\jpsi     {{\ensuremath{{\PJ\mskip -3mu/\mskip -2mu\Ppsi}}}\xspace}

\def\Y#1S{\ensuremath{\PUpsilon{(#1S)}}\xspace}

\def\FourS {{\Y4S}\xspace}

\def\proton      {{\ensuremath{\Pp}}\xspace}

\def\Lz          {{\ensuremath{\PLambda}}\xspace}

\def\LorLbar     {\kern \thebaroffset\optbar{\kern -\thebaroffset \PLambda}\xspace}

\def\Xires       {{\ensuremath{\PXi}}\xspace}
\def\Xiz         {{\ensuremath{\Xires^0}}\xspace}
\def\Xim         {{\ensuremath{\Xires^-}}\xspace}

\def\Lb           {{\ensuremath{\Lz^0_\bquark}}\xspace}

\def\Xib          {{\ensuremath{\Xires_\bquark}}\xspace}
\def\Xibz         {{\ensuremath{\Xires^0_\bquark}}\xspace}

\def\BF         {{\ensuremath{\mathcal{B}}}\xspace}

\newcommand{\decay}[2]{\mbox{\ensuremath{#1\!\to #2}}\xspace}

\def\to                 {\ensuremath{\rightarrow}\xspace}

\def\AT#1     {\ensuremath{A_{\mathrm{T}}^{#1}}\xspace}           

\def\C#1      {\ensuremath{\mathcal{C}_{#1}}\xspace}                       
\def\Cp#1     {\ensuremath{\mathcal{C}_{#1}^{'}}\xspace}                    
\def\Ceff#1   {\ensuremath{\mathcal{C}_{#1}^{\mathrm{(eff)}}}\xspace}        
\def\Cpeff#1  {\ensuremath{\mathcal{C}_{#1}^{'\mathrm{(eff)}}}\xspace}       
\def\Ope#1    {\ensuremath{\mathcal{O}_{#1}}\xspace}                       
\def\Opep#1   {\ensuremath{\mathcal{O}_{#1}^{'}}\xspace}                    

\newcommand{\aunit}[1]{\ensuremath{\text{\,#1}}}

\newcommand{\tev}{\aunit{Te\kern -0.1em V}\xspace}
\newcommand{\gev}{\aunit{Ge\kern -0.1em V}\xspace}
\newcommand{\mev}{\aunit{Me\kern -0.1em V}\xspace}
\newcommand{\kev}{\aunit{ke\kern -0.1em V}\xspace}
\newcommand{\ev}{\aunit{e\kern -0.1em V}\xspace}

\newcommand{\mevc}{\ensuremath{\aunit{Me\kern -0.1em V\!/}c}\xspace}
\newcommand{\gevc}{\ensuremath{\aunit{Ge\kern -0.1em V\!/}c}\xspace}
\newcommand{\mevcc}{\ensuremath{\aunit{Me\kern -0.1em V\!/}c^2}\xspace}
\newcommand{\gevcc}{\ensuremath{\aunit{Ge\kern -0.1em V\!/}c^2}\xspace}

\def\mm   {\aunit{mm}\xspace}

\def\fb   {\ensuremath{\aunit{fb}}\xspace}
\def\invfb   {\ensuremath{\fb^{-1}}\xspace}

\newcommand{\chisq}{\ensuremath{\chi^2}\xspace}

\def\gsim{{~\raise.15em\hbox{$>$}\kern-.85em
          \lower.35em\hbox{$\sim$}~}\xspace}
\def\lsim{{~\raise.15em\hbox{$<$}\kern-.85em
          \lower.35em\hbox{$\sim$}~}\xspace}

\def\pt         {\ensuremath{p_{\mathrm{T}}}\xspace}

\def\evtgen     {\mbox{\textsc{EvtGen}}\xspace}

\def\geant      {\mbox{\textsc{Geant4}}\xspace}

\def\photos     {\mbox{\textsc{Photos}}\xspace}

\def\pythia     {\mbox{\textsc{Pythia}}\xspace}

\def\tell1  {TELL1\xspace}
\def\ukl1   {UKL1\xspace}

\newcommand{\eg}{\mbox{\itshape e.g.}\xspace}

\newcommand{\lhcborcid}[1]{\href{https://orcid.org/#1}{\hspace*{0.1em}\raisebox{-0.45ex}{\includegraphics[width=1em]{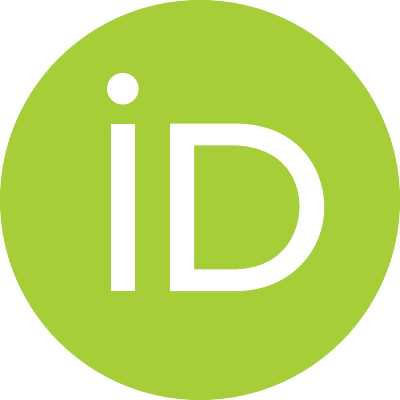}}}}

\hypersetup{
  colorlinks   = true, 
  urlcolor     = blue, 
  linkcolor    = blue, 
  citecolor    = red   
}

\ifEnableSectionTOCLinks
    \usepackage[explicit]{titlesec} 
    
    \let\oldcontentsline\contentsline
    \renewcommand

    \titleformat{\section}{\normalfont\Large\bf}{\hyperlink{tocsection.\thesection}{{\thesection} \parbox[t]{\dimexpr\textwidth-1pc}{#1}}}{1pc}{}

    \titleformat{\subsection}{\normalfont\bf}{\hyperlink{tocsubsection.\thesubsection}{{\thesubsection} \parbox[t]{\dimexpr\textwidth-1pc}{#1}}}{1pc}{}

    \titleformat{name=\section,numberless}[display]{}{}{0pt}{\normalfont\Huge\bfseries #1}
\fi

\usepackage{cite} 
\usepackage{mciteplus}

\def\LbJpsiLambda {\decay{\Lb}{\jpsi\Lz}} 
\def\BdJpsiKS     {\decay{\Bd}{\jpsi\KS}}
\def\BdJpsiKZ     {\decay{\Bd}{\jpsi\Kz}}
\def\BsJpsiKS     {\decay{\Bs}{\jpsi\KS}}
\def\BuJpsiK      {\decay{\Bp}{\jpsi\Kp}}
\def\BuJpsiPi     {\decay{\Bp}{\jpsi\pip}}
\def\XibJpsiXi    {\decay{\Xib^{+,0}}{\jpsi\Xires^{+,0}}}
\def\XibzJpsiXiz  {\decay{\Xibz}{\jpsi\Xiz}}

\def\XibJpsiLambda{\decay{\Xibz}{\jpsi\Lz}}
\def\BsMuMu       {\decay{\Bs}{\mumu}} 
 
\def\LambdaPPi    {\decay{\Lz}{\proton\pim}} 
\def\KSPiPi       {\decay{\KS}{\pip\pim}}

\def\AI           {\ensuremath{A_{\rm I}}\xspace}
\def\fracLb       {\ensuremath{f(\bquark\to\Lb)}\xspace}
\def\fracBu       {\ensuremath{f(\bquark\to\Bp)}\xspace}
\def\fracBd       {\ensuremath{f(\bquark\to\Bz)}\xspace}
\def\fracBs       {\ensuremath{f(\bquark\to\Bs)}\xspace}

\usepackage{booktabs}

\begin{document}

\renewcommand{\thefootnote}{\fnsymbol{footnote}}
\setcounter{footnote}{1}


\begin{titlepage}
\pagenumbering{roman}

\vspace*{-1.5cm}
\centerline{\large EUROPEAN ORGANIZATION FOR NUCLEAR RESEARCH (CERN)}
\vspace*{1.5cm}
\noindent
\begin{tabular*}{\linewidth}{lc@{\extracolsep{\fill}}r@{\extracolsep{0pt}}}
\ifthenelse{\boolean{pdflatex}}
{\vspace*{-1.5cm}\mbox{\!\!\!\includegraphics[width=.14\textwidth]{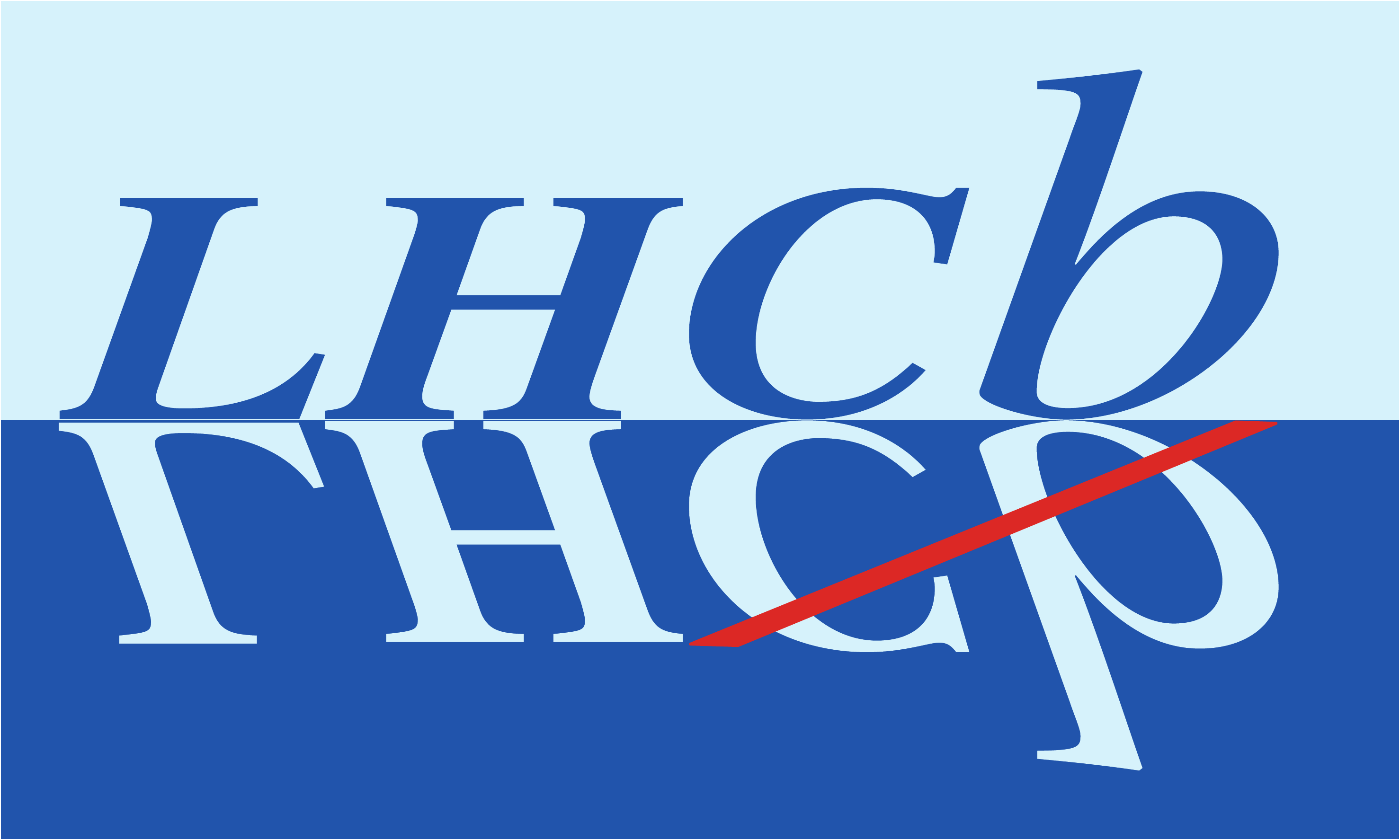}} & &}%
{\vspace*{-1.2cm}\mbox{\!\!\!\includegraphics[width=.12\textwidth]{figs/lhcb-logo.eps}} & &}%
\\
 & & CERN-EP-2025-183 \\  
 & & LHCb-PAPER-2025-035 \\  
 & & March 23, 2026 \\ 
 & & \\
\end{tabular*}

\vspace*{3.0cm}

{\normalfont\bfseries\boldmath\huge
\begin{center}
  \papertitle 
\end{center}
}

\vspace*{1.0cm}

\begin{center}
\paperauthors\footnote{Authors are listed at the end of this paper.}
\end{center}

\vspace{\fill}

\begin{abstract}
  \noindent
   This paper describes a measurement of the \LbJpsiLambda branching fraction using data collected with the LHCb experiment in proton-proton collisions from 2016 to 2018. 
   The dataset corresponds to an integrated luminosity of 5.4\invfb.
   The branching fraction is determined relative to that of \BdJpsiKS decays,
    \begin{displaymath}
    \frac{\BF(\LbJpsiLambda)}{\BF(\BdJpsiKS)} = 0.750 \pm 0.005 \pm 0.022 \pm 0.005 \pm 0.062\,,
    \end{displaymath}
   yielding $\BF(\LbJpsiLambda) = (3.34 \pm 0.02 \pm 0.10 \pm 0.08 \pm 0.28)\times 10^{-4}$, 
   where the first uncertainty is statistical, the second systematic, the third due to external inputs on branching fractions and the fourth due to the ratio of \Lb baryon and \Bd meson hadronisation fractions. 
   In addition, the isospin asymmetry between the rates of \BdJpsiKS and \BuJpsiK decays is measured to be 
   \begin{displaymath}
    A_{\rm I} = -0.0135 \pm 0.0004 \pm 0.0133\,,
    \end{displaymath}
    where the first uncertainty is statistical and the second systematic. 
\end{abstract}

\vspace*{1.5cm}

\begin{center}
  Published in JHEP 01 (2026) 159
\end{center}

\vspace{\fill}

{\footnotesize 
\centerline{\copyright~\papercopyright. \href{\paperlicenceurl}{\paperlicence}.}}
\vspace*{2mm}

\end{titlepage}


\newpage
\setcounter{page}{2}
\mbox{~}



\renewcommand{\thefootnote}{\arabic{footnote}}
\setcounter{footnote}{0}


\cleardoublepage


\pagestyle{plain} 
\setcounter{page}{1}
\pagenumbering{arabic}


\section{Introduction}
\label{sec:introduction}

Decays of \bquark hadrons to final states with \jpsi mesons provide clean experimental signatures that have been used extensively to calibrate measurements at the \B-factory experiments, the LHC and the Tevatron.
At the Tevatron and the LHC, these decays have also played an important role in normalising  event yields in analyses of rare or forbidden processes, such as the \BsMuMu decay~\cite{LHCb-PAPER-2021-007,LHCb-PAPER-2021-008,CMS:2022mgd,ATLAS:2018cur}.
The branching fractions of the \BdJpsiKS and \BuJpsiK decays are known precisely from measurements by the \babar, \belle and \cleo experiments~\cite{BELLE:2019xld,BaBar:2004htr,Belle:2017psv,Belle:2019avj,CLEO:1997ilq,CLEO:2000emb}.\footnote{
The inclusion of charge-conjugate processes is implied throughout.}
The branching fractions of \bquark-baryon decays to final states with \jpsi mesons are known much less precisely as \bquark baryons are not producible in the $B$-factory experiments. 
At the Tevatron, the \dzero~\cite{D0:2011pqa} and \cdf~\cite{CDF:1996rvy} experiments have  measured the product of the fraction of \bquark quarks that hadronise to form a \Lb baryon and the branching fraction of the \LbJpsiLambda decay, $\fracLb \times \BF(\LbJpsiLambda)$. 
The \Lb production fraction is known to be strongly \pt-dependent and different assumptions on \fracLb have a significant impact on the determination of $\BF(\LbJpsiLambda)$~\cite{Blake:2019guk}. 
The \pt dependence of \fracLb has been studied by the LHCb collaboration in $pp$ collisions at the LHC~\cite{LHCb-PAPER-2011-018,LHCb-PAPER-2018-050}, allowing for a more precise determination of $\BF(\LbJpsiLambda)$.

This paper describes a measurement of the \LbJpsiLambda branching fraction using a dataset collected by the LHCb experiment from 2016 to 2018 in proton-proton collisions at a centre-of-mass energy of $\sqrt{s} = 13\tev$.
The dataset corresponds to an integrated luminosity of 5.4\invfb.  
To account for the \pt dependence of the \bquark-hadron production, and to validate the analysis procedure, the data are analysed in bins of \pt using intervals of 1\gevc in the range $[4,14]$\gevc and 2\gevc in the range $[14,20]$\gevc. 
The observed yields of \LbJpsiLambda decays are normalised with respect to those of \BdJpsiKS decays, exploiting the similar topology of the two decays in the detector when the \KS meson and \Lz baryon are reconstructed through the decays \KSPiPi and \LambdaPPi.

A comparison of the rates of \BuJpsiK and \BdJpsiKS decays as a function of the \B-meson \pt is used to validate the reconstruction and selection of the long-lived hadrons in the LHCb detector.
In the Standard Model, it is expected that the partial widths of the \BuJpsiK and \BdJpsiKZ decays are almost identical, as the only difference between the decays is the flavour of the light spectator quark (see \eg Ref.~\cite{Lyon:2013gba}).
The resulting isospin asymmetry, 
\begin{align}
    \AI[\decay{B}{\jpsi K}] = 
    \frac{\tfrac{\tau_\Bp}{\tau_\Bz}\BF(\BdJpsiKZ) - \BF(\BuJpsiK)}
    {\tfrac{\tau_\Bp}{\tau_\Bz}\BF(\BdJpsiKZ) + \BF(\BuJpsiK)}\,,
\end{align}
where $\tau_\Bp/\tau_\Bz  = 1.076 \pm 0.004$~\cite{HFLAV21} is the ratio between the \Bp and \Bz meson lifetimes, is expected to be zero. 
Previous measurements of \AI by the \belle~\cite{BELLE:2019xld} and \belle~II~\cite{Belle-II:2022dbo} collaborations and the reinterpretation of \babar and \belle data in Ref.~\cite{Jung:2015yma} are consistent with this expectation. 
In this analysis, \AI is determined from the \BuJpsiK and \BdJpsiKS samples assuming $\BF(\BdJpsiKS) = \tfrac{1}{2}\BF(\BdJpsiKZ)$. 
Kaon regeneration in detector material, and neutral-kaon mixing modify this assumption at the per-mille level~\cite{Fetscher:1996fa}.
It is assumed that \fracBd and \fracBu are equal for $pp$ collisions at $\sqrt{s}=13\tev$. 

\section{Detector and simulation} 
\label{sec:detector}

The LHCb detector~\cite{LHCb-DP-2008-001,LHCb-DP-2014-002} is a single-arm forward spectrometer covering the pseudorapidity range $2 < \eta < 5$, designed for the study of particles containing \bquark\ or \cquark\ quarks. 
The detector elements that are particularly relevant to this analysis are: a silicon-strip vertex detector surrounding the proton-proton  interaction region that allows \cquark\ and \bquark\ hadrons to be identified from their characteristically long flight distance; a tracking system that provides a measurement of the momentum, $p$, of charged particles; two ring-imaging Cherenkov detectors (referred to as RICH\,1 and RICH\,2) that are able to discriminate between different species of charged hadrons; an electromagnetic calorimeter to reconstruct electron and photon energy deposits; a muon system to identify muons. 
Events are selected online using a trigger~\cite{LHCb-DP-2012-004,LHCb-DP-2019-001} that comprises a hardware and a software stage.
In preparation for analysis, they undergo a centralised, offline processing step to deliver physics-analysis-ready data across the entire \lhcb physics programme~\cite{Stripping}.

Simulated samples are used to determine the efficiency of the candidate selection as well as the distributions of specific sources of background. 
In the simulation, $pp$ collisions are generated using \pythia~\cite{Sjostrand:2007gs,*Sjostrand:2006za} with a specific \lhcb configuration~\cite{LHCb-PROC-2010-056}.
Decays of unstable particles are described by \evtgen~\cite{Lange:2001uf}, using \photos~\cite{davidson2015photos} to generate final-state radiation.
The interactions of the generated particles with the detector, and its response, are simulated using the \geant toolkit~\cite{Allison:2006ve, *Agostinelli:2002hh} as described in Ref.~\cite{LHCb-PROC-2011-006}.
The simulated samples are corrected to account for known differences between simulation and data. 
Corrections are applied to the simulated tracking and particle-identification performance, and to the production kinematics of the \bquark hadrons. 
The kinematic corrections are derived from \BuJpsiK decays that satisfy the candidate selection procedure described below.
It is assumed that \Bz and \Bp mesons are produced with the same kinematic distributions. 
The measured hadronisation fractions from Ref.~\cite{LHCb-PAPER-2011-018,LHCb-PAPER-2018-050} are used to derive kinematic corrections for \Lb baryons from those of \Bp mesons. 
The angular distributions of simulated \LbJpsiLambda decays are weighted to reproduce those in Ref.~\cite{LHCb-PAPER-2020-005}. 

\section{Candidate selection} 
\label{sec:selection} 

At the hardware trigger level, events are selected if they contain at least one high-\pt muon.  
In the software trigger, at least one muon is required to have a large \pt and a large impact parameter (IP) with respect to every $pp$ interaction point (PV). Further, events are required to contain a pair of opposite-charge muons that are displaced from every PV.

In the offline selection, \Bp, \Bz and \Lb candidates are formed by combining a muon pair consistent with originating from the decay of a \jpsi meson with a \Kp meson, \KS meson, or \Lz baryon candidate, respectively.
The \jpsi, \KS and \Lz candidates are required to have masses in the ranges $[3030, 3150]$\mevcc, $[468,528]$\mevcc and  $[1105,1125]$\mevcc, respectively.   
The \KS and \Lz candidates are formed from two opposite-charge tracks and are split into two categories depending on whether or not the tracks are reconstructed in the vertex detector: tracks reconstructed with (without) segments in the vertex detector are categorised as \emph{long}~(\emph{downstream}). 
Candidates in the long (downstream) category are required to have a decay vertex displaced along the beam axis from the origin of the experiment's coordinate system by less than 400\mm (2250\mm), to ensure they are well reconstructed. 
The motivation for these requirements is twofold: long-category candidates with displacements larger than 400\mm leave only a small number of hits in the vertex detector, making their reconstruction efficiency highly sensitive to detector inefficiencies, while downstream-category candidates with displacements beyond 2250\mm are affected by material interactions at the interface between RICH\,1 and the tracking system. 
The \Kp candidate is required to be positively identified as a kaon by the detector's particle-identification systems. 
No particle-identification requirements are applied to the tracks used to form the \KS and \Lz candidates, as the narrow mass window requirements around the known \KS and \Lz masses~\cite{PDG2024} reduce the level of cross-feed background from \KS mesons identified as \Lz baryons, and vice versa, to a manageable level. 
All reconstructed tracks are required to have a significant IP with respect to every PV in the event. 
The $\pip\pim$, $\proton\pim$  and $\mumu$ candidates are required to be consistent with originating from vertices corresponding to the decays of the \KS meson, \Lz baryon or \jpsi meson, respectively. 
The decay products of the \bquark hadron are also required to be consistent with originating from a single point.  
The \bquark-hadron candidate is required to be displaced from every PV in the event and be consistent with originating from a PV.
To improve the resolution on the reconstructed \bquark-hadron mass, a kinematic fit is performed~\cite{Hulsbergen:2005pu} that  constrains the \jpsi mass and \KS or \Lz mass to their known values~\cite{PDG2024}, and constrains the \bquark hadron to originate from its associated PV, defined as the PV with the smallest IP with respect to the \bquark-hadron candidate. 

Gradient-boosted decision-tree classifiers~\cite{Chen:2016btl} are used to further suppress combinatorial background, where tracks from multiple different sources are combined, in the \Bz and \Lb long-category datasets. 
This is not necessary for the downstream-category datasets, which already have good signal purity. 
A separate classifier is trained for each dataset, using \Bz and \Lb candidates from simulation as a signal proxy and data candidates with a \bquark-hadron mass in the high-mass range $[5900,7000]$\mevcc as a background proxy. 
The classifier uses information on the displacement of the \KS/\Lz and the \bquark-hadron candidates from their associated PVs, the \KS/\Lz \pt, the vertex-fit quality of the \bquark-hadron candidate and the angle between its momentum vector and the vector connecting its production and decay vertices.  
The working points of the classifiers are chosen to maximise the expected statistical significances of the \Bz and \Lb signals in the data. 
The expected signal yield at each working point is obtained by scaling the yield without any classifier requirement by the relative efficiency of the working point in simulation. 
The expected background yield is determined from a fit to the  upper-mass sideband region of the data. 
After applying the full selection procedure, 99\% of the selected events contain only a single \bquark-hadron candidate. 
If more than one candidate is reconstructed in an event, a single candidate is selected arbitrarily. 

In addition to combinatorial background, the other sources of background in the selected samples are: 
\LbJpsiLambda (\BdJpsiKS) decays that are  reconstructed as \BdJpsiKS (\LbJpsiLambda) decays in the \Bz (\Lb) sample; 
\BuJpsiPi decays where the pion is misidentified as a kaon in the \Bp sample; 
\BsJpsiKS decays in the \Bz sample;
and \XibJpsiXi decays, where the $\Xires^{+,0}$ baryon decays to $\Lz\pi^{+,0}$ and the pion is not reconstructed, in the \Lb sample. 
It is also possible for \Xibz baryons to decay directly to $\jpsi\Lz$, but the branching fraction of this process is sufficiently small, $\BF(\XibJpsiLambda)/\BF(\XibzJpsiXiz) = (8.2 \pm 2.1 \pm 0.9) \times 10^{-3}$~\cite{LHCb-PAPER-2019-039}, that its contribution can be safely neglected.

\section{Signal yield determination} 
\label{sec:fitting} 

\begin{figure}[!htb]
    \centering
    \includegraphics[width=0.65\linewidth]{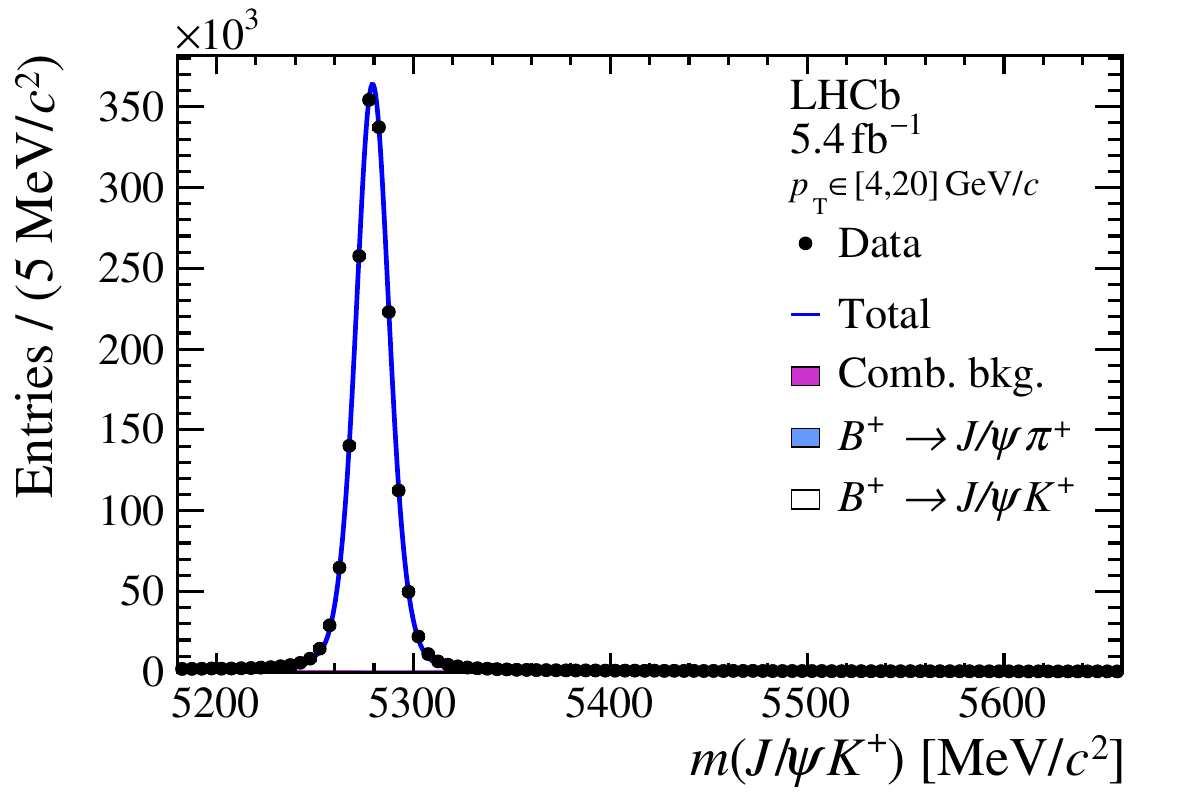} \\
    \includegraphics[width=0.65\linewidth]{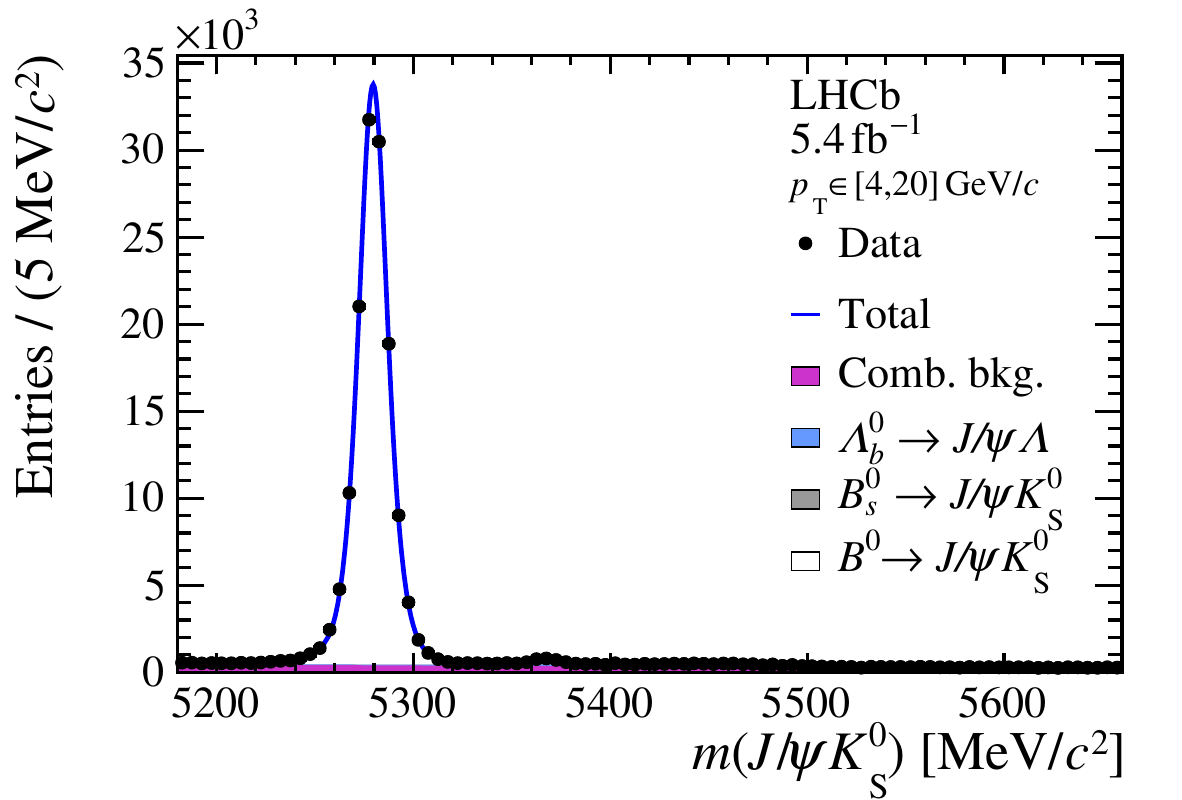} \\
    \includegraphics[width=0.65\linewidth]{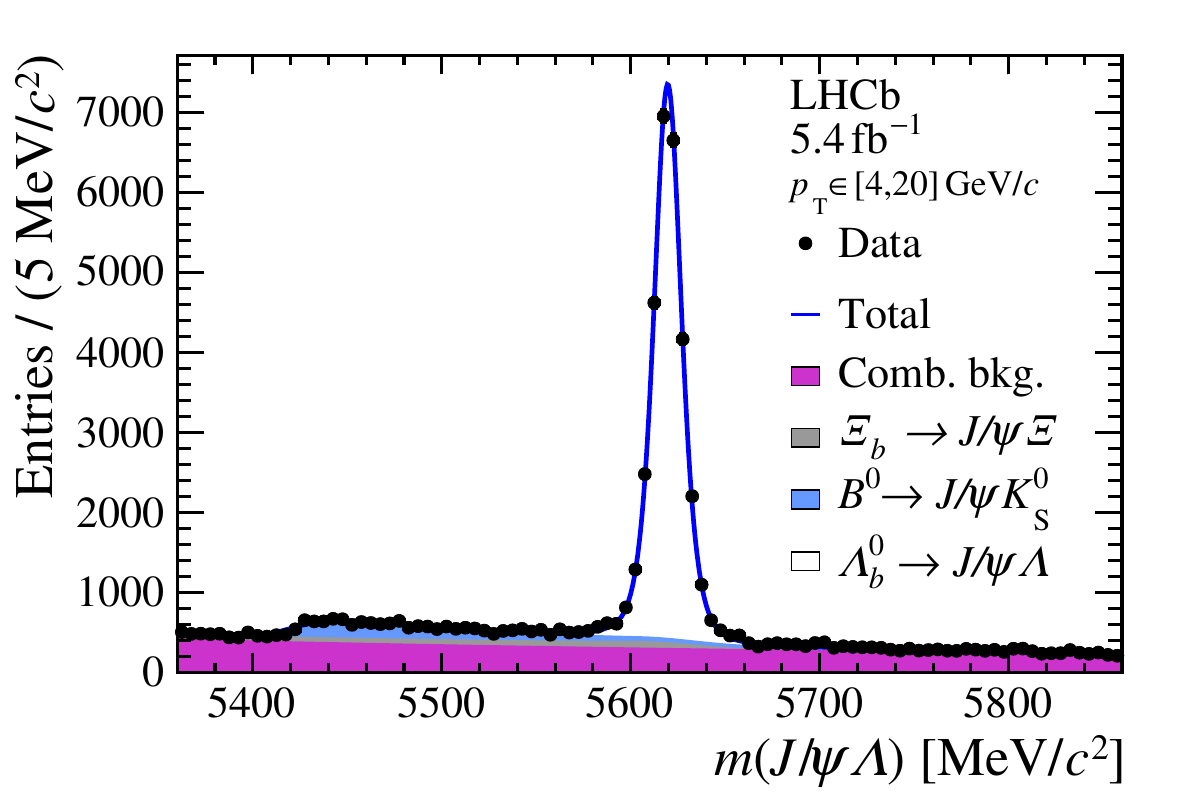}
    \caption{
    Mass distributions of selected (top) $\jpsi\Kp$, (middle) $\jpsi\KS$ and (bottom) $\jpsi\Lz$   candidates obtained by combining the data in the \pt range [4, 20]\gevc. 
    The line and shaded regions indicate the result of the fits and are obtained by summing the results of the fits to the individual \pt bins. 
    }
    \label{fig:combined}
\end{figure}

The yields of the \BuJpsiK, \BdJpsiKS and \LbJpsiLambda decays are determined in bins of \bquark-hadron \pt by performing extended unbinned maximum-likelihood fits to the reconstructed $\jpsi\Kp$, $\jpsi\KS$ and $\jpsi\Lz$ mass distributions, respectively.  
The fits are performed using the {\textsc RooFit} software package~\cite{roofit}. 
The mass distributions for the different \pt bins are shown in Figs.~\ref{fig:Kp}--\ref{fig:Lambda:downstream} in the Appendix. 
The distributions obtained by combining the \pt bins are shown in Fig.~\ref{fig:combined}. 
The \Bz and \Lb decay samples are fit simultaneously to take into account cross-feed from \BdJpsiKS decays reconstructed in the  \LbJpsiLambda sample and vice versa. 
The long and downstream samples are included separately in the simultaneous fit.

In the fits, the \Bp, \Bz and \Lb signals are described using Hypatia functions~\cite{Santos:2013gra} with tail parameters fixed from simulation. 
The peak positions and widths of the functions are allowed to vary in the fits to data separately in each bin of the \bquark-hadron \pt. 
Combinatorial background is described using exponential functions. 
The cross-feed backgrounds from \Bz decays in the \Lb sample and \Lb decays in the \Bz sample are described using Johnson $S_U$ distributions~\cite{JohnsonSU}.  
Their shape parameters are fixed from simulated \BdJpsiKS and \LbJpsiLambda decays. 
The parameters of the signal and background shapes are allowed to differ between \pt bins and decay modes. 
The yield of the \Lb (\Bz) cross-feed background in the \Bz (\Lb) samples is constrained from the corresponding yield of \Lb (\Bz) decays in the \Lb (\Bz) samples and the efficiency to correctly/incorrectly reconstruct the \Lb (\Bz) candidates from simulation. 

The background from \BuJpsiPi decays in the \Bp sample is described by a Gaussian function with power-law substituted tails on the left- and right-hand side of the distribution~\cite{Skwarnicki:1986xj}, with shape parameters fixed from simulated \BuJpsiPi decays. 
The yield of this background is constrained by the yield of the \Bp signal, the known branching-fraction ratio $\BF(\BuJpsiPi)/\BF(\BuJpsiK)$~\cite{PDG2024}, and the relative efficiency to reconstruct and select \BuJpsiPi and \BuJpsiK decays in simulation.

The background from \BsJpsiKS in the \Bz sample is modelled using the same Hypatia function describing the \Bz signal, but with a peak position shifted by the known \Bs--\Bz mass difference of $87.37$\mevcc~\cite{PDG2024}. 
The yield of this background is constrained relative to that of the \Bz signal using the branching-fraction ratio $\BF(\BsJpsiKS)/\BF(\BdJpsiKS)$ from Ref.~\cite{LHCB-PAPER-2015-005}, the production fraction ratio \mbox{$\fracBs/\fracBd$} from Ref.~\cite{LHCb-PAPER-2018-050} and the relative efficiency to reconstruct and select \Bs and \Bz decays.
The latter is determined from simulation.

The background from \XibJpsiXi decays in the \Lb sample is described by a Johnson $S_U$ distribution with parameters determined from simulation. 
Due to the long lifetime of the \Xim and \Xiz baryons, the contribution of these backgrounds to the long-track samples is very small and is neglected. 
The contribution in the downstream-track sample is also suppressed by the long lifetime. The yield of \XibJpsiXi decays is expected to be ${\cal O}(1\%)$ of that of the \LbJpsiLambda signal based on Ref.~\cite{LHCb-PAPER-2019-039}, and is allowed to vary freely in the fit.

\section{Branching fraction and isospin asymmetry determination} 
\label{sec:branching} 

\begin{figure}[!tb]
    \centering
    \includegraphics[width=0.48\linewidth]{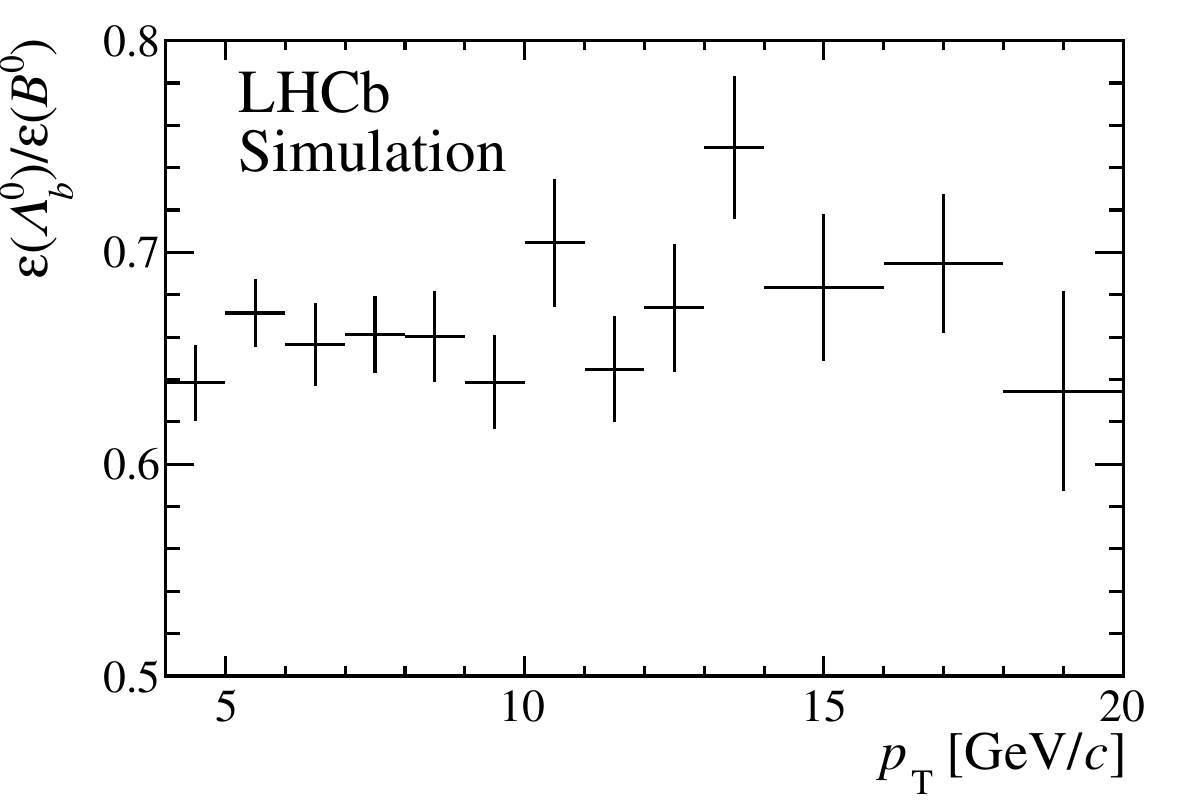}
    \includegraphics[width=0.48\linewidth]{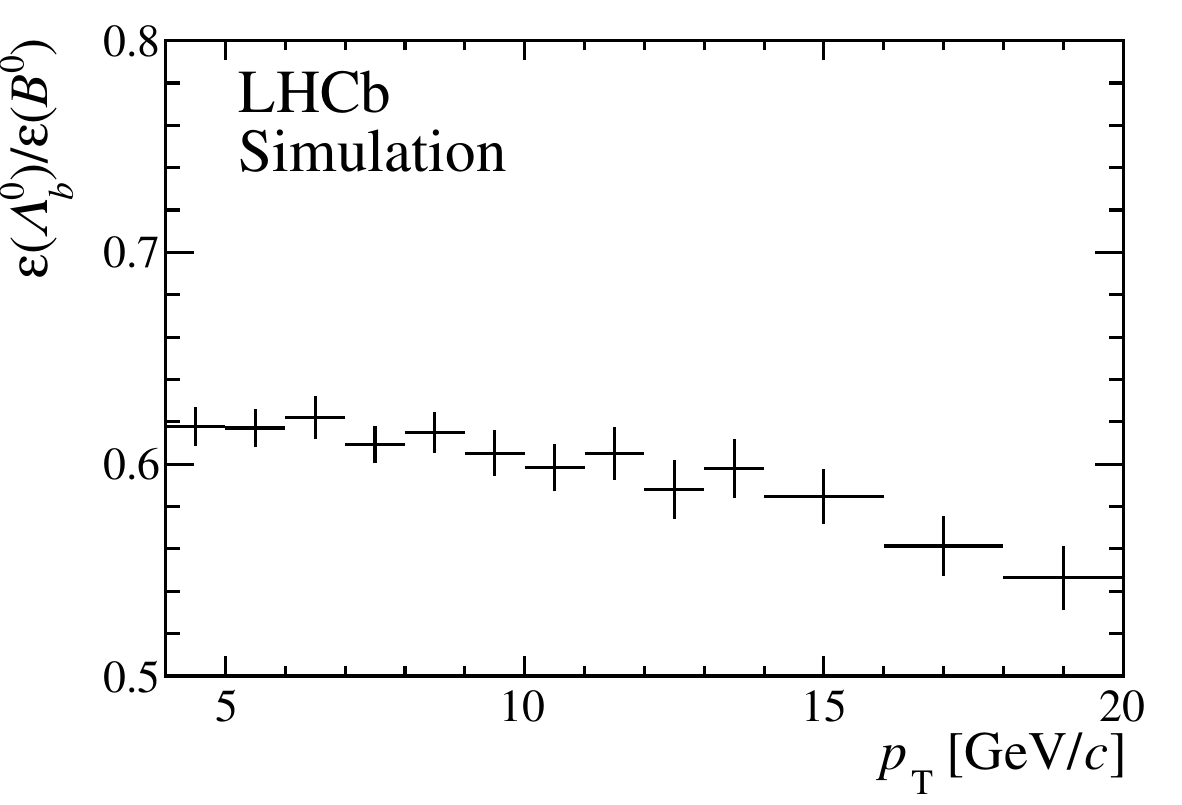}
    \caption{
    Efficiency ratio between \LbJpsiLambda and \BdJpsiKS decays determined from corrected simulation samples in intervals of \bquark-hadron transverse momentum for the (left) long-track and (right) downstream-track samples.
    }
    \label{fig:efficiency}
\end{figure}

The ratio of the yields of the \LbJpsiLambda and \BdJpsiKS decays in a \bquark-hadron \pt interval can be expressed in terms of branching fractions and production fractions as 
\begin{align}
   \frac{N[\Lb]}{N[\Bz]} = 
   \frac{\fracLb}{\fracBd} 
   \frac{\BF(\LbJpsiLambda)}{\BF(\BdJpsiKS)}  
   \frac{\BF(\LambdaPPi)}{\BF(\KSPiPi)} 
   \frac{\epsilon[\Lb]}{\epsilon[\Bz]}\,.
\end{align}
Here, $\epsilon$ is the efficiency to reconstruct and select the given decay determined from simulation. 
The efficiencies are \pt-dependent mainly due to the long lifetimes of the \Lz and \KS hadrons, and a correlation between the \bquark-hadron \pt and the typical boost of final-state particles along the beam axis. 
The ratio $\epsilon[\Lb]/\epsilon[\Bz]$ is less \pt dependent and is shown separately for the long and downstream categories in Fig.~\ref{fig:efficiency}.

The known values of the \KS and \Lz branching fractions are \mbox{$\BF(\KSPiPi) = 0.6920 \pm 0.0005$} and \mbox{$\BF(\LambdaPPi) = 0.641 \pm 0.005$}~\cite{PDG2024}, respectively.
The production fraction of \Lb baryons in $pp$ collisions has previously been studied by the LHCb collaboration at $\sqrt{s} = 13\tev$ using semileptonic decays in Ref.~\cite{LHCb-PAPER-2018-050}.  
The ratio of \fracLb and the sum of \fracBd and \fracBu is well described by the function $r(\pt)=A( p_1 +  \text{exp}( p_2 + p_3 \cdot \pt ))$, with $A = 1.000\pm 0.061$, $p_1 = (7.93 \pm 1.41)\times 10^{-2}$, $p_2 = -1.022 \pm 0.047$ and $p_3 = -0.107 \pm 0.002\gev^{-1} c$. 
The ratio does not show any significant dependence on pseudorapidity~\cite{LHCb-PAPER-2018-050}. 
Assuming isospin symmetry in high-energy $pp$ collisions, $\fracLb/\fracBd$ is twice the value given by this function.

The \LbJpsiLambda branching fraction is determined by fitting the ratio of the \Lb and \Bz yields as a function of \pt simultaneously between the long and downstream categories.  
In this fit, the efficiency ratio, the parameter $A$, and the external branching fractions of the \Lz and \KS decays are allowed to vary around their measured values with their uncertainties as a Gaussian prior, while the parameters $p_1$, $p_{2}$ and $p_{3}$ of $r(\pt)$ are constrained according to the covariance matrix reported in Ref.~\cite{LHCb-PAPER-2018-050}.
The analysis is validated using pseudoexperiments generated with yields that are Poisson distributed as appropriate for the respective dataset. 
The pseudoexperiments are found to provide an unbiased estimate of the ratio of the \LbJpsiLambda and \BdJpsiKS branching fractions.
The isospin asymmetry is determined in bins of \pt from the efficiency-corrected yields of the \BuJpsiK and \BdJpsiKS decays in data. 

\section{Systematic uncertainties} 
\label{sec:systematics}

The analysis procedure is designed to minimise sources of systematic uncertainty in the measurement of $\BF(\LbJpsiLambda)/\BF(\BdJpsiKS)$. 
The dominant sources of uncertainty on the measurement arise from the knowledge of $\fracLb/\fracBd$. 
A systematic uncertainty is also assigned on the external measurements of the \KSPiPi and \LambdaPPi branching fractions.
The estimated sources of systematic uncertainty on $\BF(\LbJpsiLambda)/\BF(\BdJpsiKS)$ and \AI are summarised in Table~\ref{tab:systematics} and are described below. 
The statistical uncertainties on the branching fraction ratio and \AI measurements are determined by fixing all of the external parameters in the fit to their central values. 

In the determination of the branching-fraction ratio from the yields of \LbJpsiLambda and \BdJpsiKS decays, $\fracLb/\fracBd$ is evaluated at the bin centre of each \bquark-hadron \pt bin. 
In reality there will be a distribution of candidates in \pt over the bin. 
Taking instead the value of $\fracLb/\fracBd$ one-third or two-thirds of the way across the bin changes the branching-fraction ratio by 1.4\%.  

Two sources of systematic uncertainty are related to the determination of the efficiencies from the simulation samples. The first of these is due to assumptions made in the matching of reconstructed candidates and true decays in the simulation, for which new efficiency values are derived from the simulated samples with a looser set of association criteria between detector energy deposits and the true particle trajectories. The second source of uncertainty is due to the finite size of the simulated samples, and is determined by comparing the baseline fit result to the one obtained fixing the efficiency ratio to its central value in the fit. 

A systematic uncertainty is estimated on the choice of models used for determining the signal yields from data in Sec.~\ref{sec:fitting}. The mass distributions are instead fit using a signal function with a Gaussian core and power-law substituted tails on the left- and right-hand side of the distribution. The difference in signal yields between the baseline fit results and those with the alternative fit functions is used to assign a systematic uncertainty. 

Material interactions of \Kpm, \KS (\Kz and \Kzb) and \Lz hadrons in the detector could lead to a biased estimate of $\BF(\LbJpsiLambda)/\BF(\BdJpsiKS)$, or \AI,  if the detector material or hadronic cross-sections are not correctly described by simulation. 
For reference, in simulation 3\% (3\%) of \KS (\Lz) hadrons interact before they could decay in the long sample and 5\% (9\%) before they could decay in the downstream sample.
The most relevant material interactions happen in the vertex detector and in RICH\,1. 
The amount of material in the vertex detector is known at the level of $\pm6\%$~\cite{LHCb-DP-2018-002, LHCb-DP-2014-001}. 
The material in RICH\,1 is well described by the simulation. 
Dedicated studies are performed using simulated samples of kaons and \Lz baryons, comparing the results of the \geant toolkit~\cite{Allison:2006ve, *Agostinelli:2002hh} to measurements of hadronic cross-sections on various materials reported in Ref.~\cite{PDG2024}. 
To assess a systematic uncertainty, the survival probabilities of different particles are determined in the simulation when varying the amount of material in the vertex detector and when varying the hadronic cross-sections between different models in the \geant toolkit. 

The $\BF(\LbJpsiLambda)/\BF(\BdJpsiKS)$ and \AI measurements are determined with a simultaneous fit that includes long and downstream samples. 
The ratio of \Lb and \Bz yields as a function of \pt is found to be in good agreement between the two samples (see Sec.~\ref{sec:results}). 
Comparing the efficiency-corrected yields of the individual decays between the long and downstream samples, the yields of \LbJpsiLambda decays are compatible, but the yields of \BdJpsiKS decays differ by 3.6\%. 
Half of the difference on the values of $\BF(\LbJpsiLambda)/\BF(\BdJpsiKS)$ obtained separately from the long and downstream samples is assigned as a systematic uncertainty. 
The same approach is used to assign a systematic uncertainty on \AI. 

Systematic uncertainties are estimated for the corrections applied to simulation to reproduce the data, namely due to the finite size of various calibration samples, and the choice of binning used for determining the corrections. For the measurement of $\BF(\LbJpsiLambda)/\BF(\BdJpsiKS)$, these systematic uncertainties are completely dominated by corrections on the production kinematics of the \bquark hadrons. 
For the measurement of \AI, the systematic uncertainties due to the corrections applied to simulation are negligible because it is assumed that \Bz and \Bp mesons have the same kinematic distributions. 
However, additional sources of systematic uncertainty arise due to the need to reconstruct an additional track for the \BdJpsiKS decay and the use of particle-identification requirements to select the \Kp meson. They are determined taking into account the individual contributions to differences in track reconstruction efficiency between data and simulation.
The efficiency for a kaon to satisfy the particle-identification requirements used in the reconstruction of \Bp candidates is determined from data using samples of \decay{\Dstarp}{\Dz\pip} decays, with \decay{\Dz}{\Km\pip}, as a function of the kaon momentum and the multiplicity of charged particles in the event~\cite{LHCb-DP-2018-001}.
The measurement of \AI is also sensitive to the knowledge of the lifetime ratio $\tau_{\Bp}/\tau_{\Bz}$, which is taken from Ref.~\cite{HFLAV21}. 

\begin{table}[!tb]
    \caption{
    Summary of sources of relative systematic uncertainty on the $\BF(\LbJpsiLambda)/\BF(\BdJpsiKS)$ measurement and absolute systematic uncertainty on the \AI measurement. All uncertainties are treated as uncorrelated.
    }
    \label{tab:systematics}
    \centering
    \begin{tabular}{r|cc}
    \toprule
    Source  & $\dfrac{\BF(\LbJpsiLambda)}{\BF(\BdJpsiKS)}$ [\%] & $A_{\rm I}$ [$\times 10^{-2}$] \\
    \midrule
    $f(\bquark\to\Lb)/f(\bquark\to\Bz)$ & 8.27 &  -- \\ 
    \KS and \Lz branching fractions & 0.78 & 0.01 \\
    $f(\bquark\to\Lb)/f(\bquark\to\Bz)$ \pt-binning & 1.40 &  -- \\ 
    Truth matching & 1.40 & 0.60 \\ 
    Simulation sample sizes & 1.20 & 0.04 \\ 
    Mass line shapes & 0.10 & 0.10 \\ 
    Material interactions & 0.50 & 0.35 \\ 
    Long/downstream compatibility & 1.80 & 0.85 \\
    Data-simulation corrections & 0.60 & -- \\ 
    Pion track reconstruction efficiency & -- & 0.50 \\ 
    Kaon particle identification & -- & 0.50 \\ 
    \bquark-hadron lifetimes & -- & 0.18 \\ 
    \bottomrule
    \end{tabular}
\end{table}

\section{Results} 
\label{sec:results} 

\begin{figure}[!tb]
    \centering
    \includegraphics[width=0.6\linewidth]{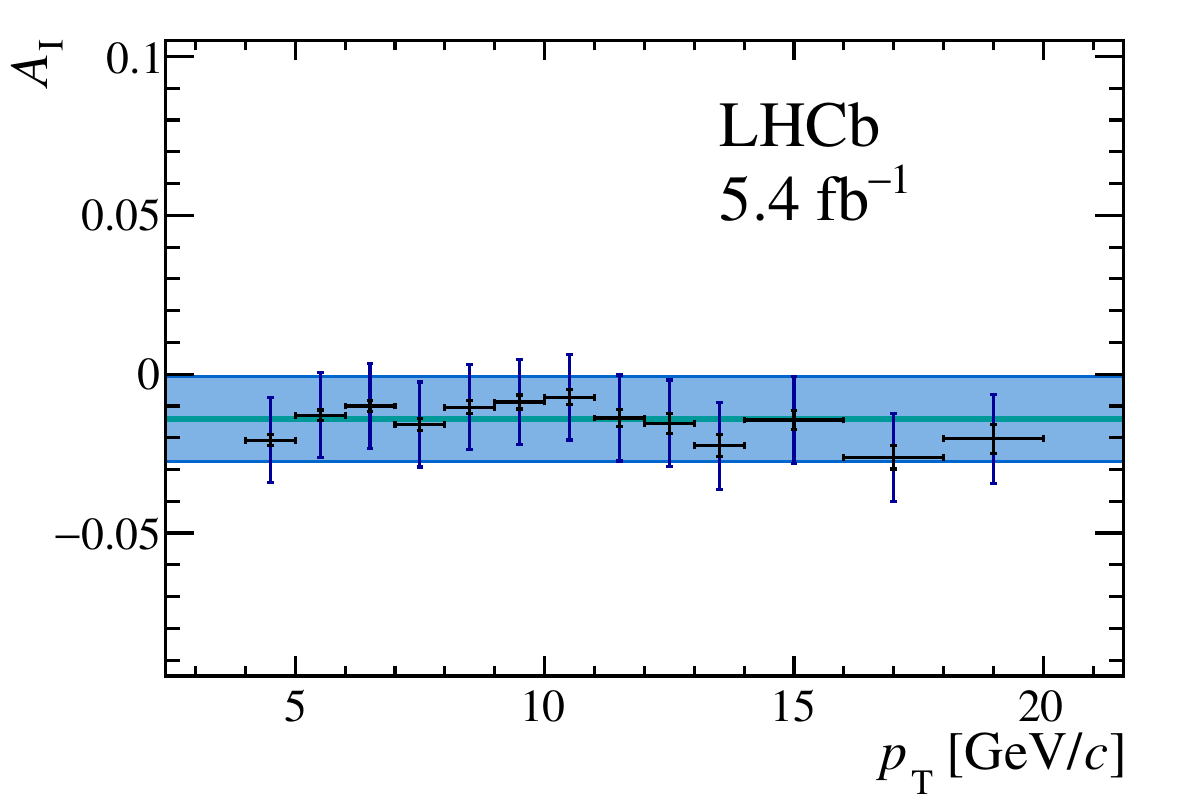}
    \caption{
    Isospin asymmetry between \BdJpsiKZ and \BuJpsiK decays in bins of \B-meson \pt for the long and downstream-track samples averaged.  The inner (outer) bars indicate the statistical (total) uncertainty on the data points. The green (blue) band indicates the average isospin asymmetry across the \pt bins and its statistical (total) uncertainty.  
    }
    \label{fig:isospin}
\end{figure}

Figure~\ref{fig:isospin} shows the measured isospin asymmetry between \Bz and \Bp decays, in bins of the \B-meson \pt, averaging the value of the \BdJpsiKS branching fraction between the long- and downstream-track categories. 
The measured asymmetry shows no significant \pt dependence and the two distributions are found to be compatible at the percent level. 
There is also no evidence for any significant deviation from the expectation that $\AI\approx 0$.
An average value of \AI is obtained by performing a \chisq fit to the measurements in intervals of \pt in the long- and downstream-track samples, under the assumption of isospin symmetry in the production of \Bz and \Bp mesons. 
The resulting value of \AI is 
\begin{displaymath}
A_{\rm I} = -0.0135 \pm 0.0004 \pm 0.0133\,,
\end{displaymath}
where the first uncertainty is statistical and the second systematic. 
This \AI value is consistent with recent measurements of the same quantity by other experiments~\cite{BELLE:2019xld,Belle-II:2022dbo}. 
Assuming instead isospin symmetry between decays, the ratio of production cross-sections of \Bp and \Bz mesons is determined to be 
\begin{displaymath}
    \fracBu/\fracBd = 1.027\pm 0.001 \pm 0.027
\end{displaymath}
where the first uncertainty is statistical and the second systematic. 

\begin{figure}[!htb]
    \centering
    \includegraphics[width=0.48\linewidth]{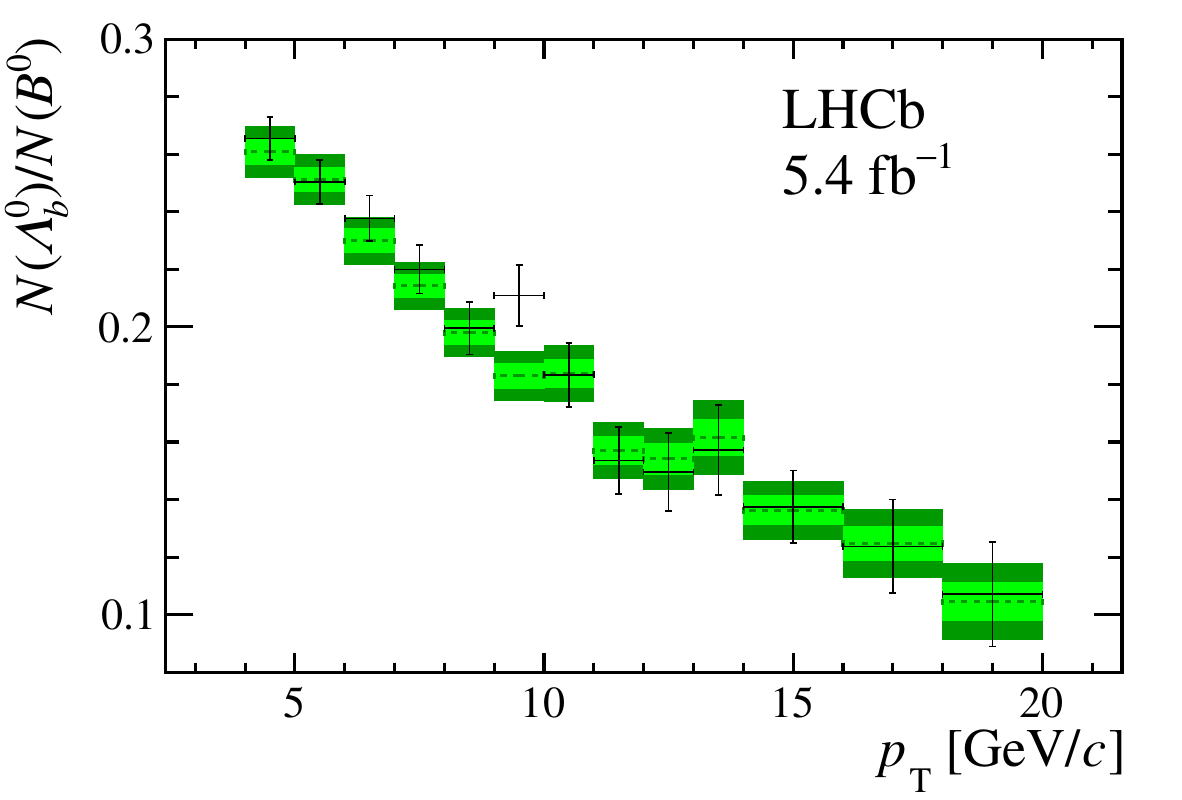}
    \includegraphics[width=0.48\linewidth]{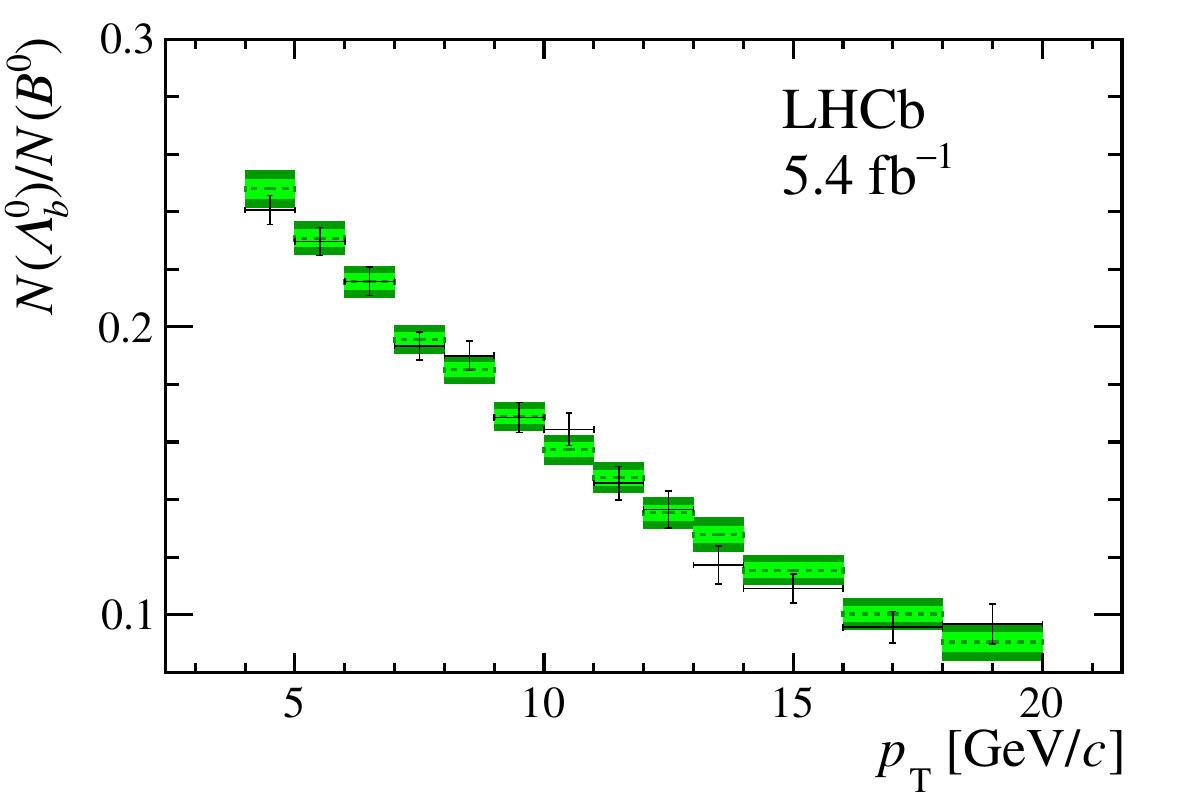}
    \caption{
    Ratio of yields between \LbJpsiLambda and \BdJpsiKS decays in bins of \bquark-hadron \pt for the (left) long-track and (right) downstream-track samples. 
    The dashed line shows the result of the fit described in the text.  
    The inner and outer shaded regions correspond to 68\% and 95\% confidence intervals, respectively.
    These regions include contributions from the statistical uncertainty on the event yields and uncertainties on the external parameters entering the branching-fraction ratio fit. 
    }
    \label{fig:ratio}
\end{figure}

Figure~\ref{fig:ratio} shows the measured ratio between the yields of \LbJpsiLambda and \BdJpsiKS decays in bins of \pt. 
The data for the long- and downstream-track samples provide consistent results considering the ratio of efficiencies to select and reconstruct \Lb and \Bz signal in those categories.
The fit to the ratio of yields in bins of \pt gives 
\begin{displaymath}
    \frac{\BF(\LbJpsiLambda)}{\BF(\BdJpsiKS)} = 0.750 \pm 0.005 \pm 0.022 \pm 0.005 \pm 0.062\,,
\end{displaymath}
where the first uncertainty is statistical, the second systematic, the third due to external inputs on $\BF(\KSPiPi)$ and $\BF(\LambdaPPi)$ and the fourth due to  $\fracLb/\fracBd$. 
The fit has a \chisq of 23.9 with 25 degrees of freedom. 

Using the world average value of $\BF(\BdJpsiKS) = (4.45\pm 0.11) \times 10^{-4}$ from Ref.~\cite{PDG2024},  the branching fraction is determined to be
\begin{displaymath}
    \BF(\LbJpsiLambda) = (3.34 \pm 0.02 \pm 0.10 \pm 0.08 \pm 0.28 )\times 10^{-4}\,,
\end{displaymath}
where the third uncertainty also includes the uncertainty on $\BF(\BdJpsiKS)$.
Note that the value used for $\BF(\BdJpsiKS)$ averages measurements with slightly different assumptions on the relative production of $\Bz\Bzb$ and $\Bp\Bm$ mesons in \FourS decays. 
However, a reinterpretation of older measurements of the branching fraction, taking possible production asymmetries into account, gives a very consistent value~\cite{Jung:2015yma}.

\section{Summary} 
\label{sec:summary} 

A measurement of the branching-fraction ratio between \LbJpsiLambda and \BdJpsiKS decays has been performed. The measurement uses a dataset corresponding to an integrated luminosity of 5.4\invfb collected in proton-proton collisions at $\sqrt{s}=13\tev$ by the LHCb experiment. 
The ratio $\BF(\LbJpsiLambda)/\BF(\BdJpsiKS)$ is found to be $0.750\pm 0.066$, combining all sources of uncertainty in quadrature.  
This value can be compared to the ratio of lifetimes between \Lb baryons and \Bz mesons, $\tau_\Lb/\tau_\Bz = 0.964\pm 0.007$~\cite{HFLAV21,PDG2024}. 
Accounting for the 50\% probability for a \Kz meson to produce a \KS, the partial width of the \LbJpsiLambda decay is found to be approximately 39\% of that of the \BdJpsiKZ decay. 
The resulting \LbJpsiLambda branching fraction is compatible with estimates based on perturbative QCD and QCD factorisation approaches obtained in Refs.~\cite{Rui:2023fiz} and \cite{Hsiao:2021mlp}, respectively. 
The partial widths of the \BuJpsiK and \BdJpsiKZ decays are compatible at the percent level.

\section*{Acknowledgements}
%
%
\noindent We express our gratitude to our colleagues in the CERN
accelerator departments for the excellent performance of the LHC. We
thank the technical and administrative staff at the LHCb
institutes.
We acknowledge support from CERN and from the national agencies:
ARC (Australia);
CAPES, CNPq, FAPERJ and FINEP (Brazil); 
MOST and NSFC (China); 
CNRS/IN2P3 (France); 
BMBF, DFG and MPG (Germany); 
INFN (Italy); 
NWO (Netherlands); 
MNiSW and NCN (Poland); 
MCID/IFA (Romania); 
MICIU and AEI (Spain);
SNSF and SER (Switzerland); 
NASU (Ukraine); 
STFC (United Kingdom); 
DOE NP and NSF (USA).
We acknowledge the computing resources that are provided by ARDC (Australia), 
CBPF (Brazil),
CERN, 
IHEP and LZU (China),
IN2P3 (France), 
KIT and DESY (Germany), 
INFN (Italy), 
SURF (Netherlands),
Polish WLCG (Poland),
IFIN-HH (Romania), 
PIC (Spain), CSCS (Switzerland), 
and GridPP (United Kingdom).
We are indebted to the communities behind the multiple open-source
software packages on which we depend.
Individual groups or members have received support from
Key Research Program of Frontier Sciences of CAS, CAS PIFI, CAS CCEPP, 
Fundamental Research Funds for the Central Universities,  and Sci.\ \& Tech.\ Program of Guangzhou (China);
Minciencias (Colombia);
EPLANET, Marie Sk\l{}odowska-Curie Actions, ERC and NextGenerationEU (European Union);
A*MIDEX, ANR, IPhU and Labex P2IO, and R\'{e}gion Auvergne-Rh\^{o}ne-Alpes (France);
Alexander-von-Humboldt Foundation (Germany);
ICSC (Italy); 
Severo Ochoa and Mar\'ia de Maeztu Units of Excellence, GVA, XuntaGal, GENCAT, InTalent-Inditex and Prog.~Atracci\'on Talento CM (Spain);
SRC (Sweden);
the Leverhulme Trust, the Royal Society and UKRI (United Kingdom).


\clearpage
\section*{Appendix}

\appendix

\begin{figure}[!htb] 
    \centering
    \includegraphics[width=0.32\linewidth]{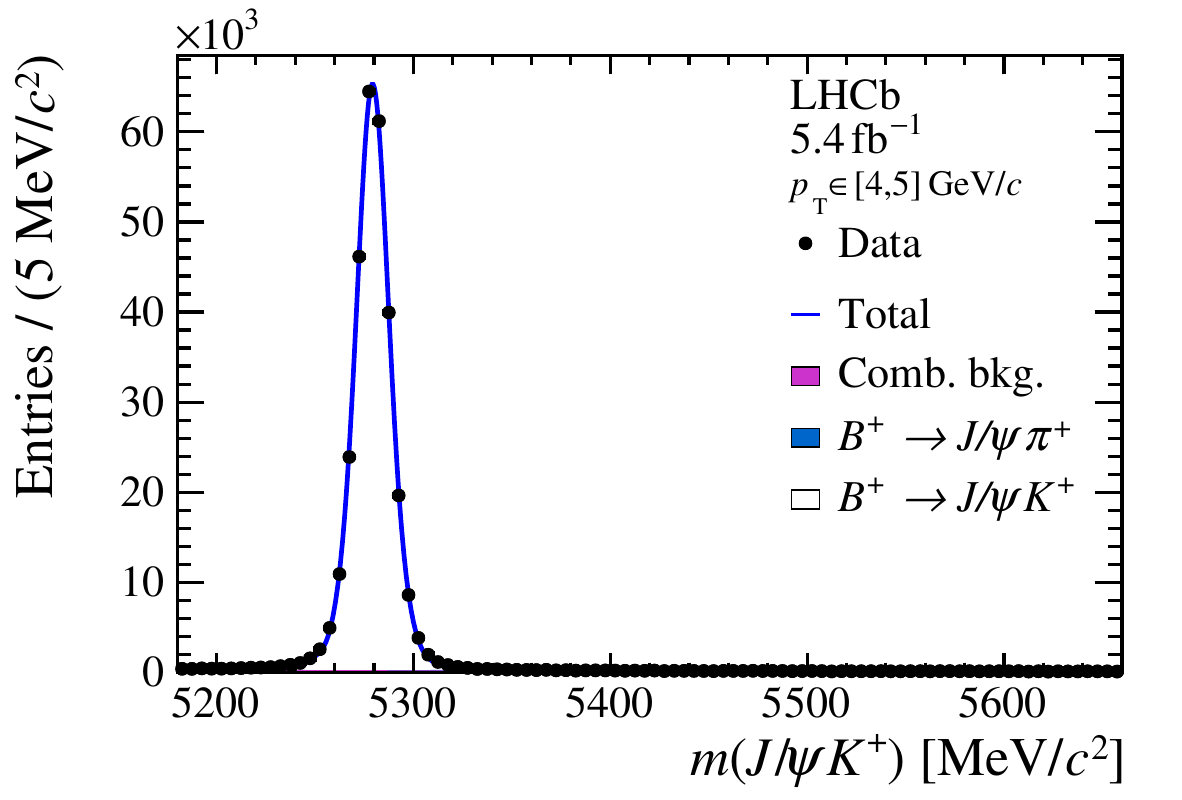} 
    \includegraphics[width=0.32\linewidth]{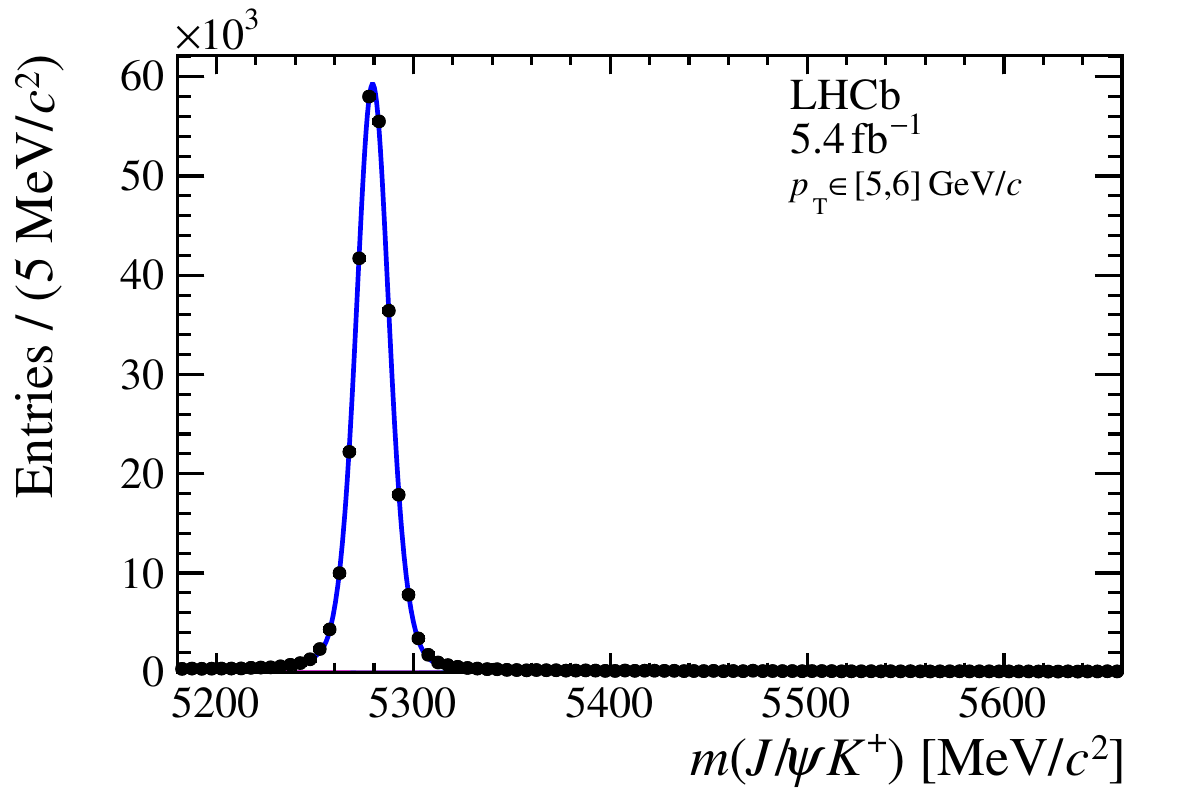} 
    \includegraphics[width=0.32\linewidth]{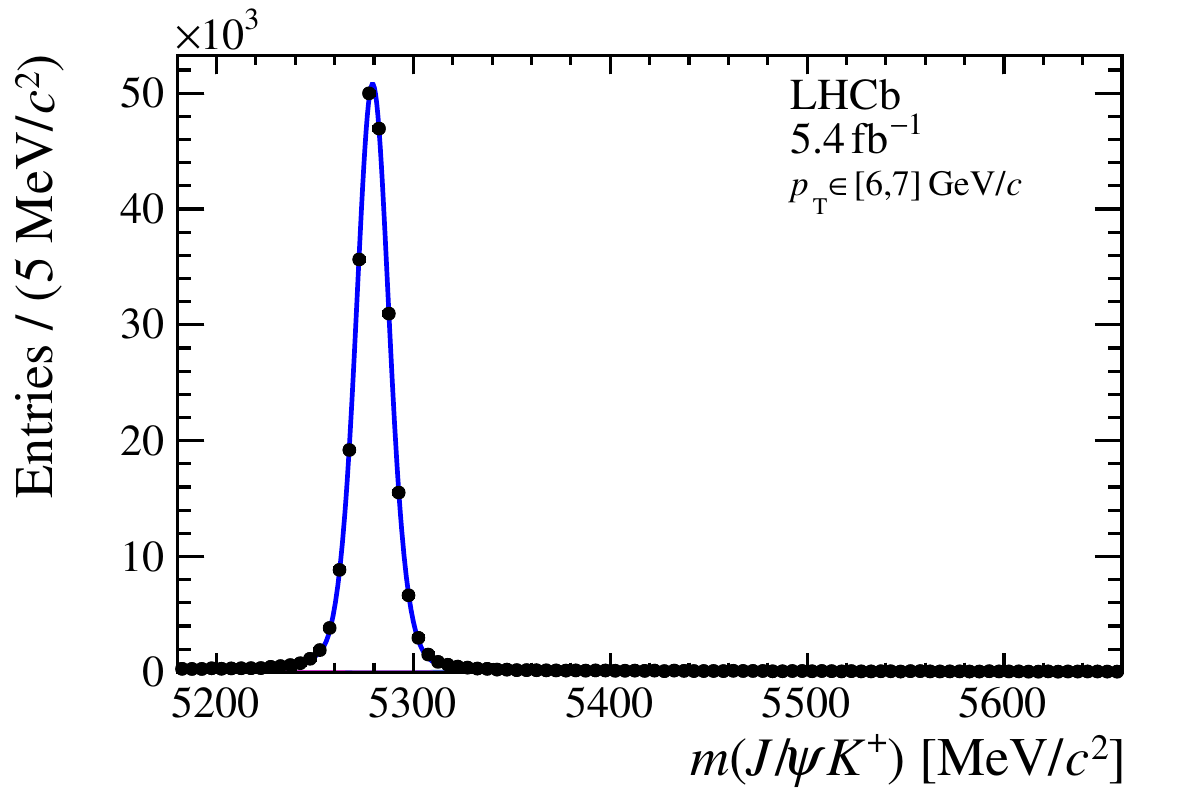} \\
    \includegraphics[width=0.32\linewidth]{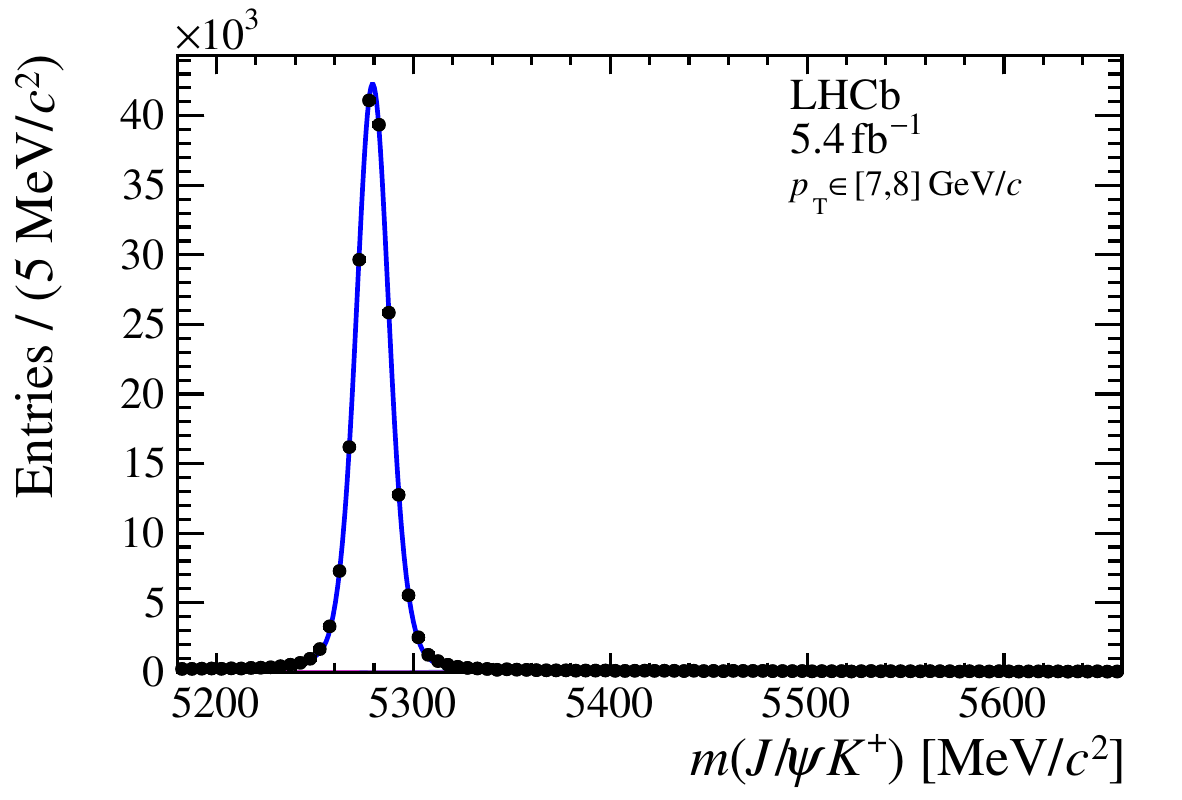} 
    \includegraphics[width=0.32\linewidth]{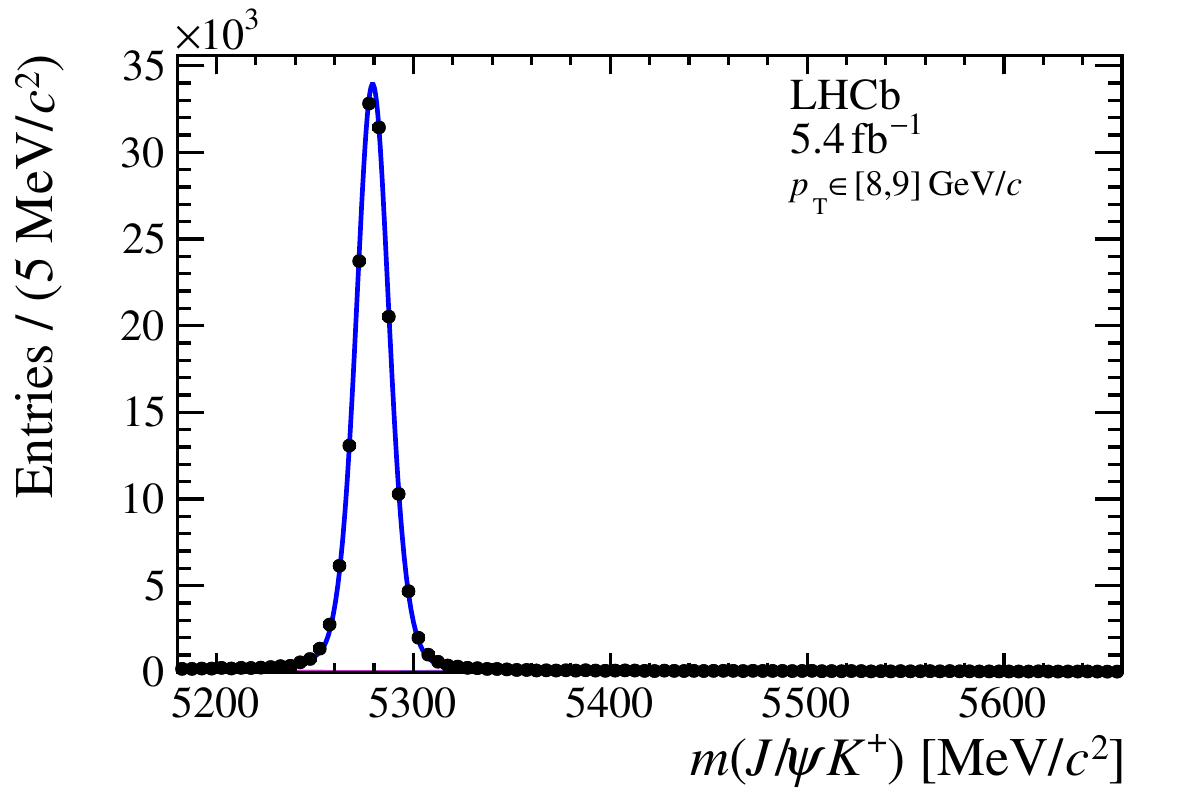} 
    \includegraphics[width=0.32\linewidth]{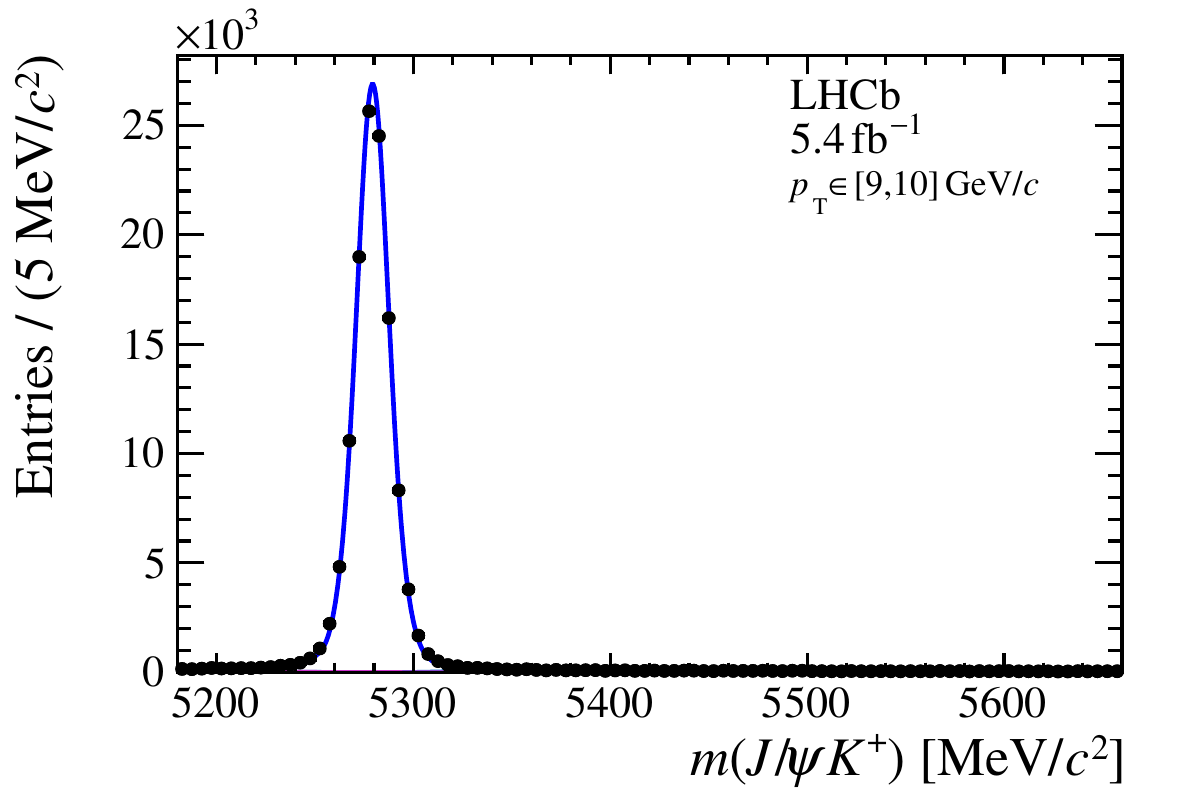} \\
    \includegraphics[width=0.32\linewidth]{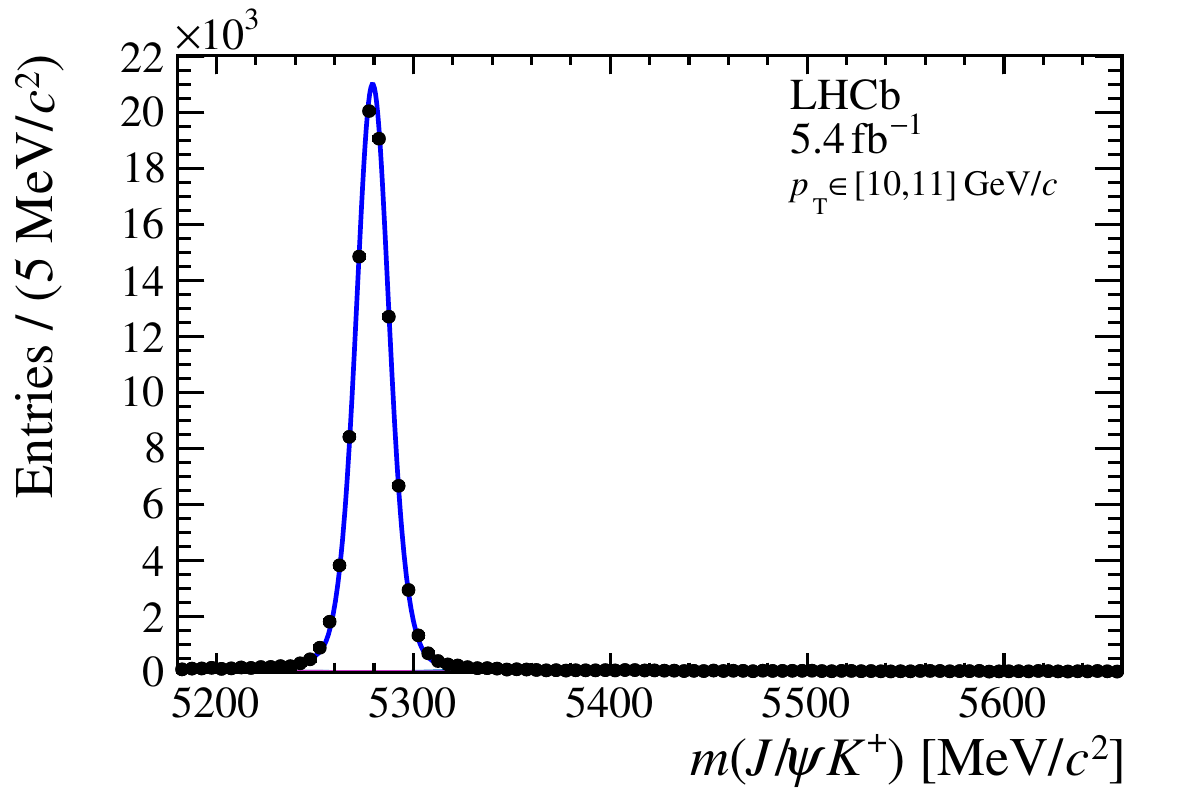} 
    \includegraphics[width=0.32\linewidth]{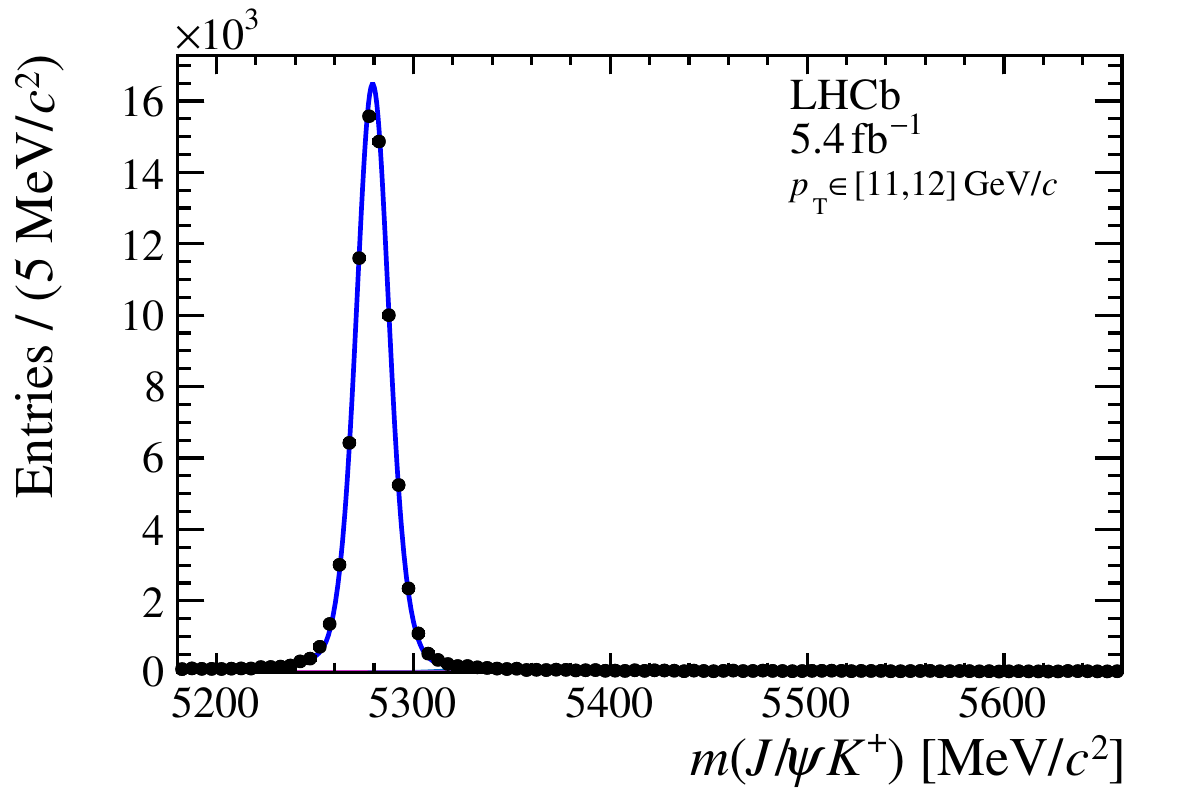} 
    \includegraphics[width=0.32\linewidth]{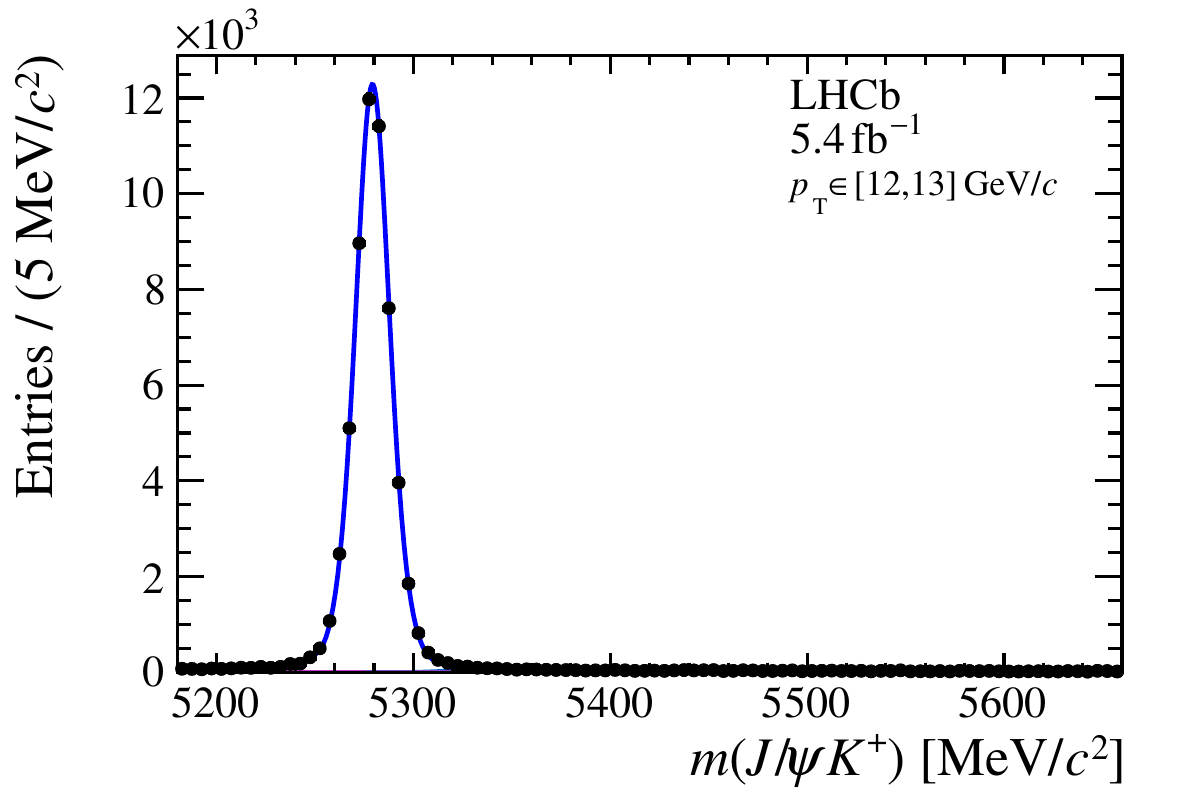} \\
    \includegraphics[width=0.32\linewidth]{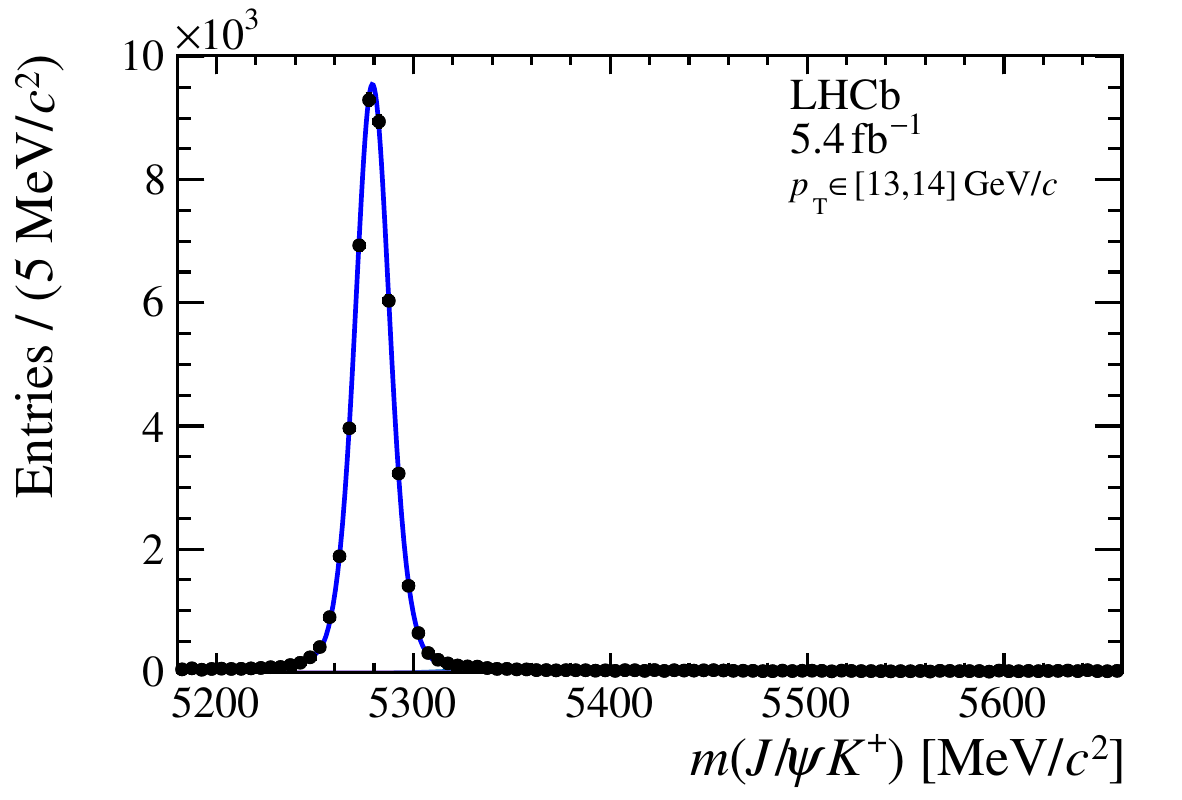} 
    \includegraphics[width=0.32\linewidth]{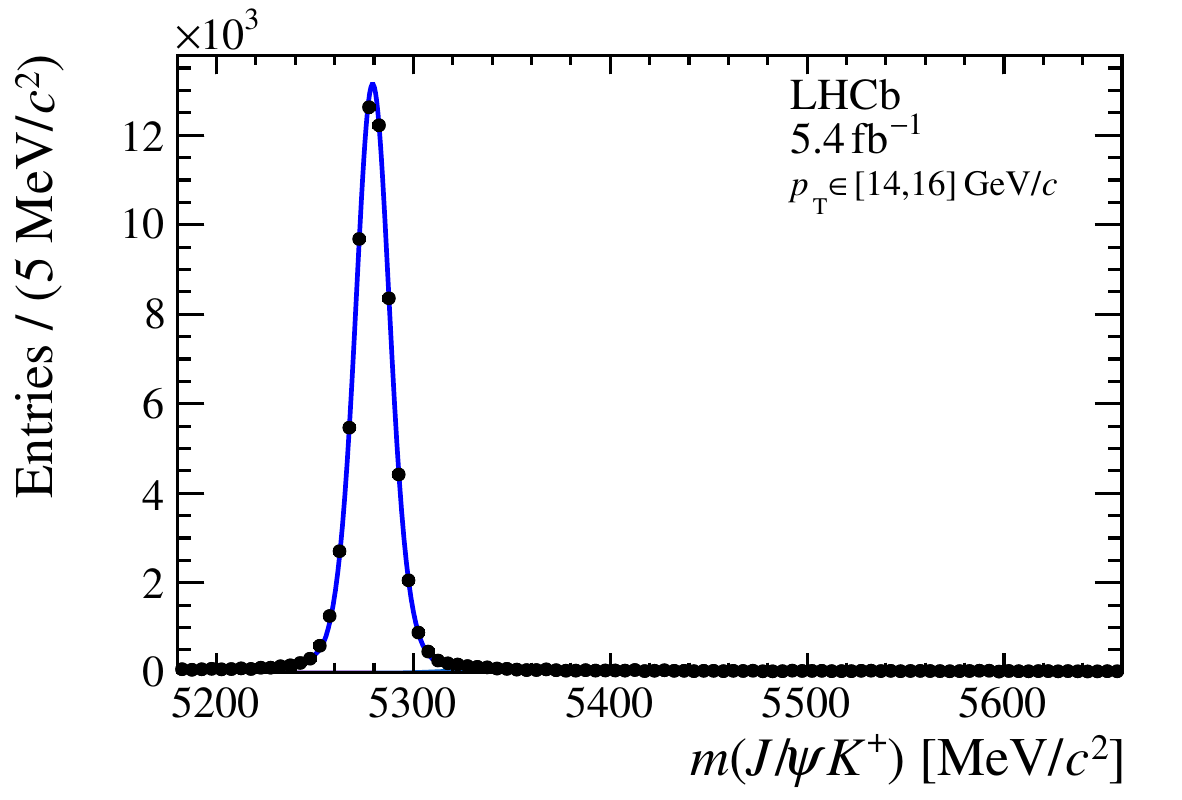} 
    \includegraphics[width=0.32\linewidth]{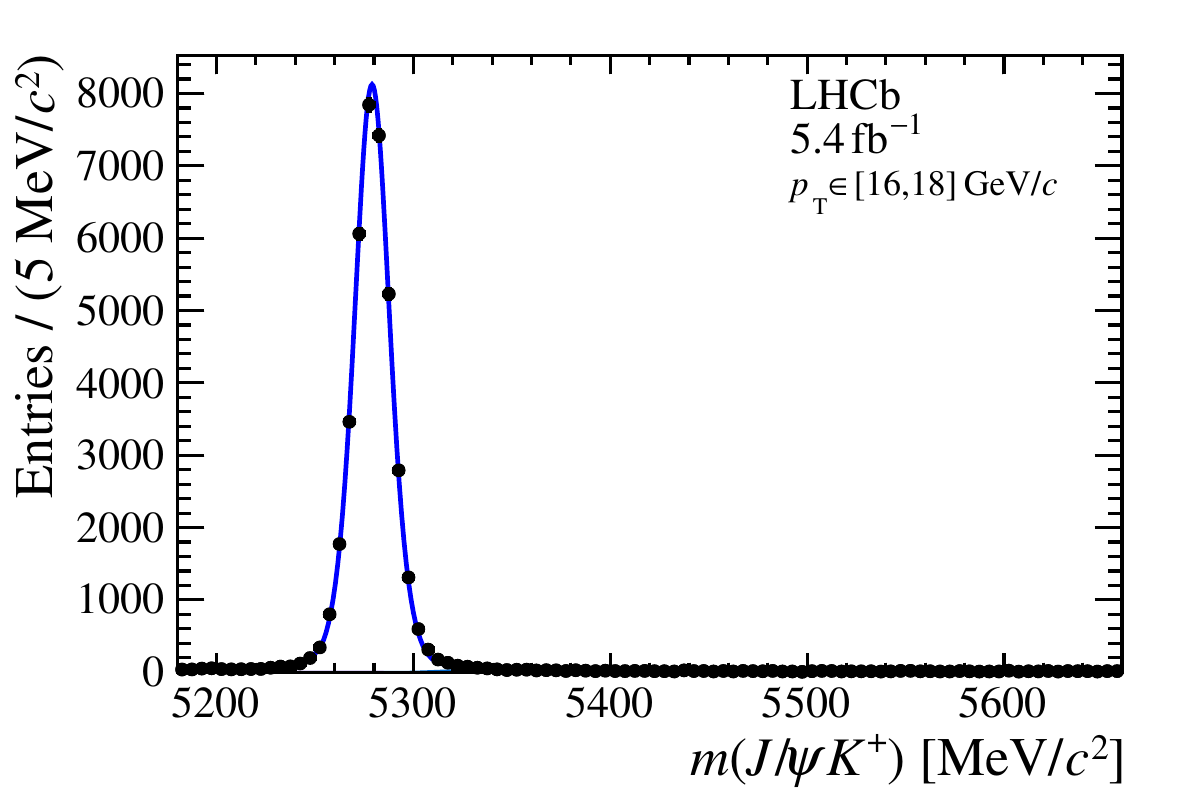} \\
    \includegraphics[width=0.32\linewidth]{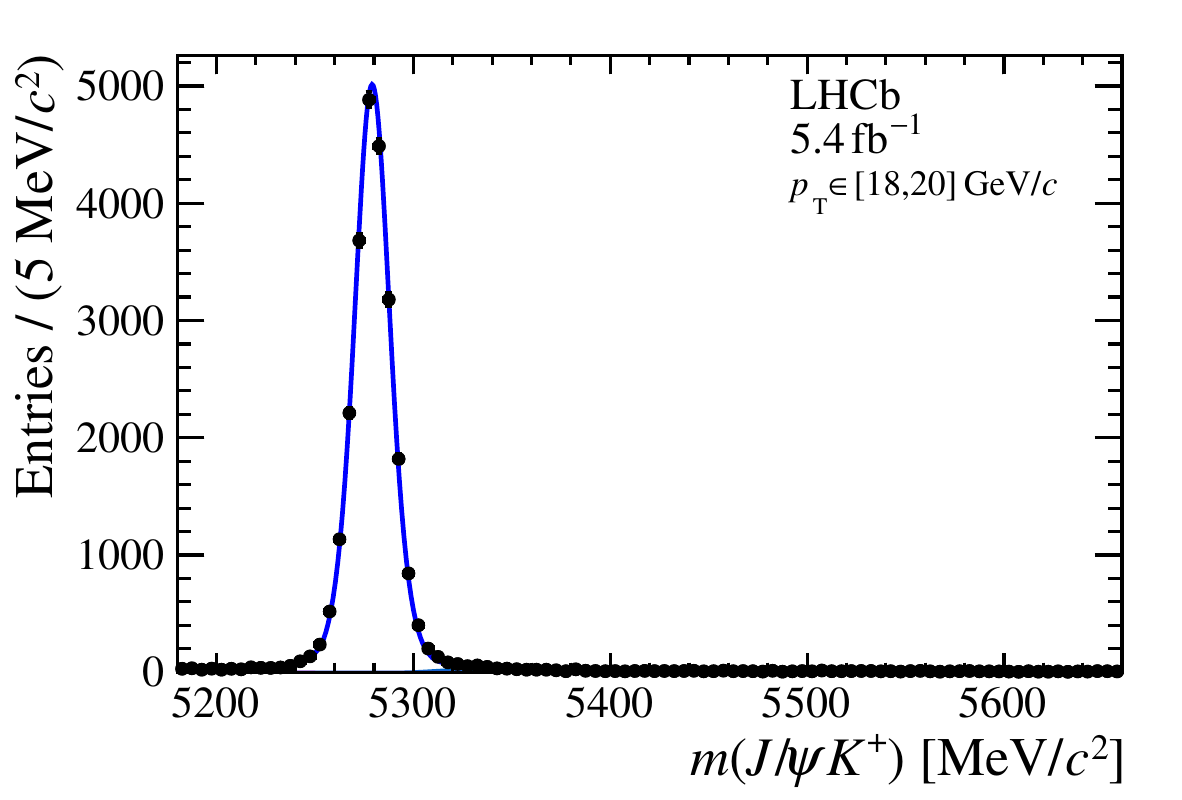} 
    \caption{
    Reconstructed $\jpsi\Kp$ mass distribution of selected \BuJpsiK candidates in bins of \Bp meson \pt. 
    The line and shaded regions show the result of the fit described in the text. 
    }
    \label{fig:Kp}
\end{figure}

\begin{figure}[!htb] 
    \centering
    \includegraphics[width=0.32\linewidth]{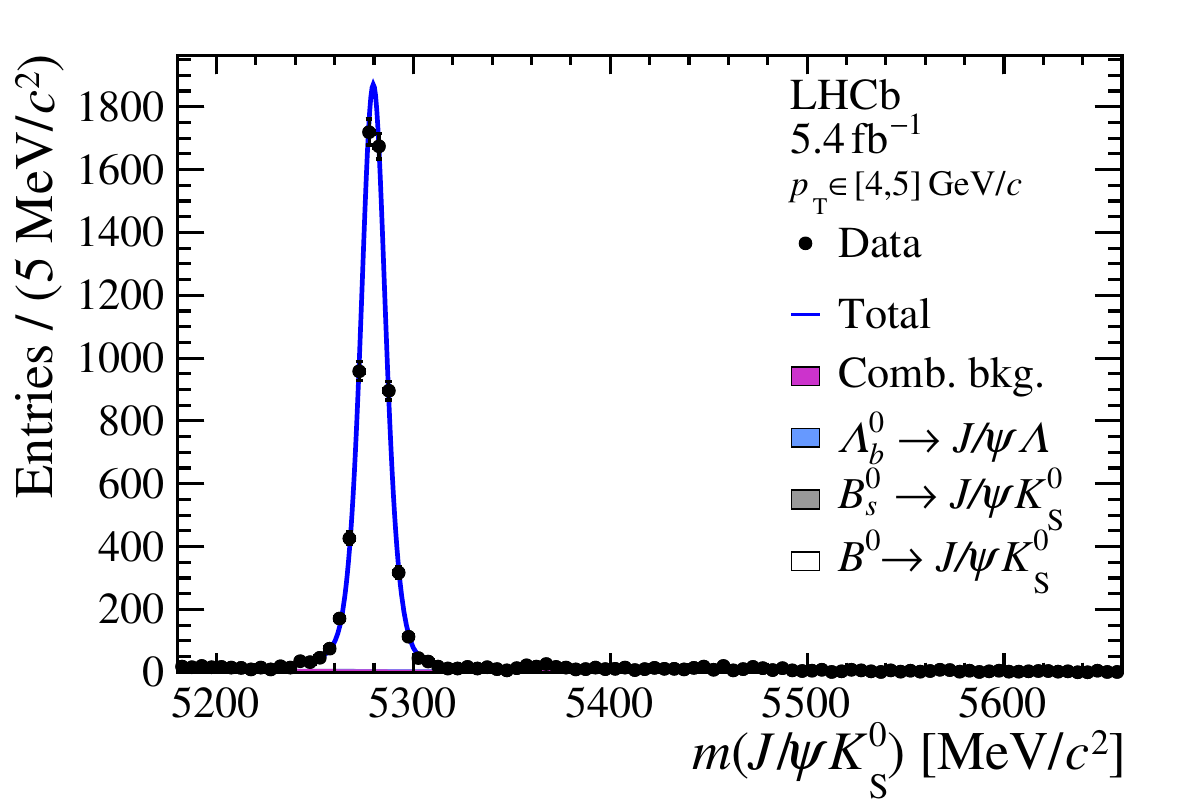} 
    \includegraphics[width=0.32\linewidth]{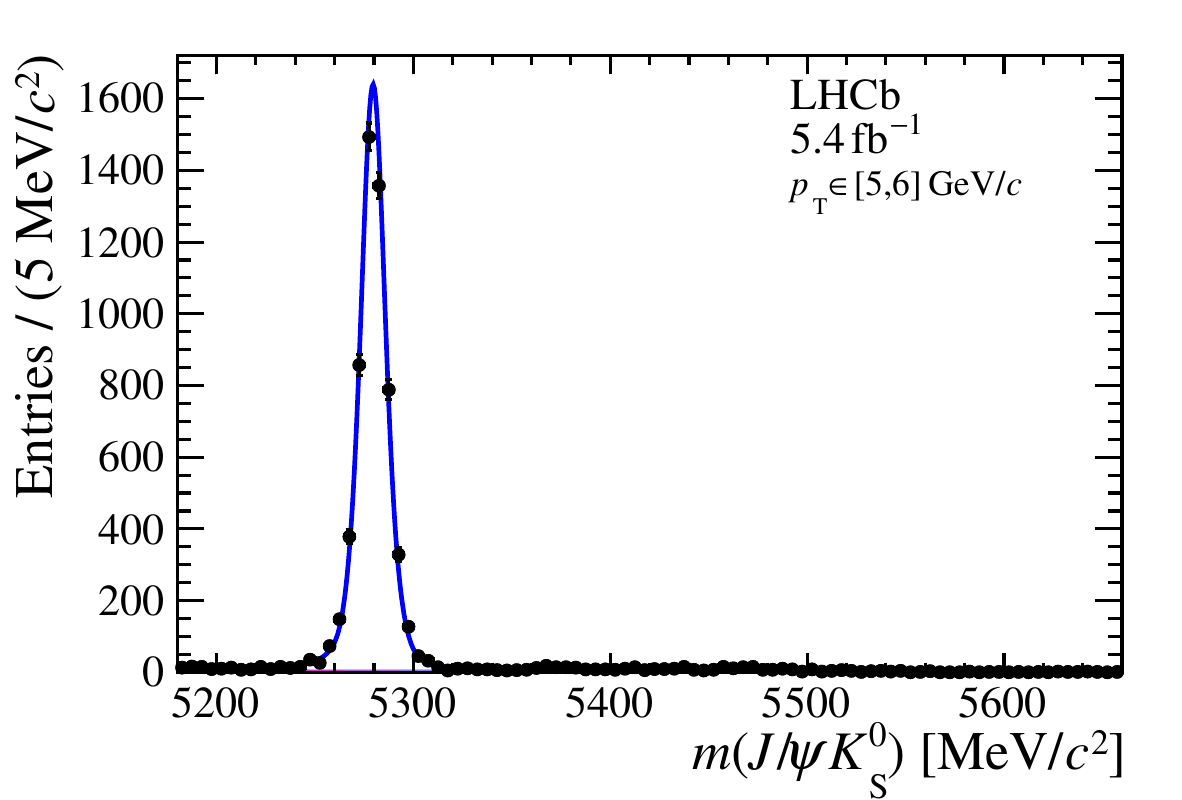} 
    \includegraphics[width=0.32\linewidth]{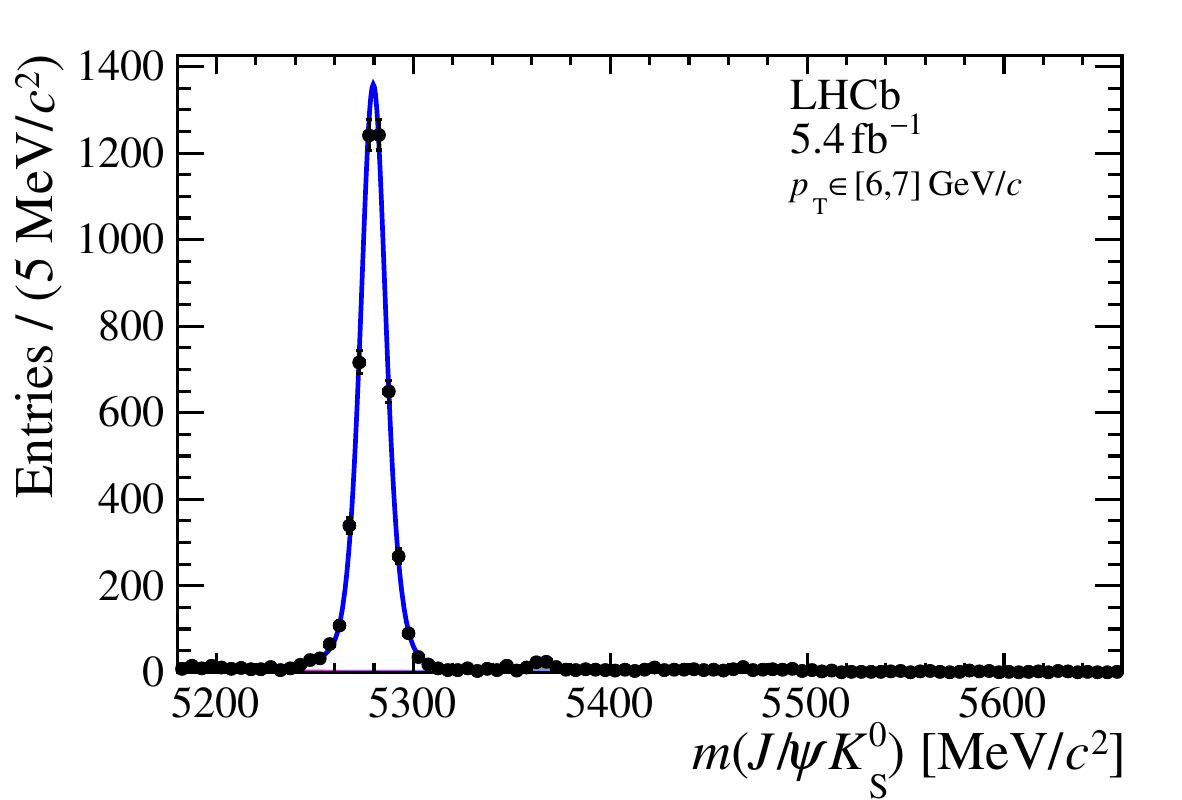} \\
    \includegraphics[width=0.32\linewidth]{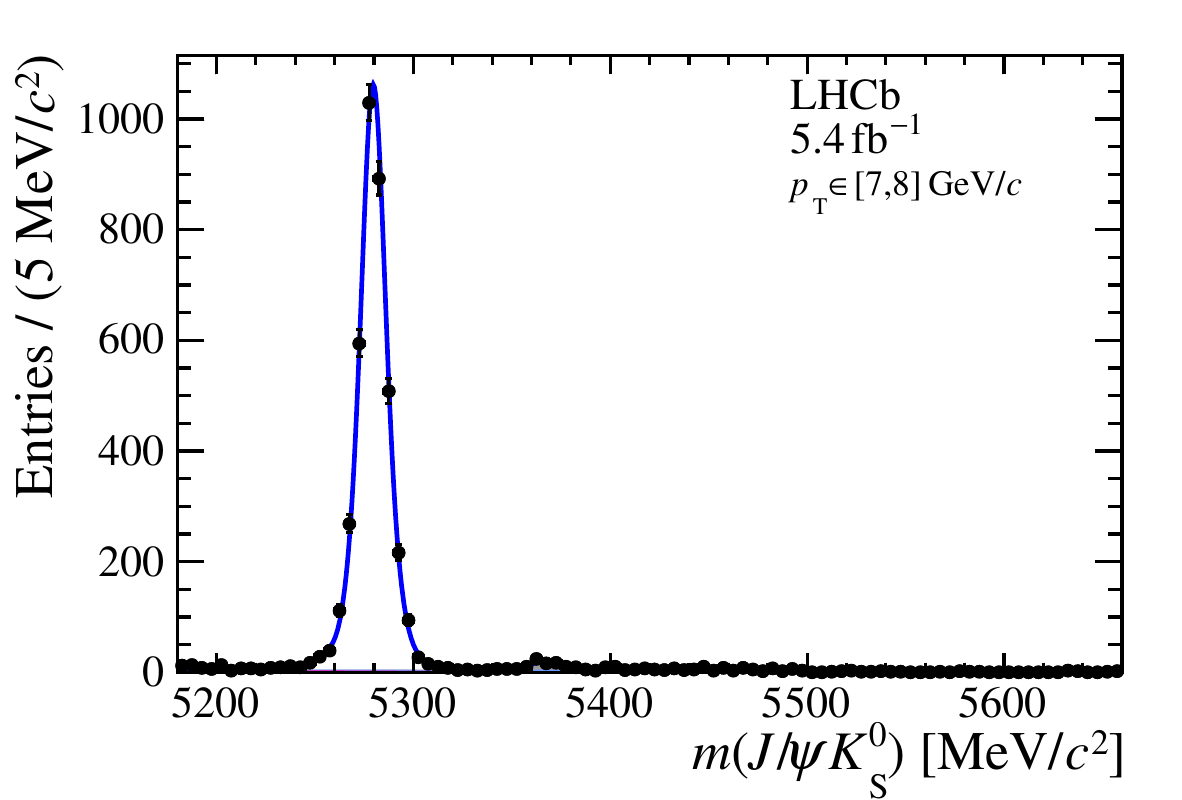} 
    \includegraphics[width=0.32\linewidth]{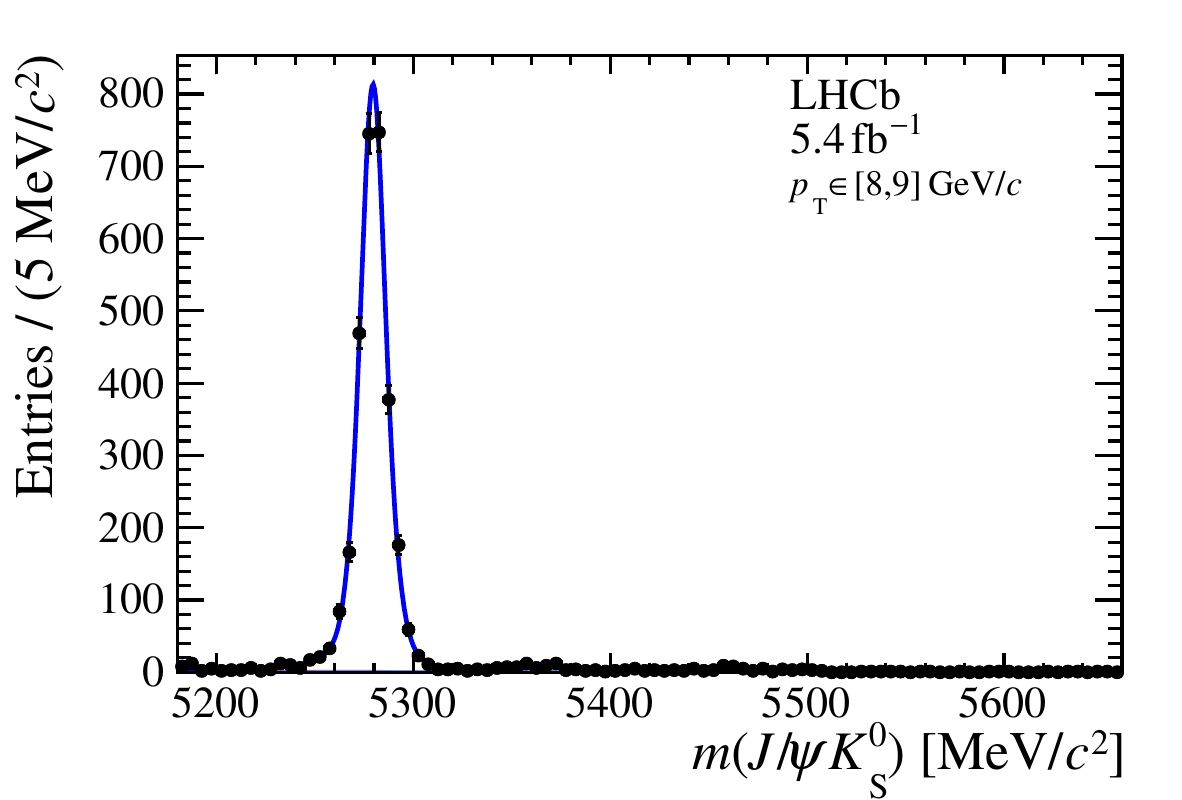} 
    \includegraphics[width=0.32\linewidth]{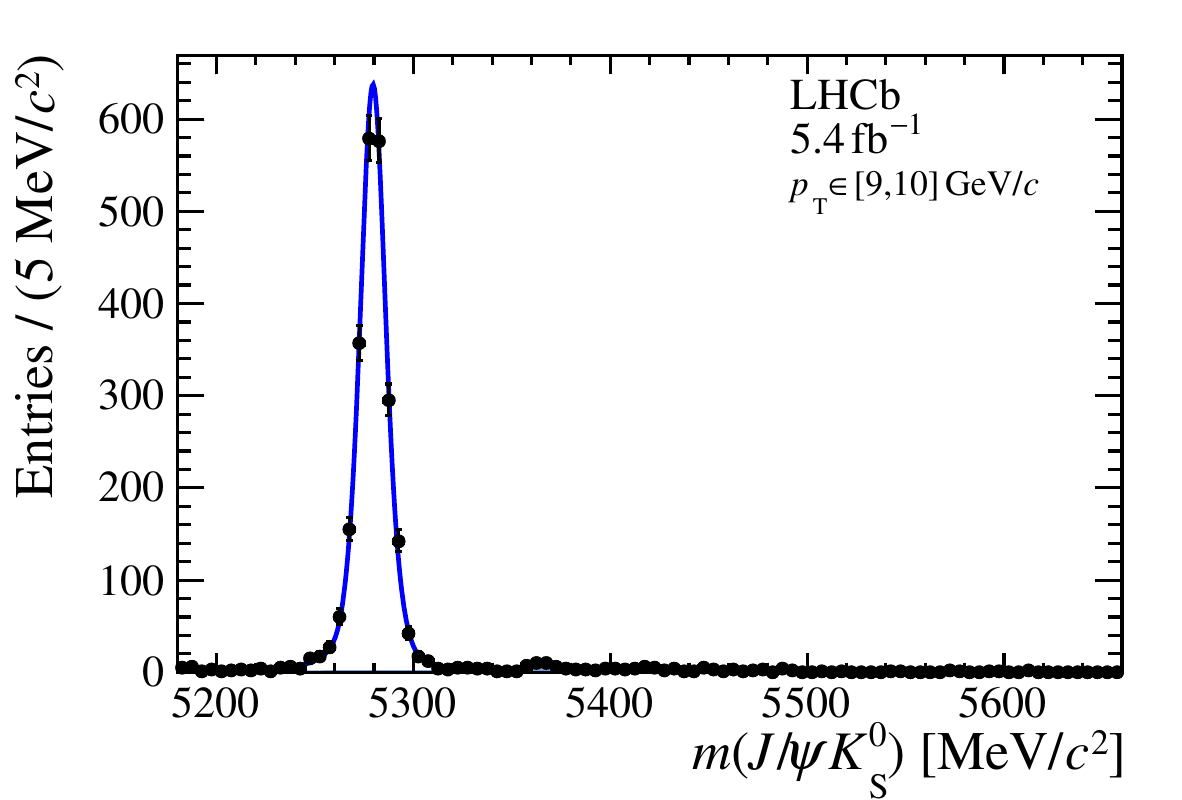} \\
    \includegraphics[width=0.32\linewidth]{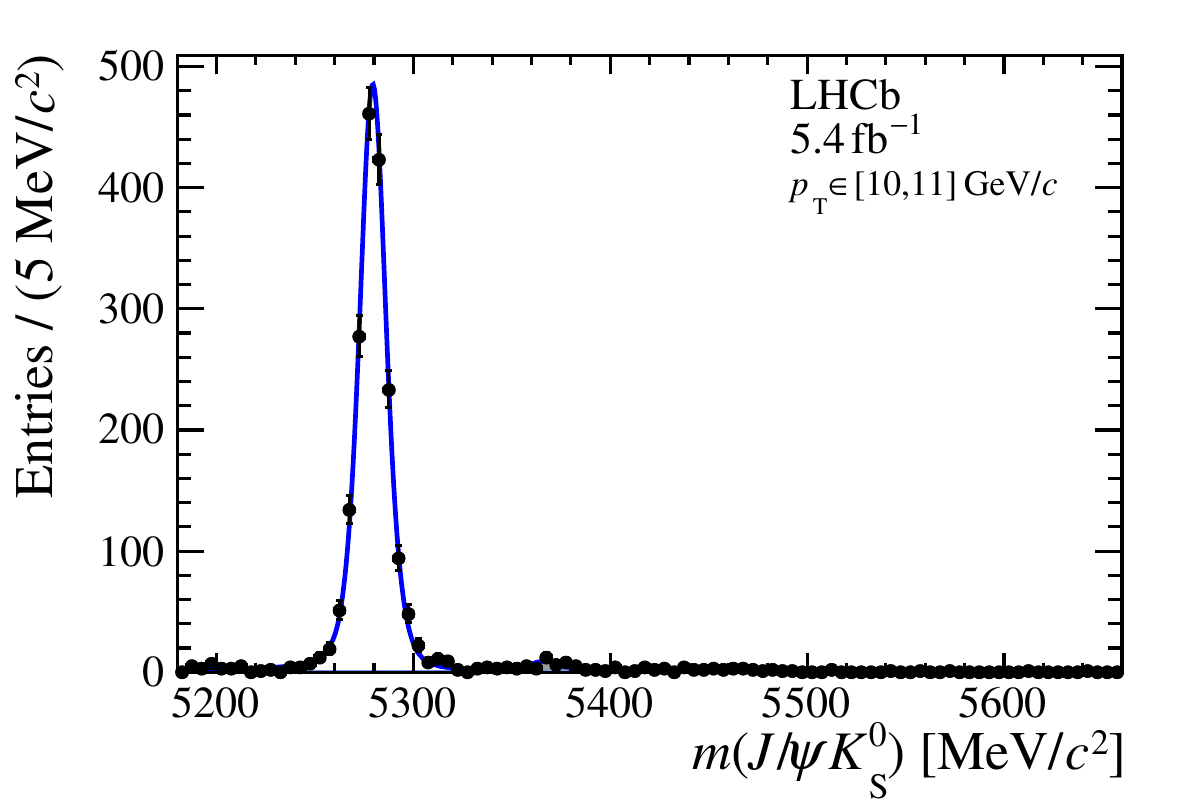} 
    \includegraphics[width=0.32\linewidth]{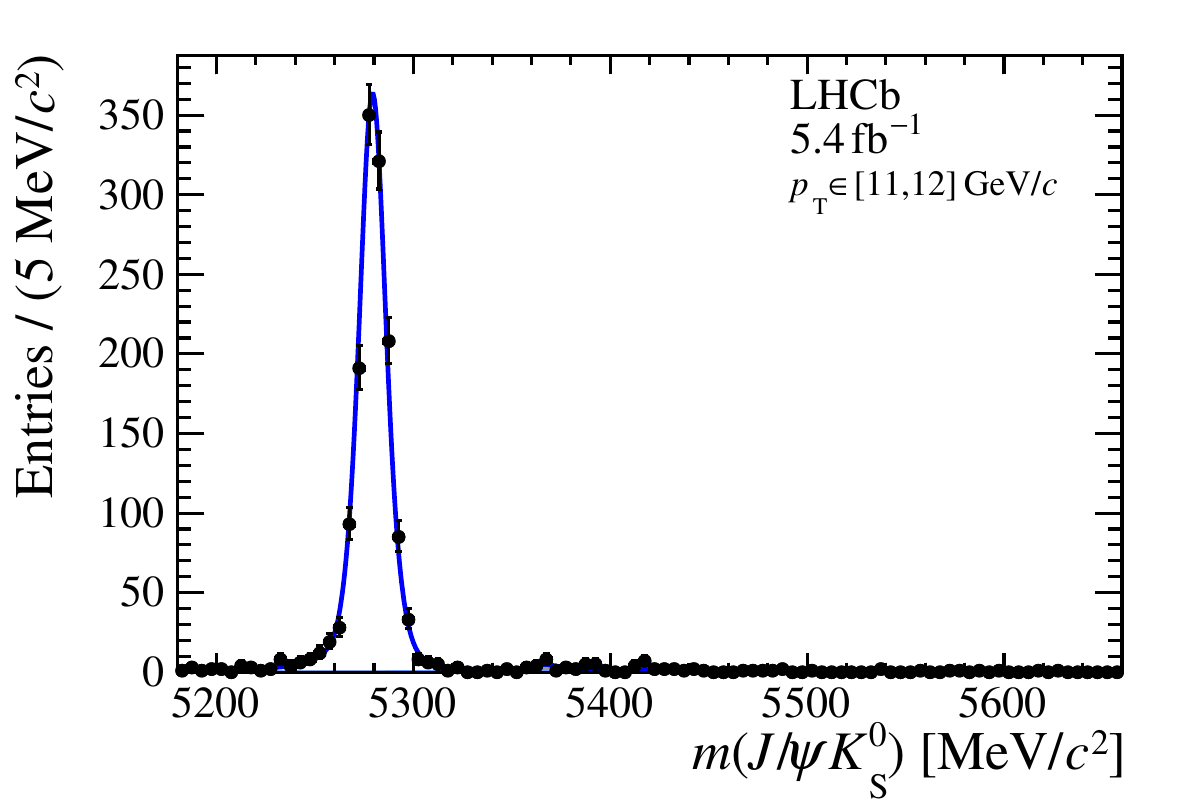} 
    \includegraphics[width=0.32\linewidth]{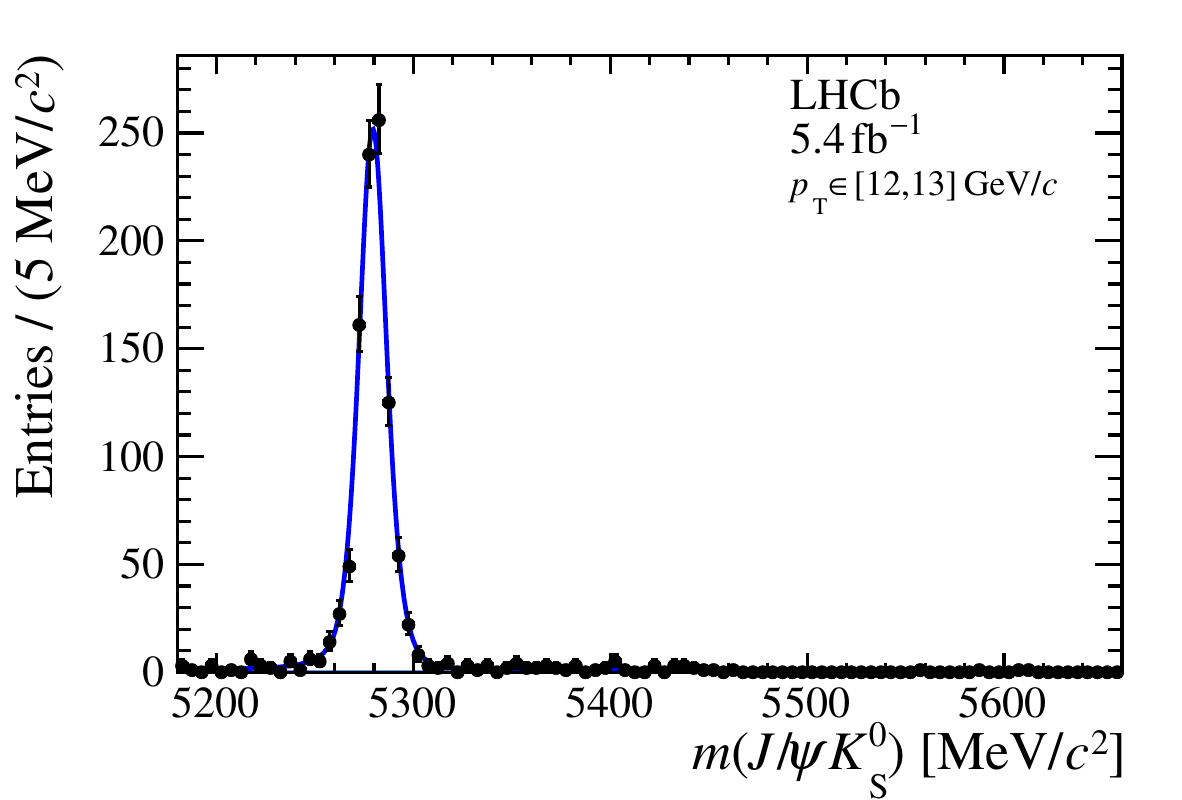} \\
    \includegraphics[width=0.32\linewidth]{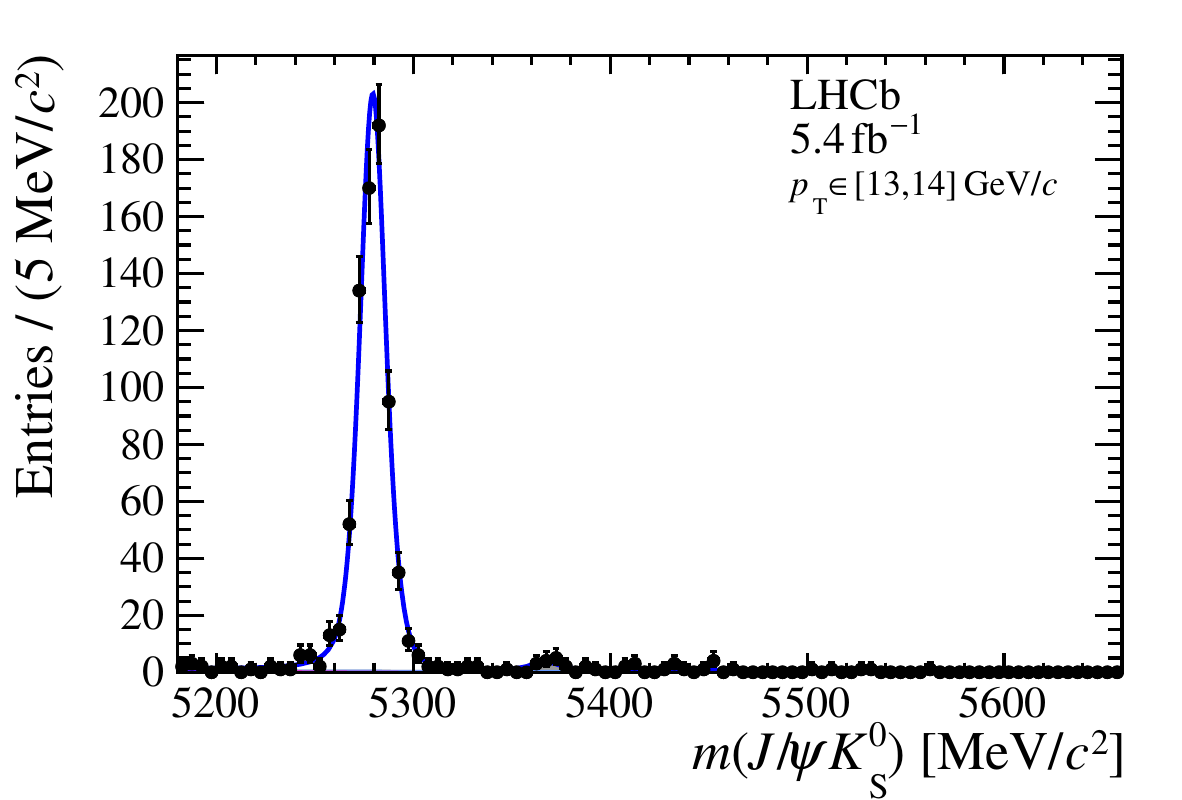} 
    \includegraphics[width=0.32\linewidth]{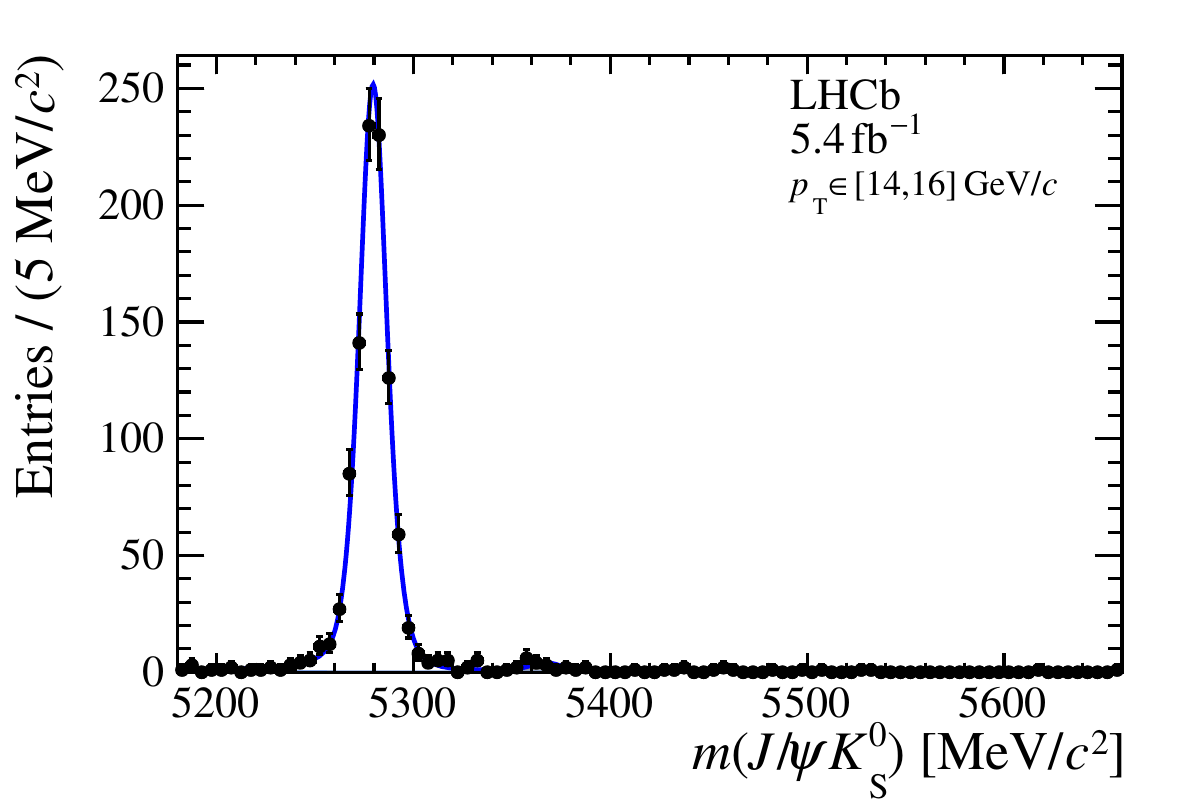} 
    \includegraphics[width=0.32\linewidth]{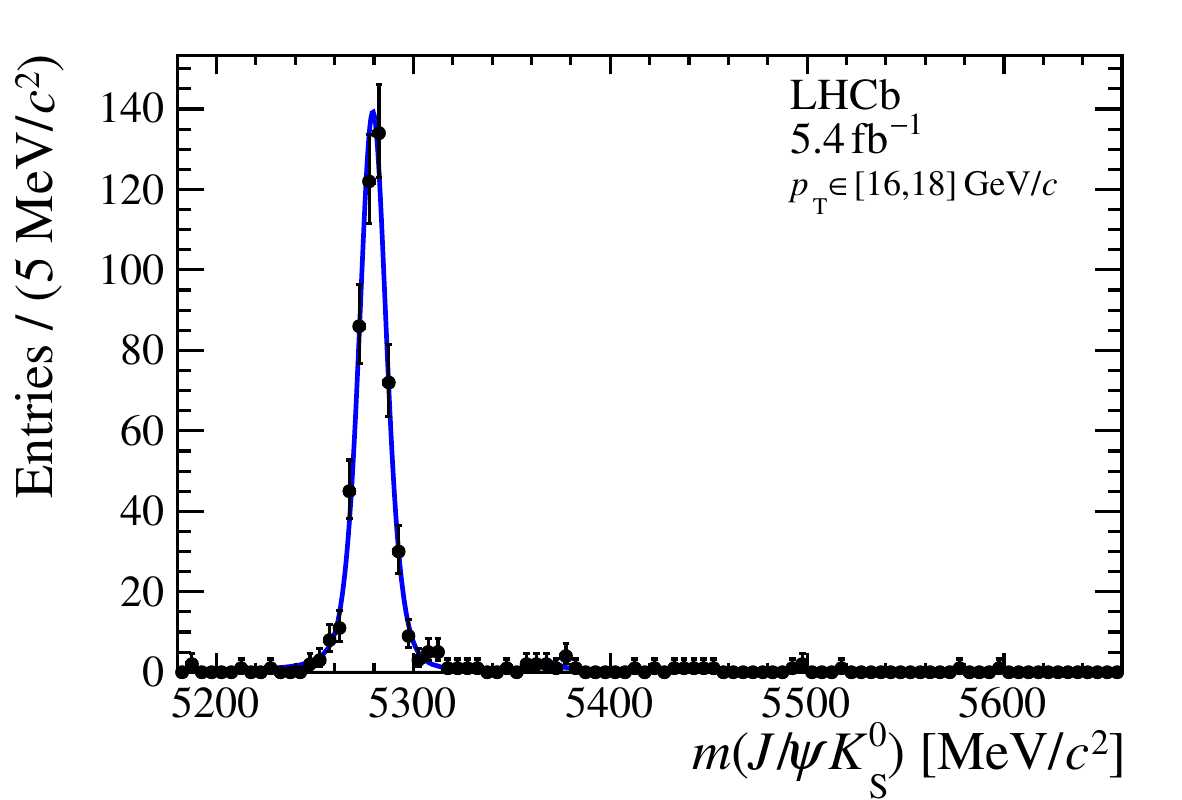} \\
    \includegraphics[width=0.32\linewidth]{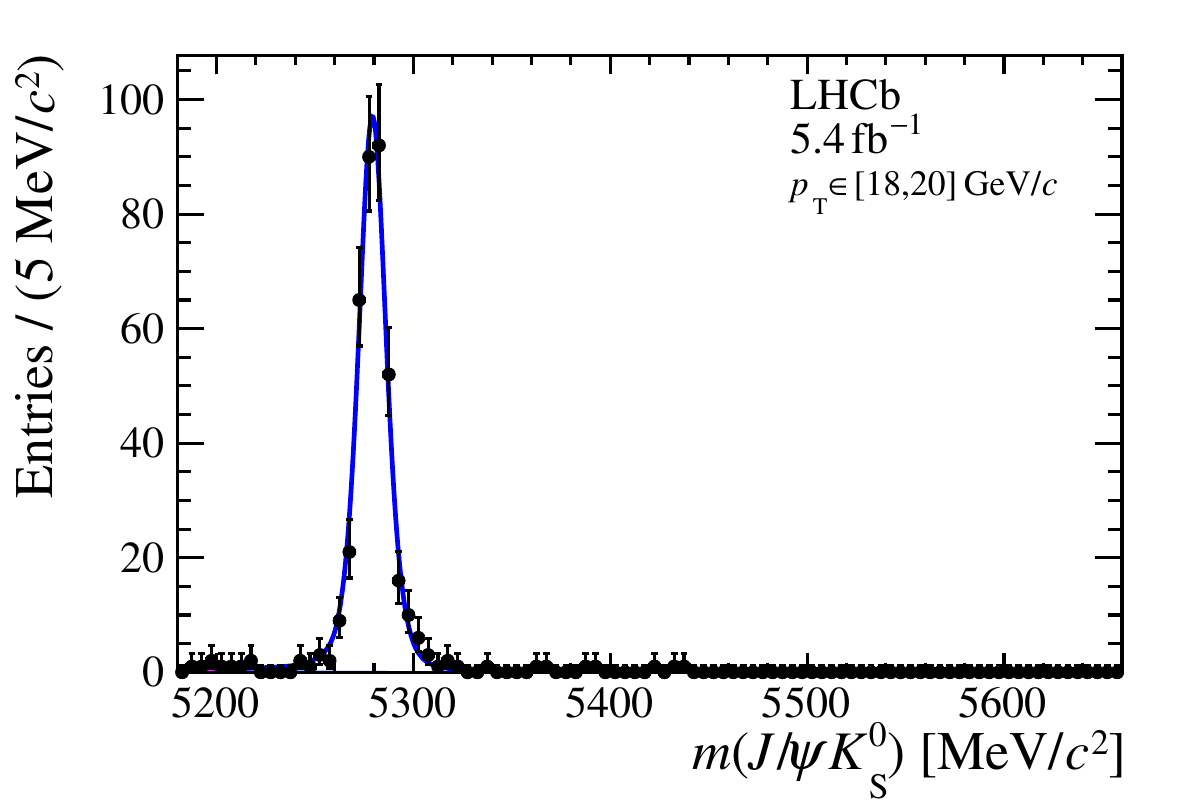} 
    \caption{
    Reconstructed $\jpsi\KS$ mass distribution of selected \BdJpsiKS candidates in bins of \Bz meson \pt, reconstructed from long-category tracks. 
    The line and shaded regions show the result of the fit described in the text. 
    }
    \label{fig:KS:long}
\end{figure}

\begin{figure}[!htb] 
    \centering
    \includegraphics[width=0.32\linewidth]{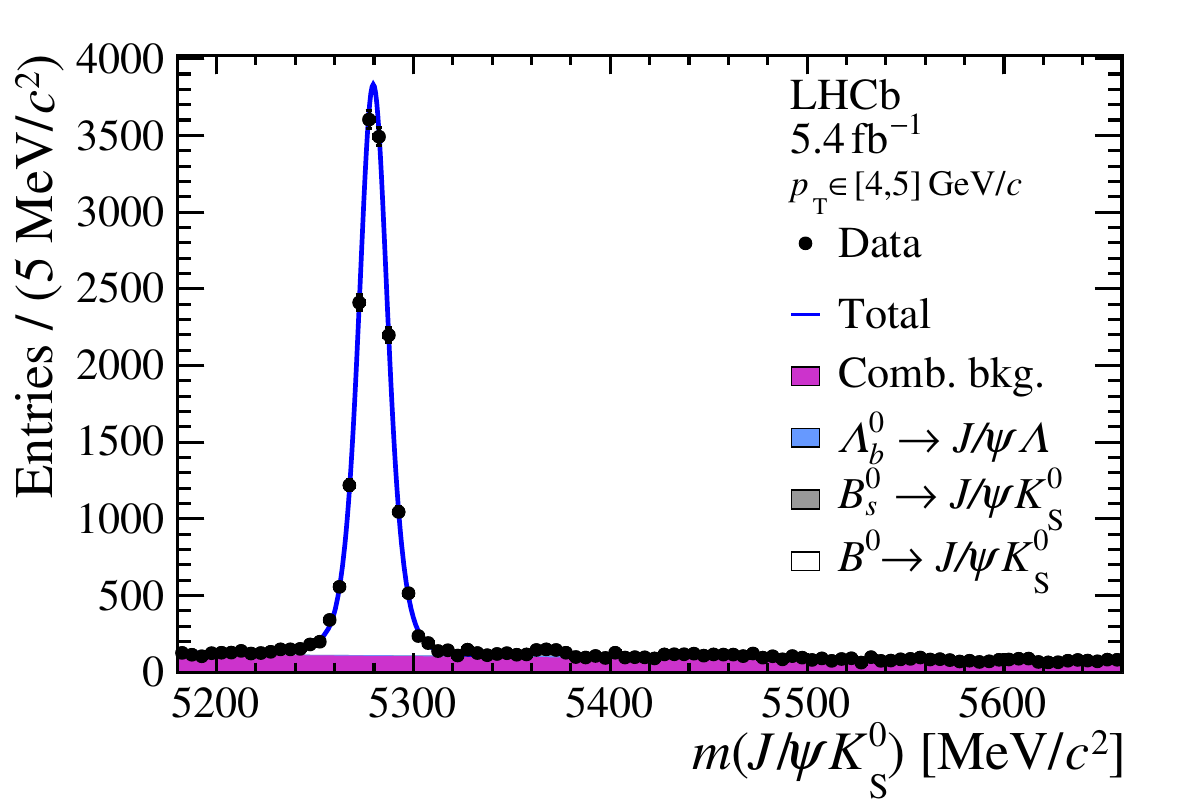} 
    \includegraphics[width=0.32\linewidth]{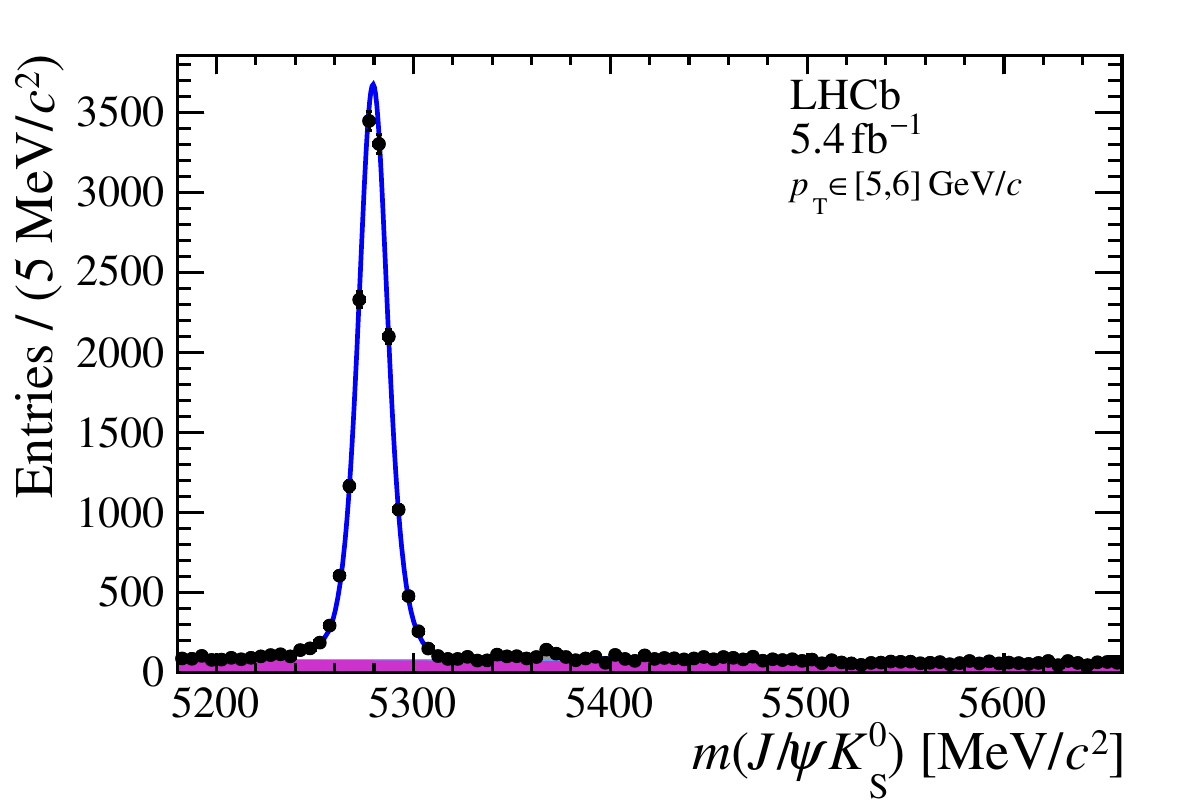} 
    \includegraphics[width=0.32\linewidth]{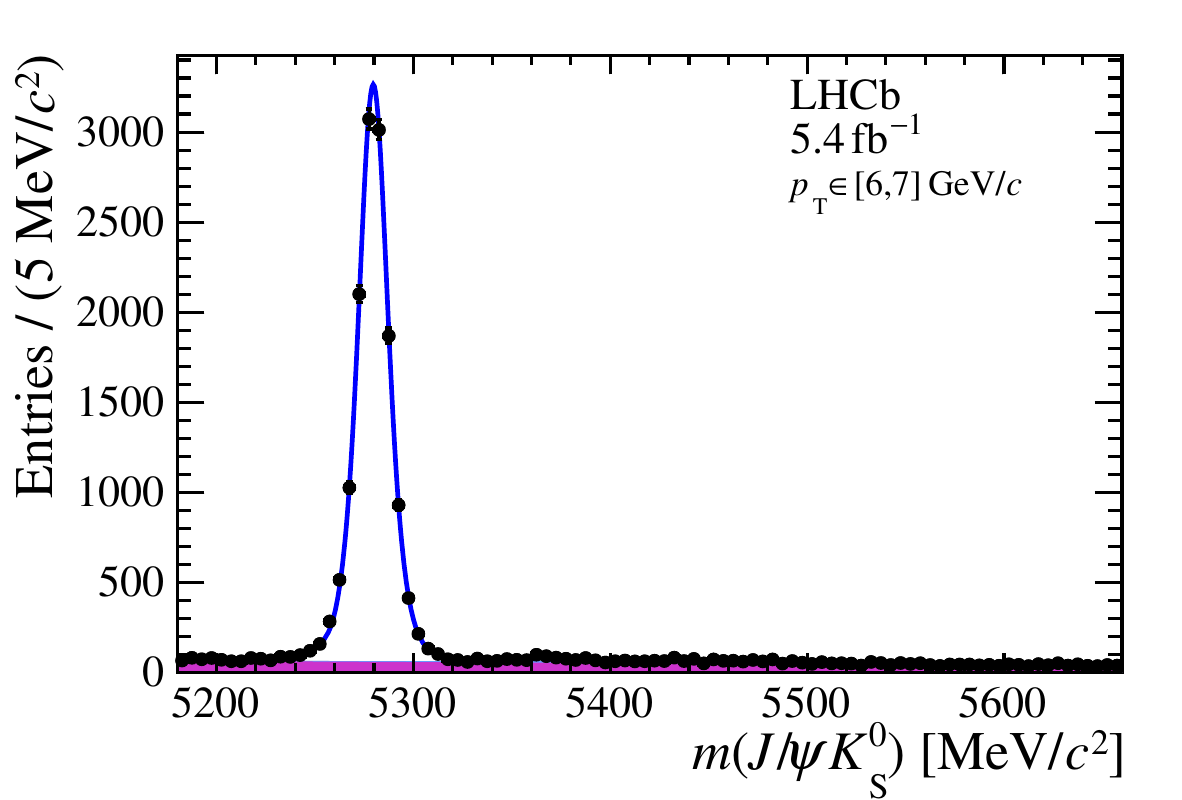} \\
    \includegraphics[width=0.32\linewidth]{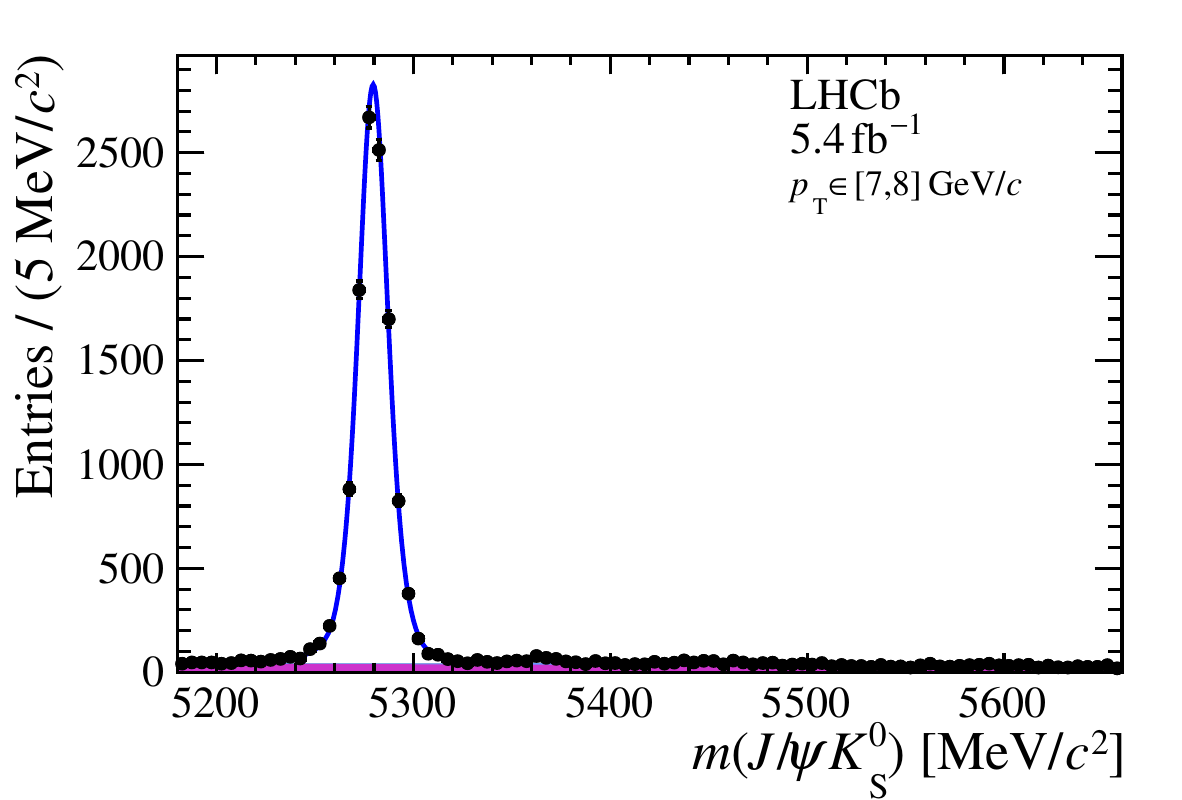} 
    \includegraphics[width=0.32\linewidth]{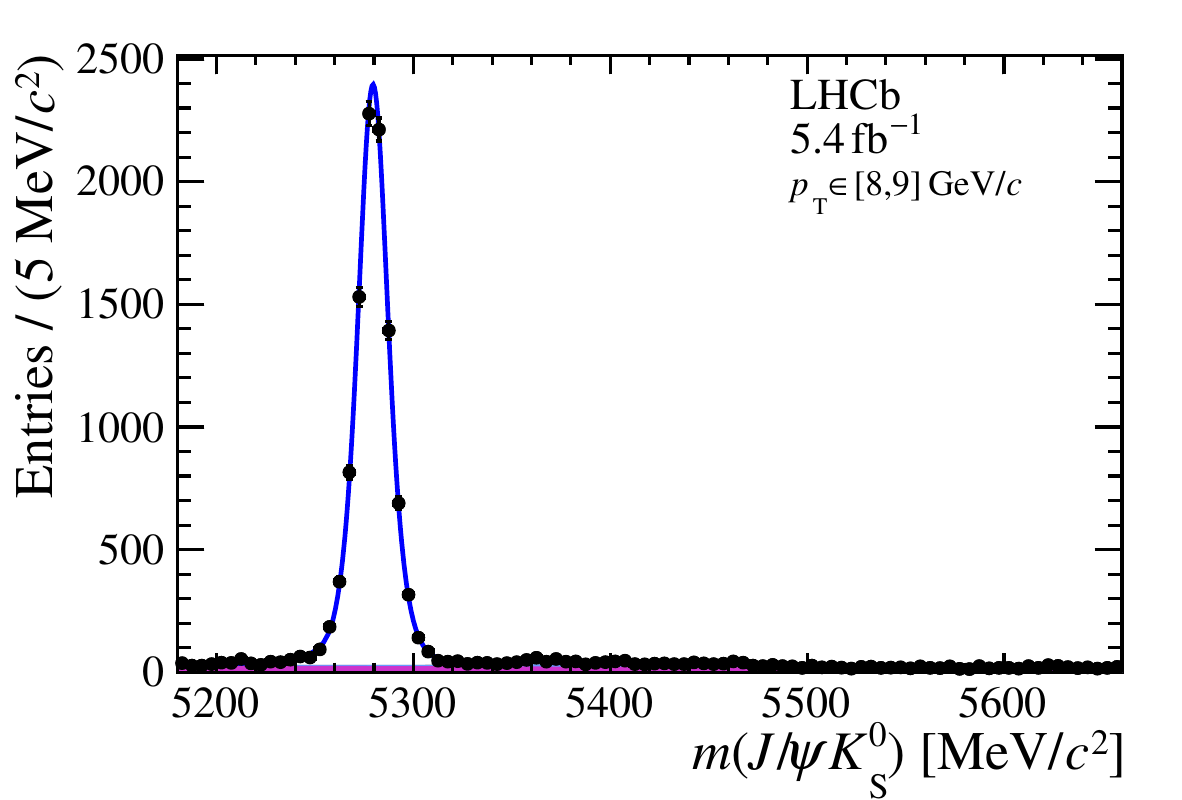} 
    \includegraphics[width=0.32\linewidth]{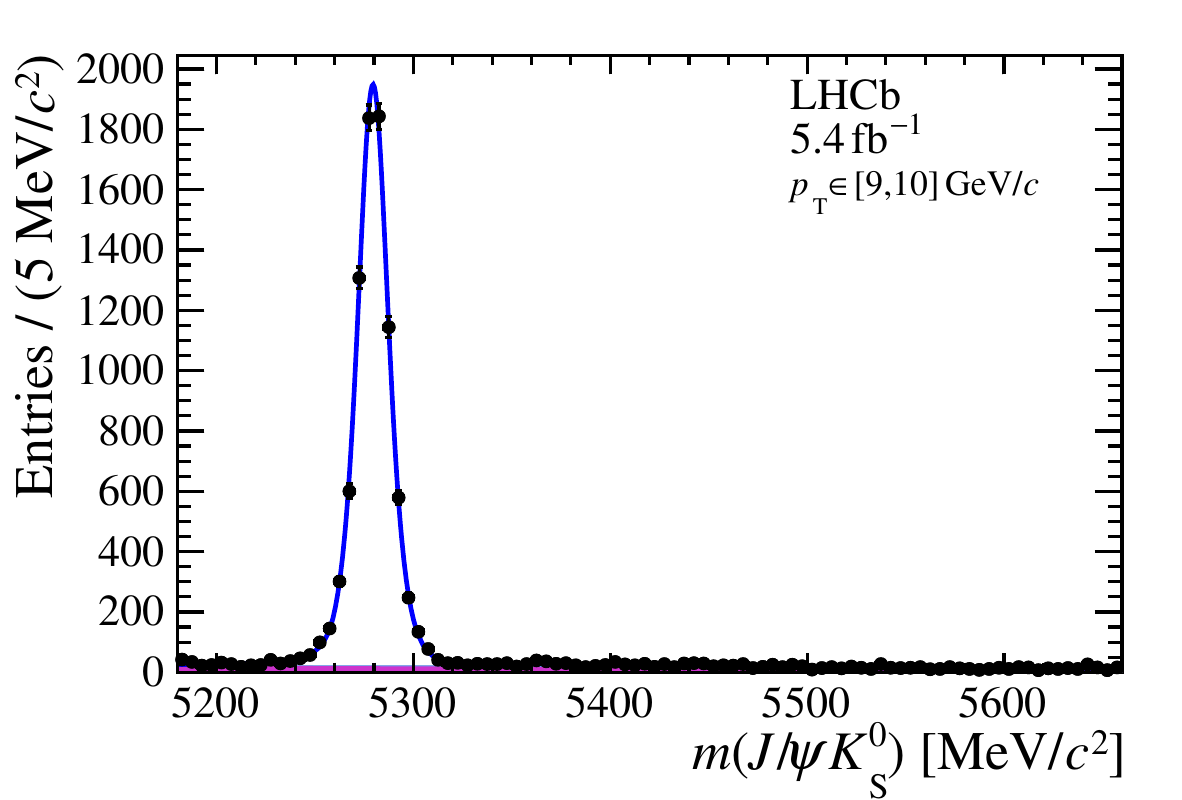} \\
    \includegraphics[width=0.32\linewidth]{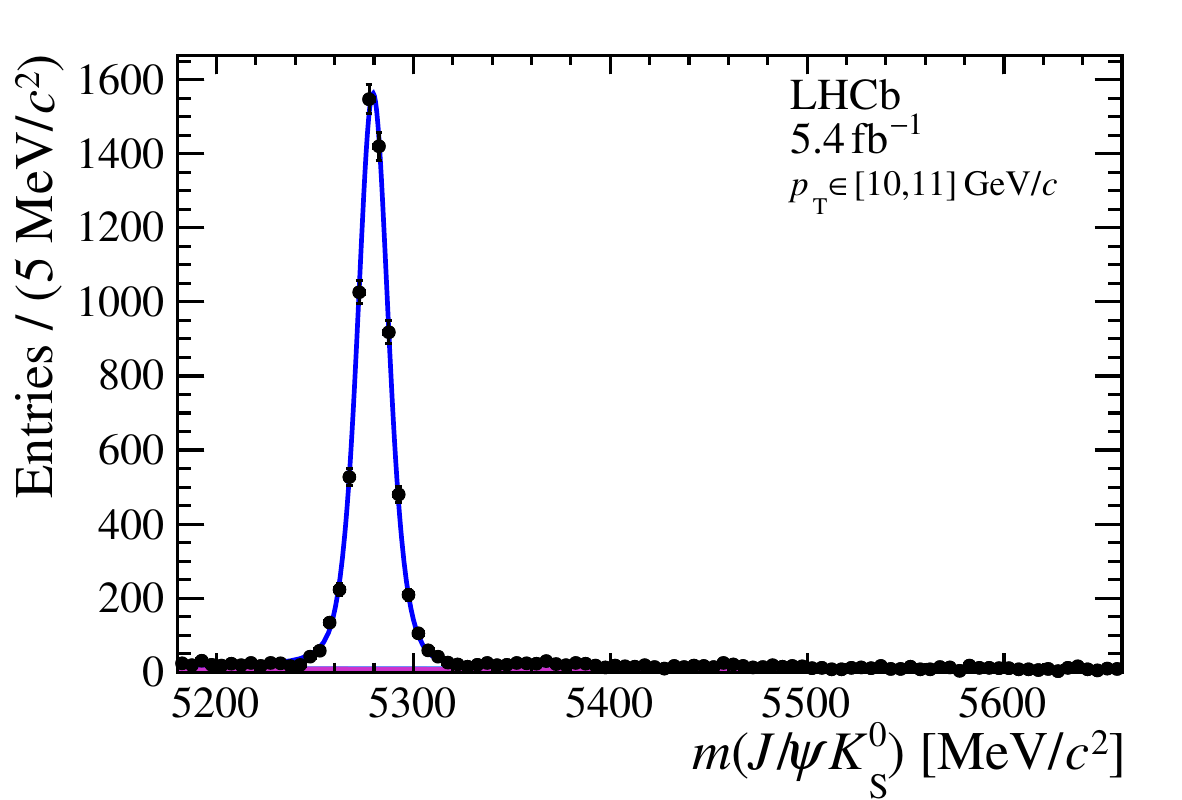} 
    \includegraphics[width=0.32\linewidth]{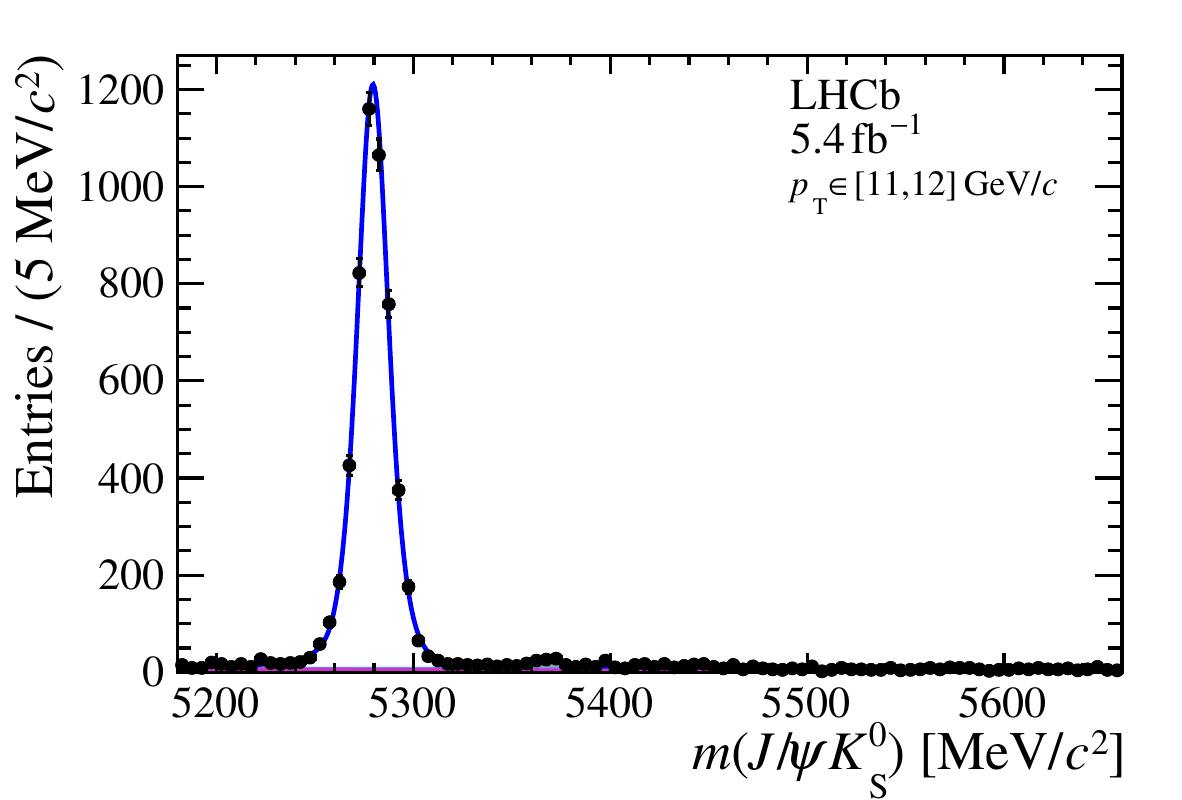} 
    \includegraphics[width=0.32\linewidth]{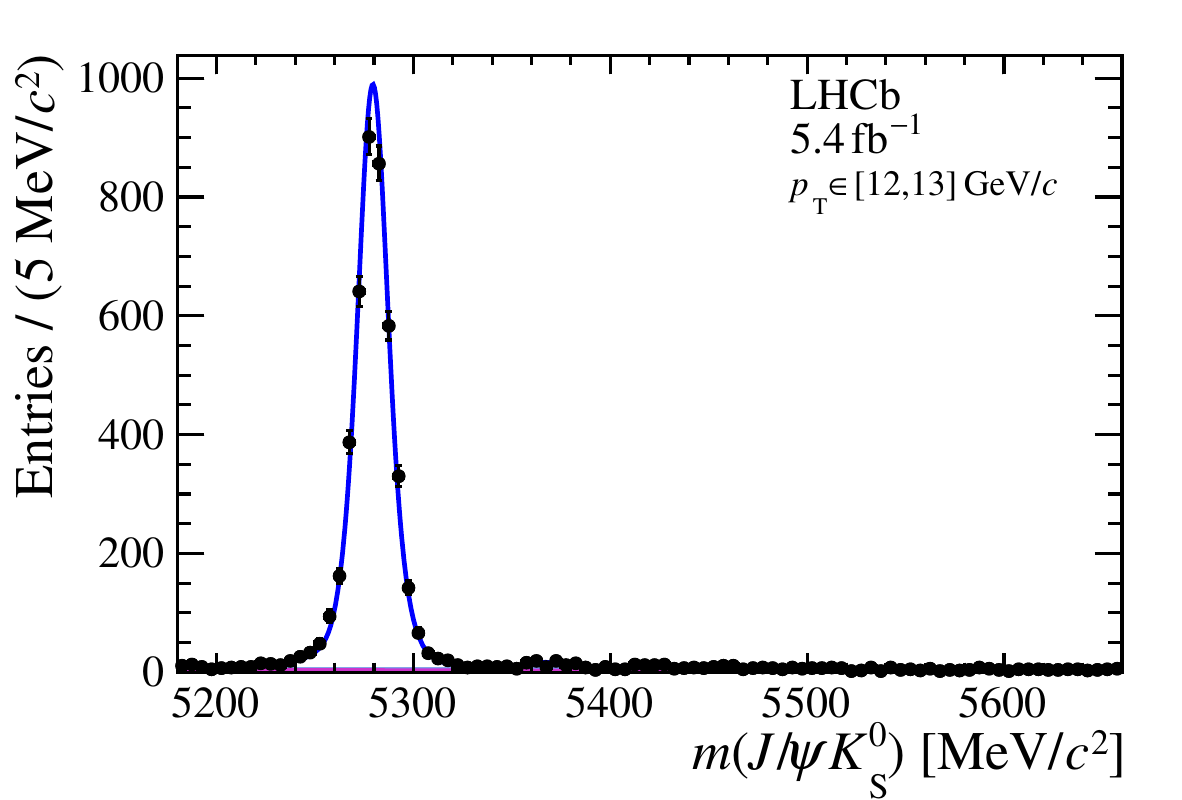} \\
    \includegraphics[width=0.32\linewidth]{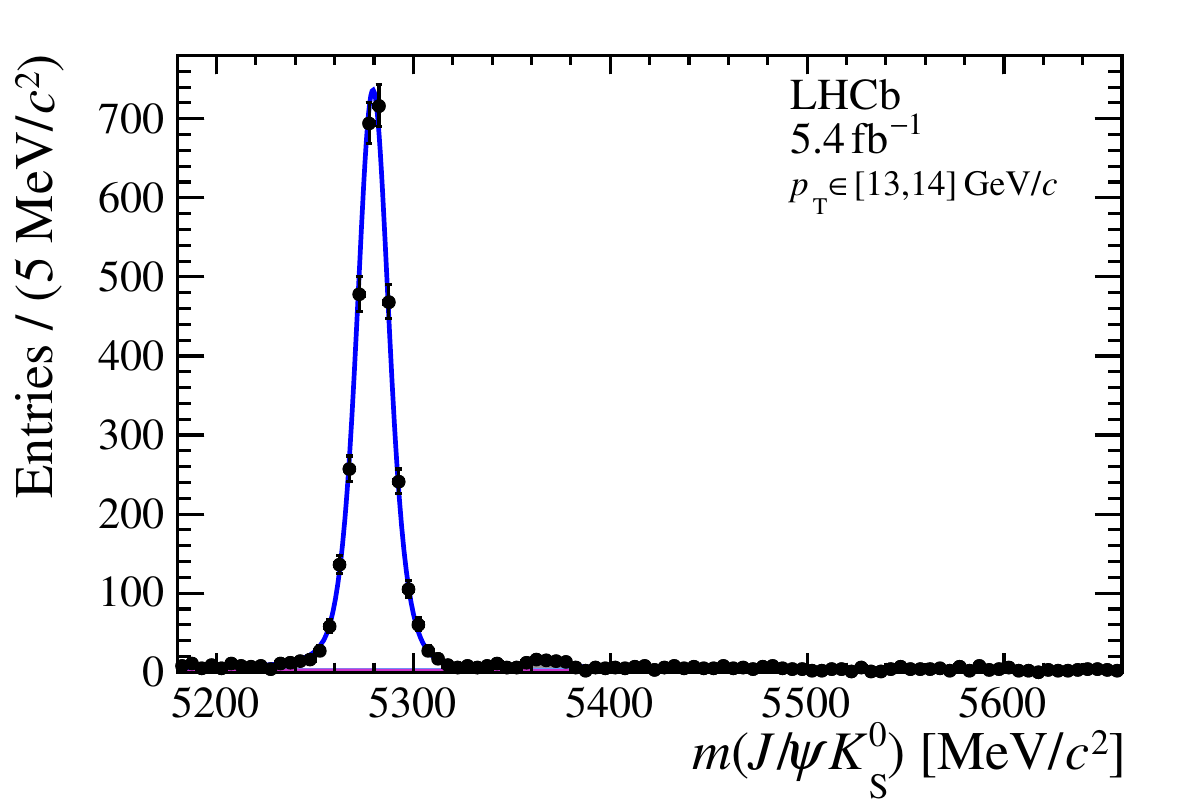} 
    \includegraphics[width=0.32\linewidth]{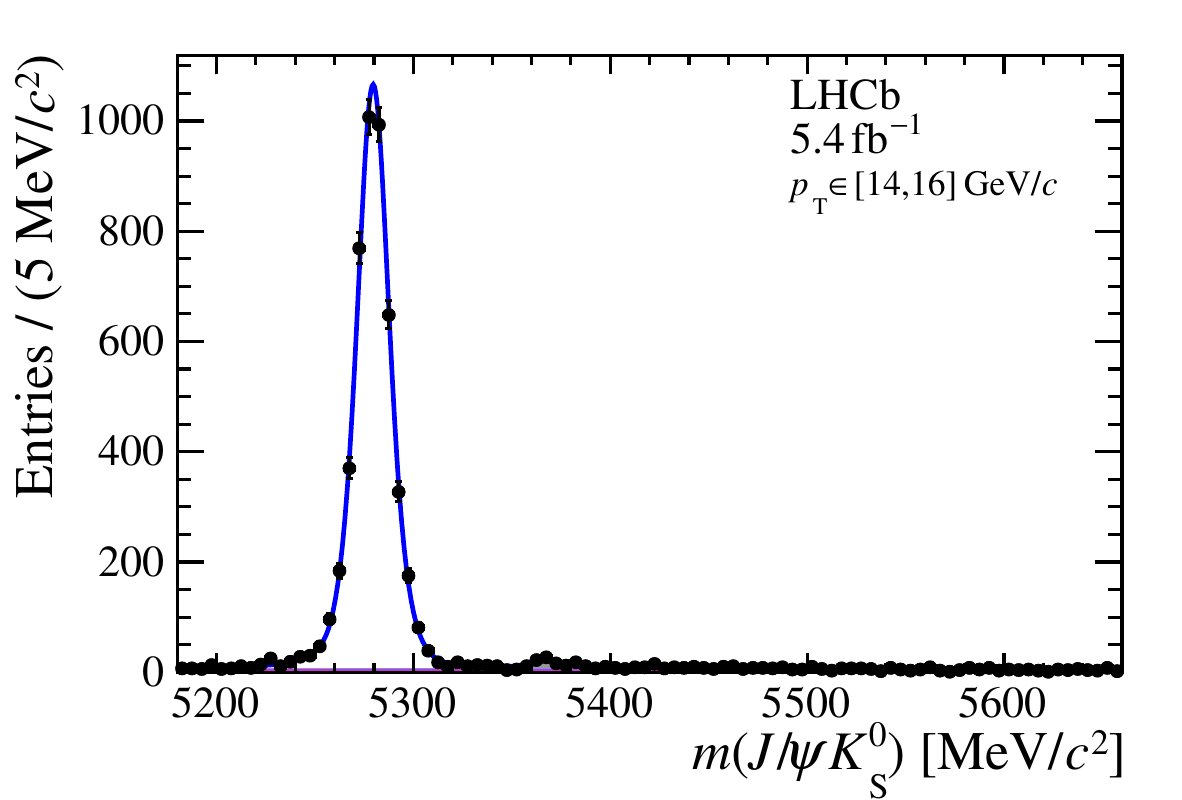} 
    \includegraphics[width=0.32\linewidth]{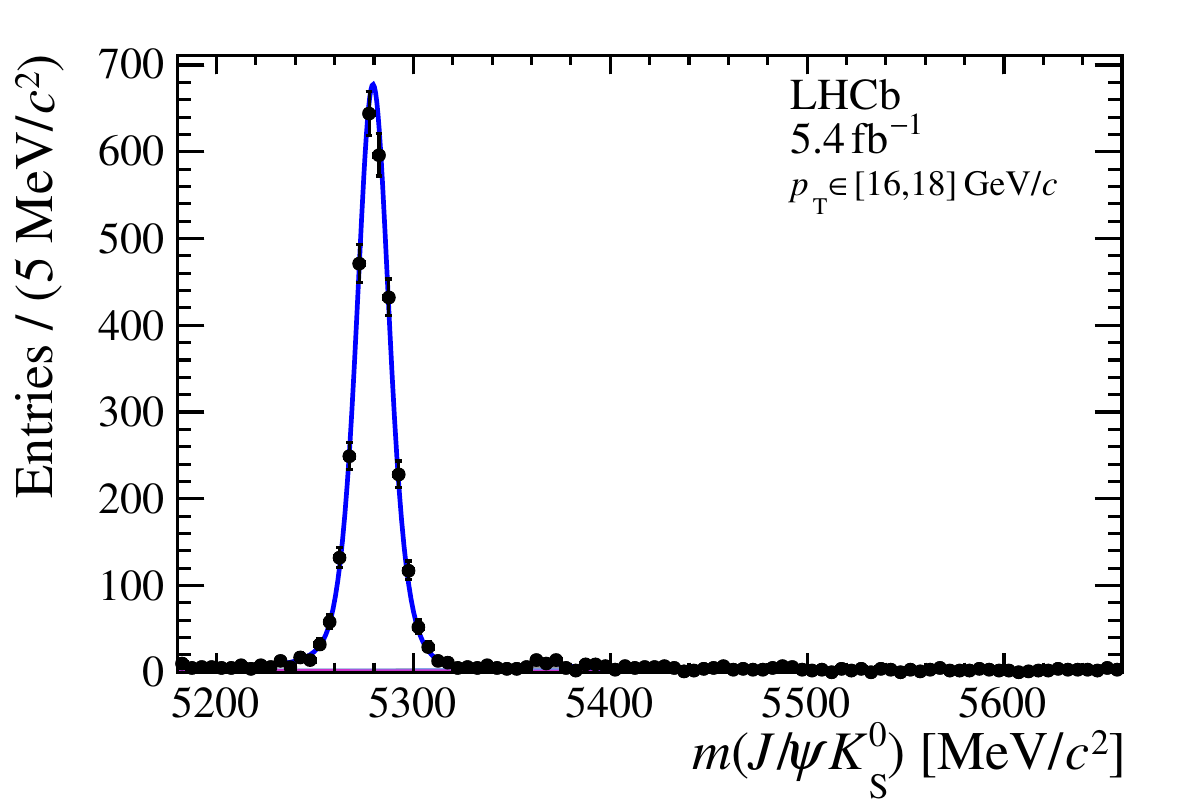} \\
    \includegraphics[width=0.32\linewidth]{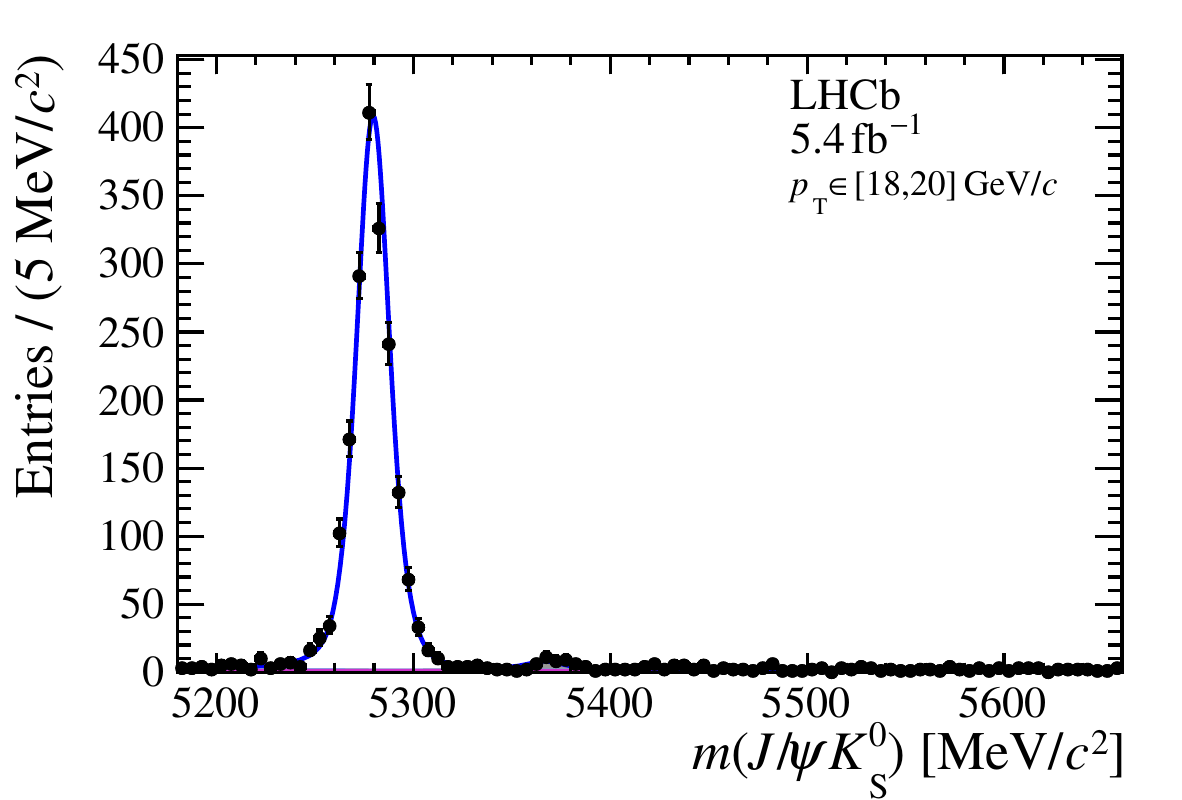} 
    \caption{
    Reconstructed $\jpsi\KS$ mass distribution of selected \BdJpsiKS candidates in bins of \Bz meson \pt, reconstructed from downstream-category tracks. 
    The line and shaded regions show the result of the fit described in the text. 
    }
    \label{fig:KS:downstream}
\end{figure}

\begin{figure}[!htb] 
    \centering
    \includegraphics[width=0.32\linewidth]{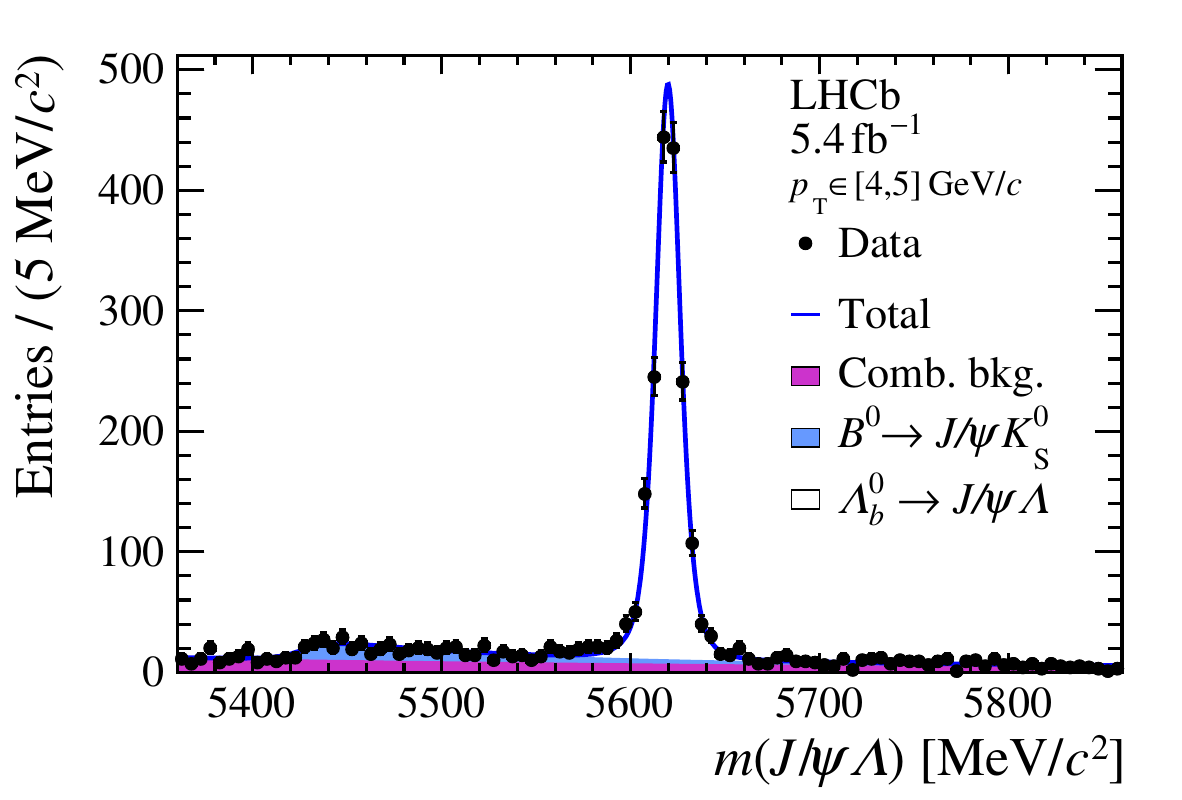} 
    \includegraphics[width=0.32\linewidth]{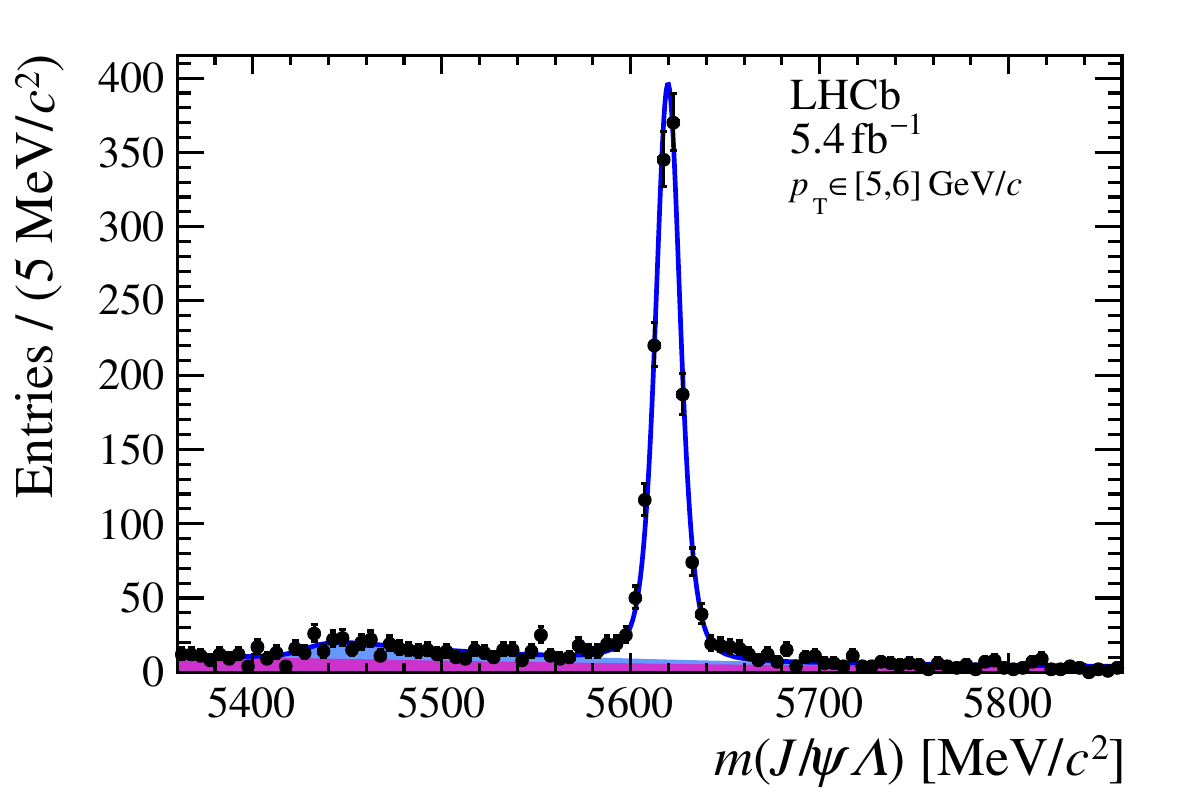} 
    \includegraphics[width=0.32\linewidth]{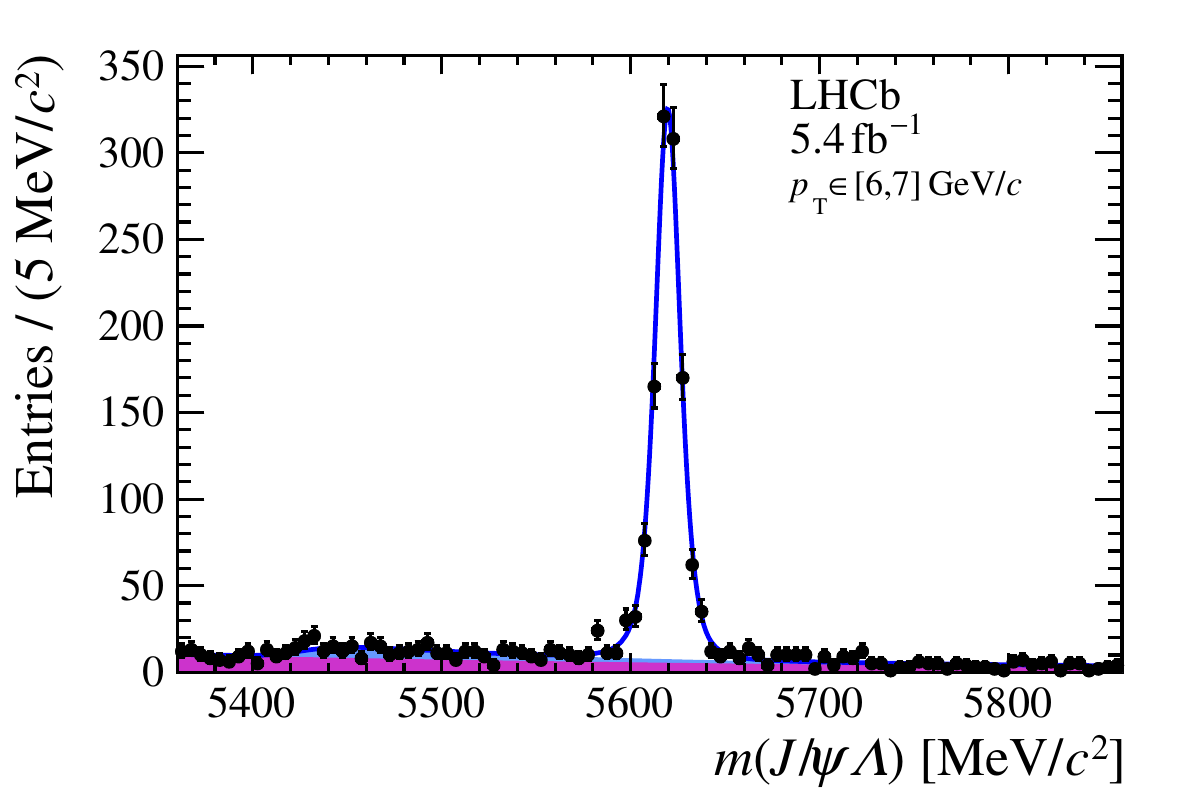} \\
    \includegraphics[width=0.32\linewidth]{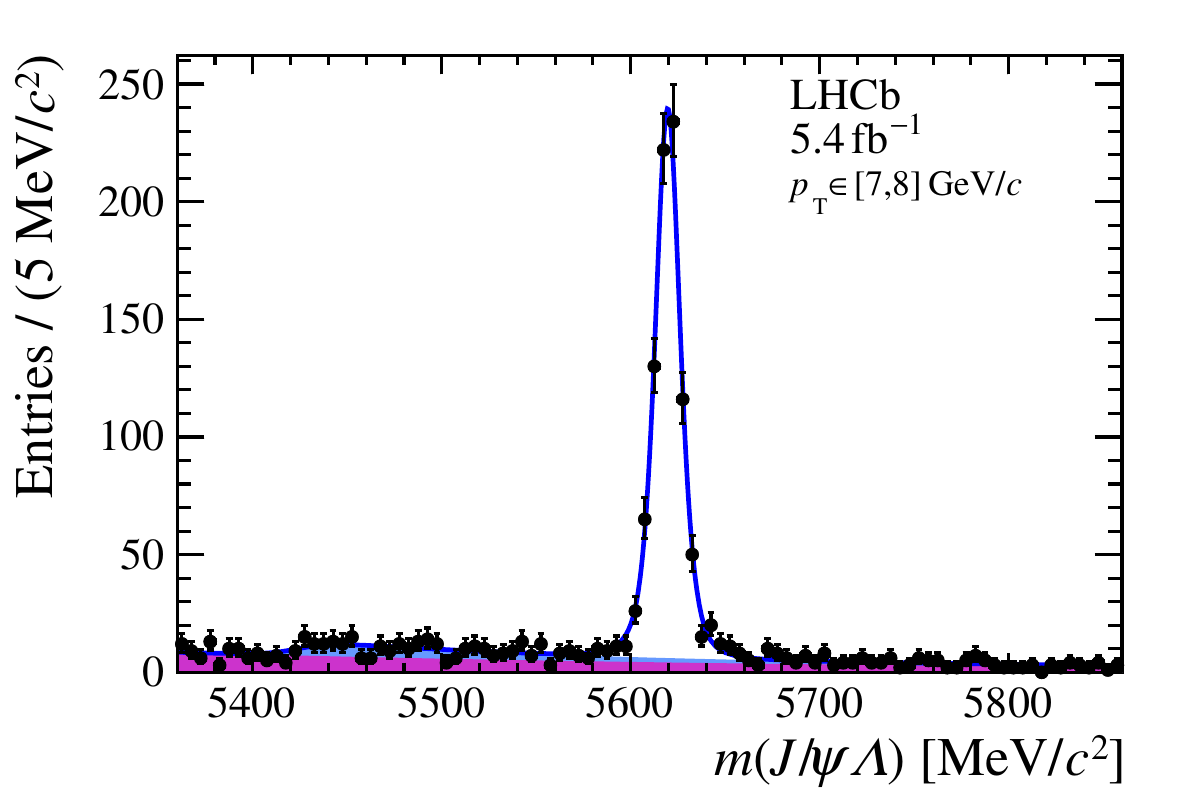} 
    \includegraphics[width=0.32\linewidth]{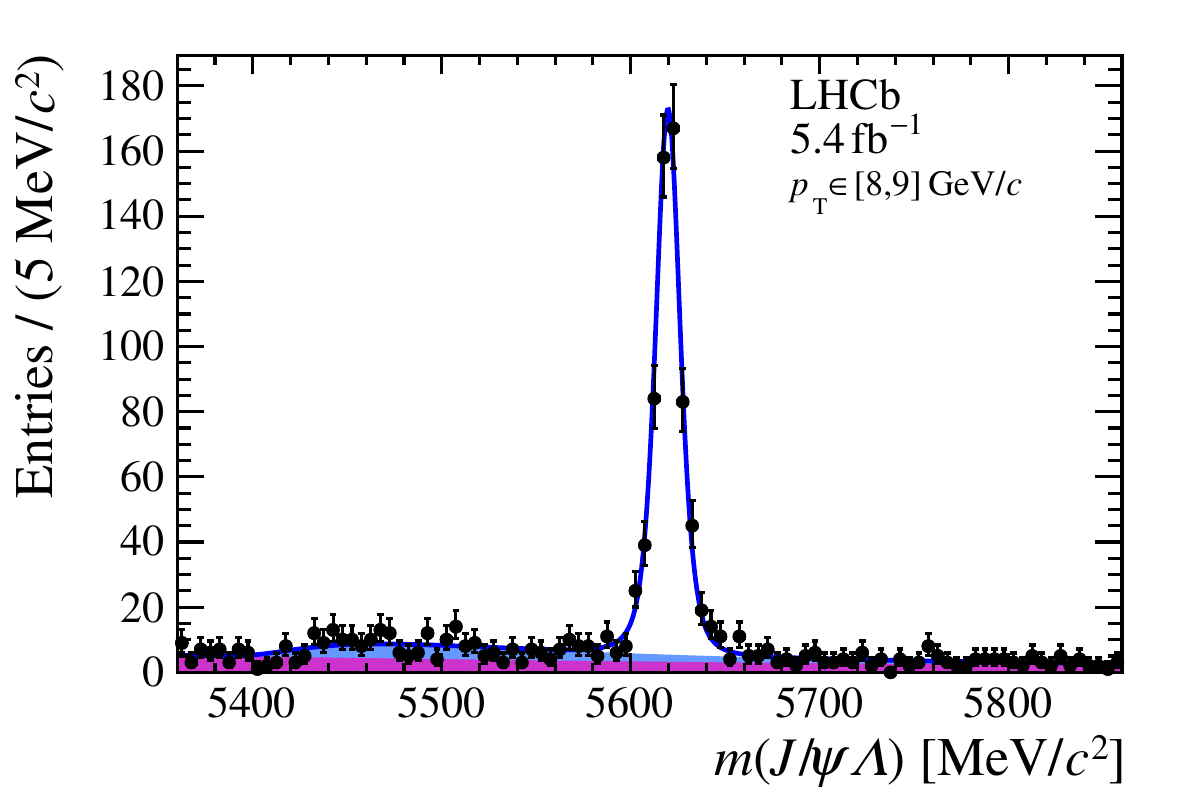} 
    \includegraphics[width=0.32\linewidth]{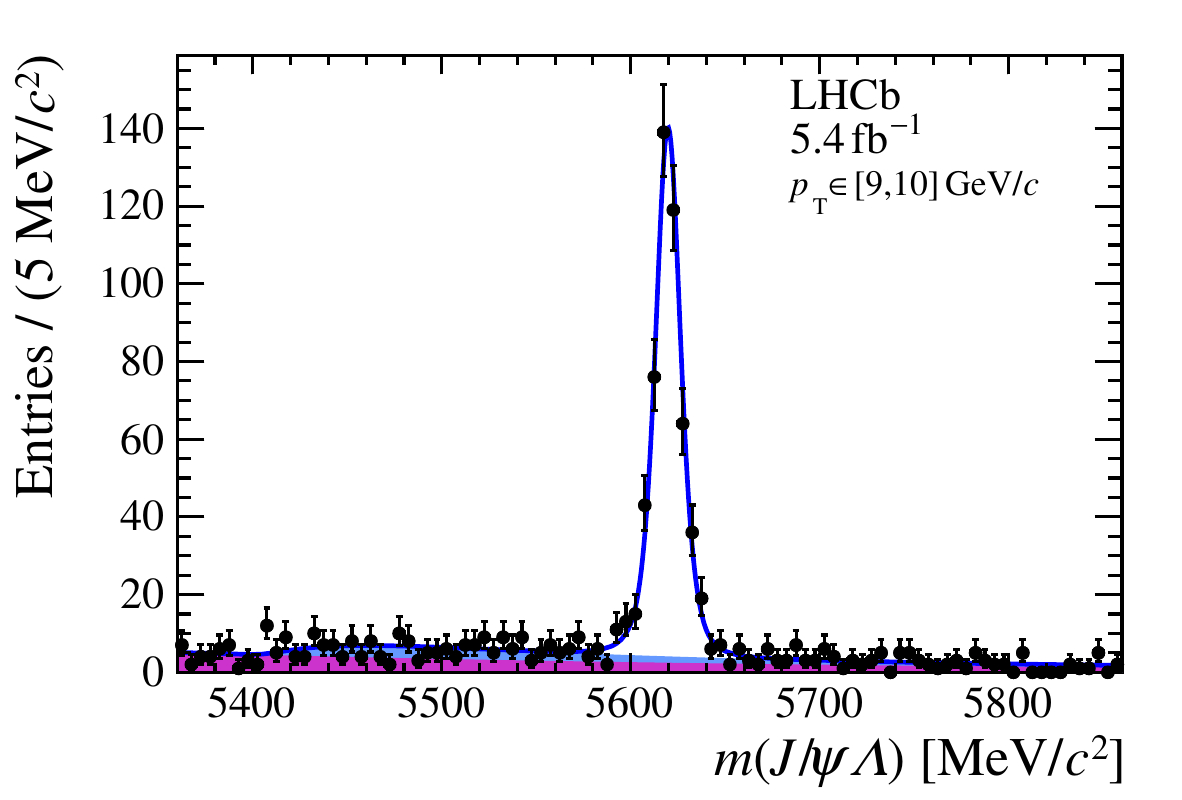} \\
    \includegraphics[width=0.32\linewidth]{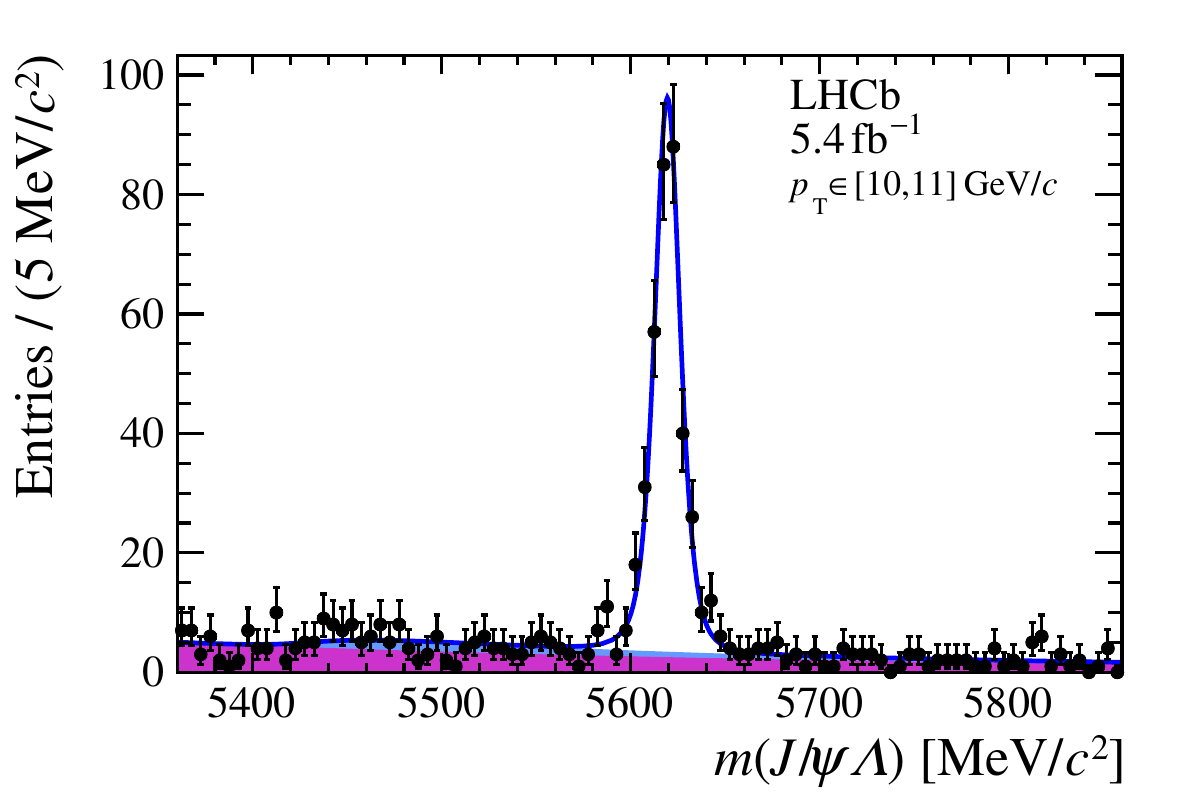} 
    \includegraphics[width=0.32\linewidth]{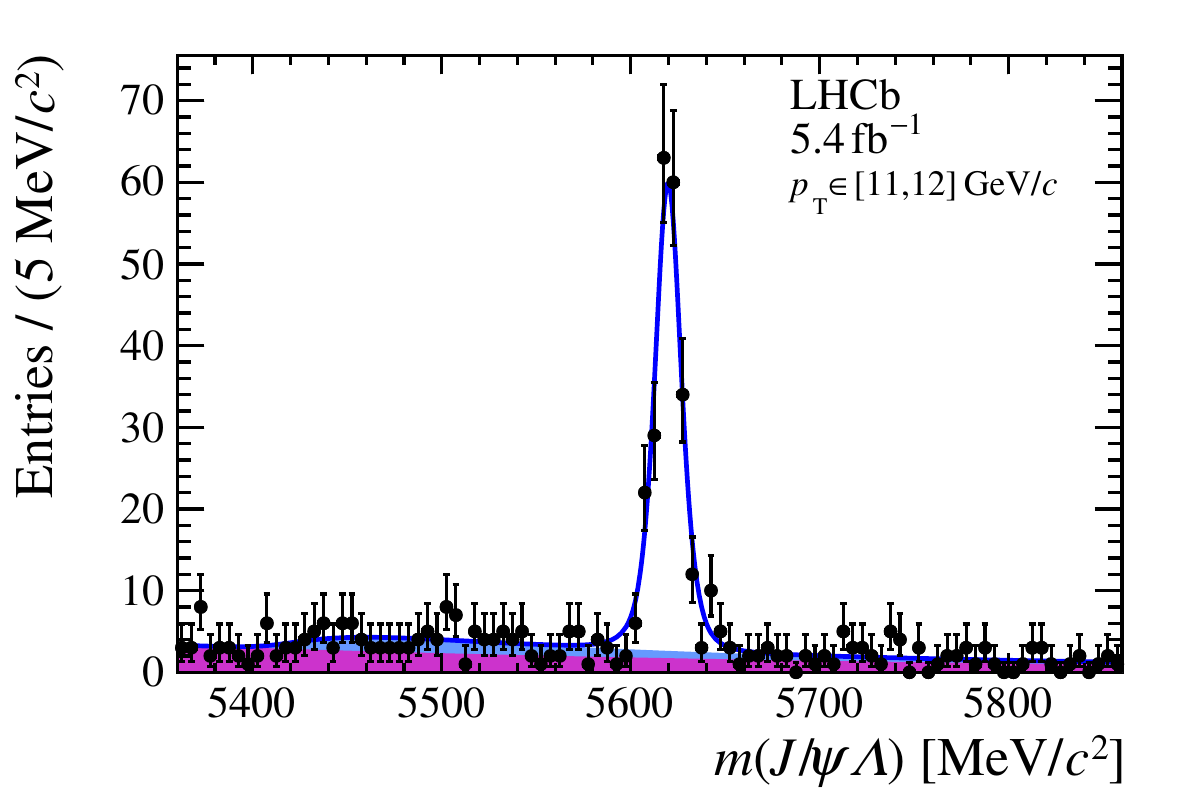} 
    \includegraphics[width=0.32\linewidth]{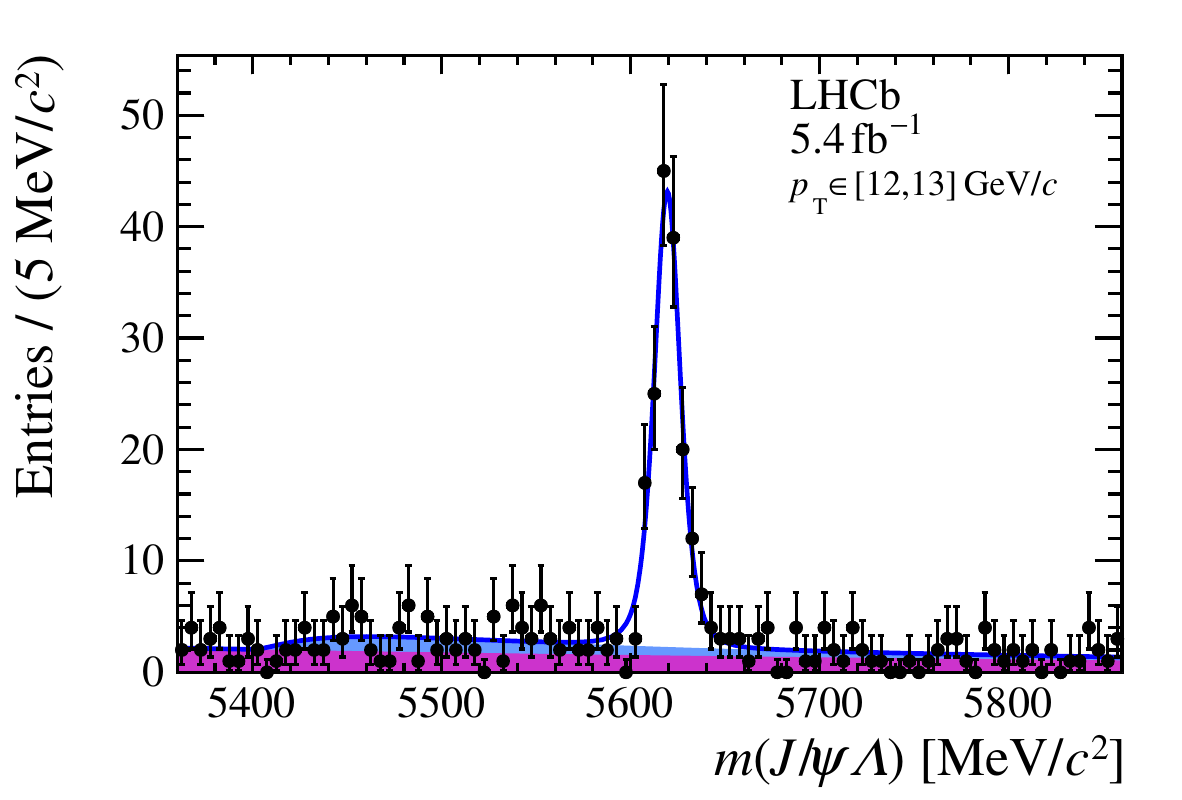} \\
    \includegraphics[width=0.32\linewidth]{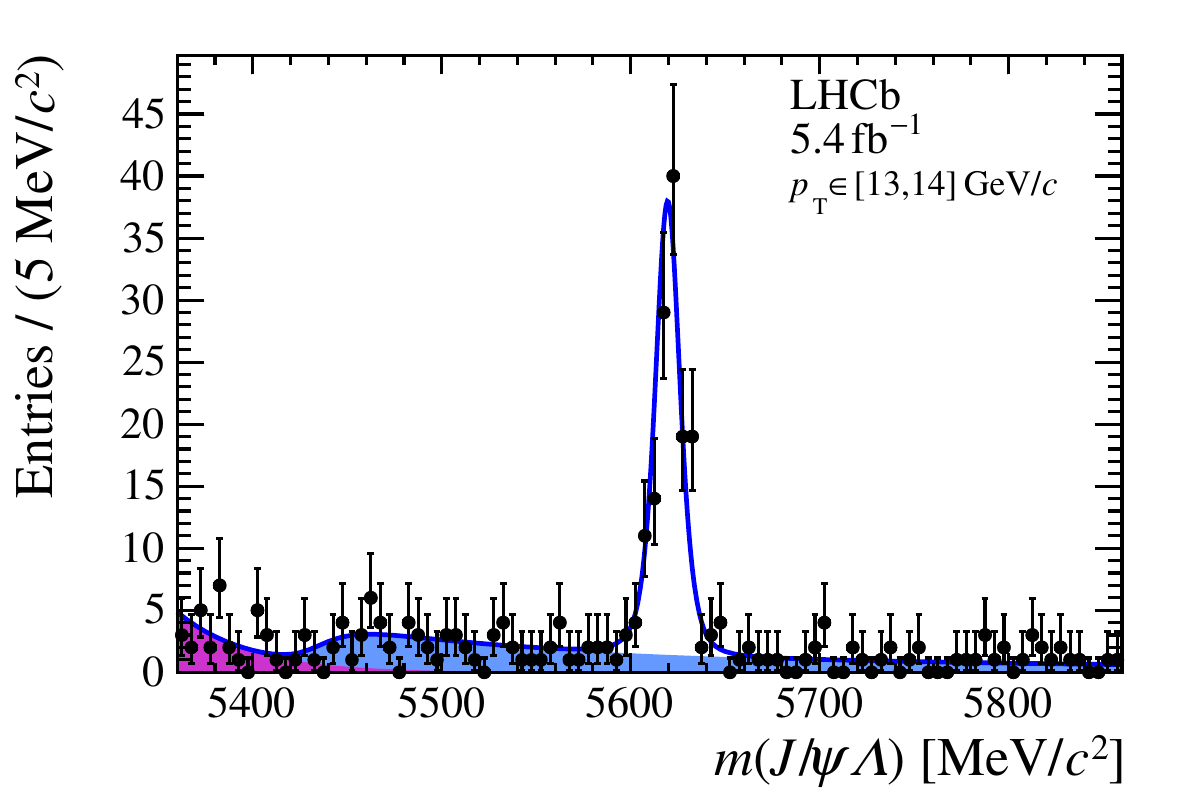} 
    \includegraphics[width=0.32\linewidth]{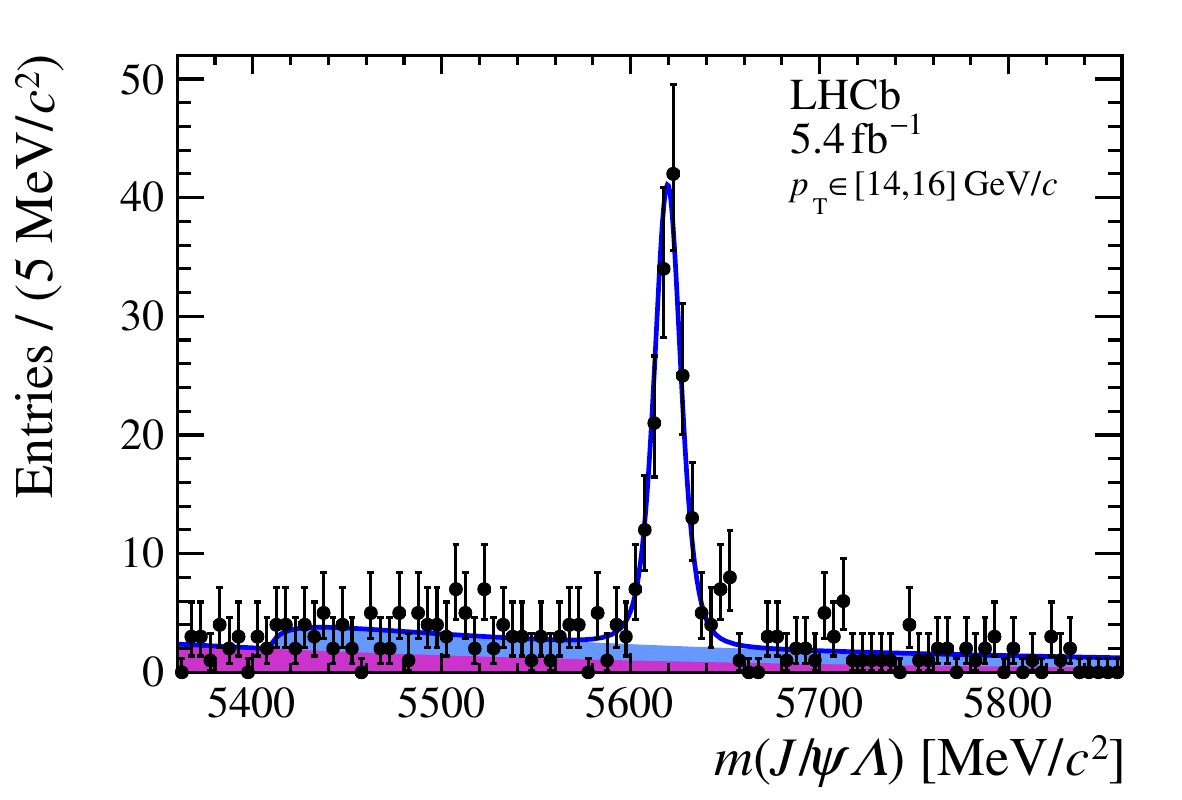} 
    \includegraphics[width=0.32\linewidth]{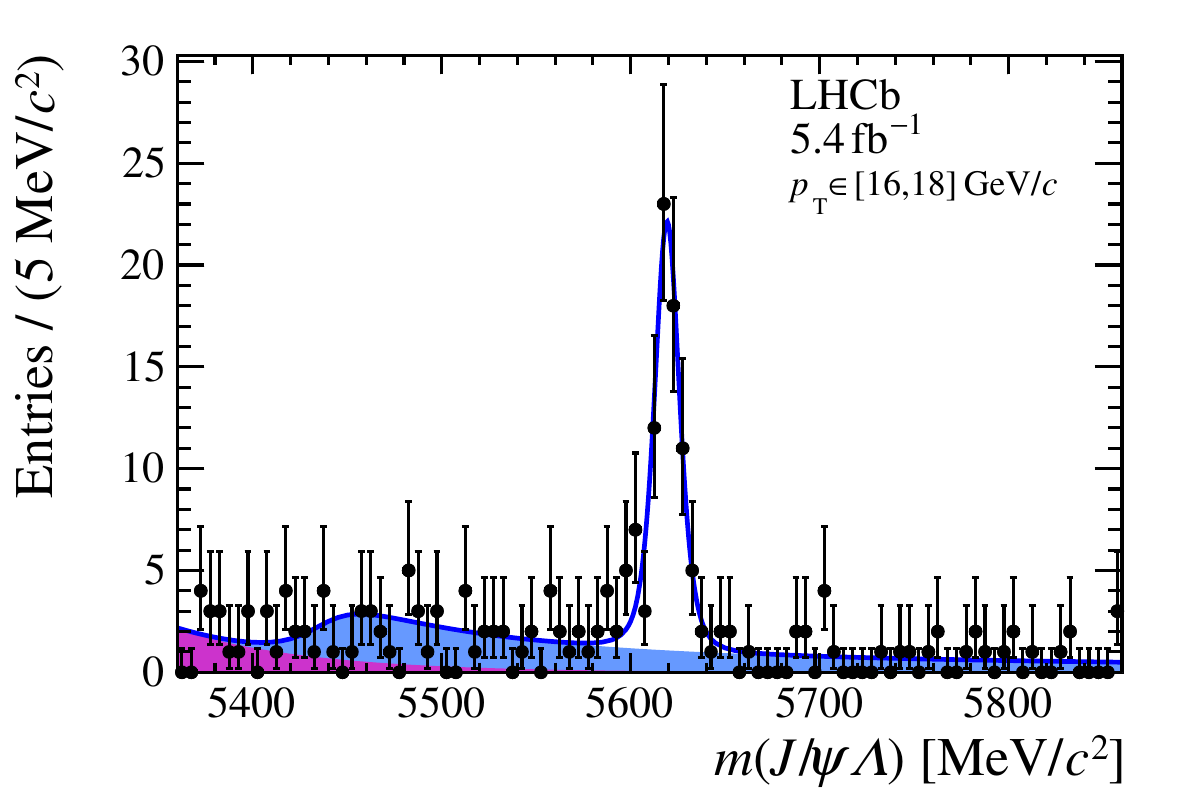} \\
    \includegraphics[width=0.32\linewidth]{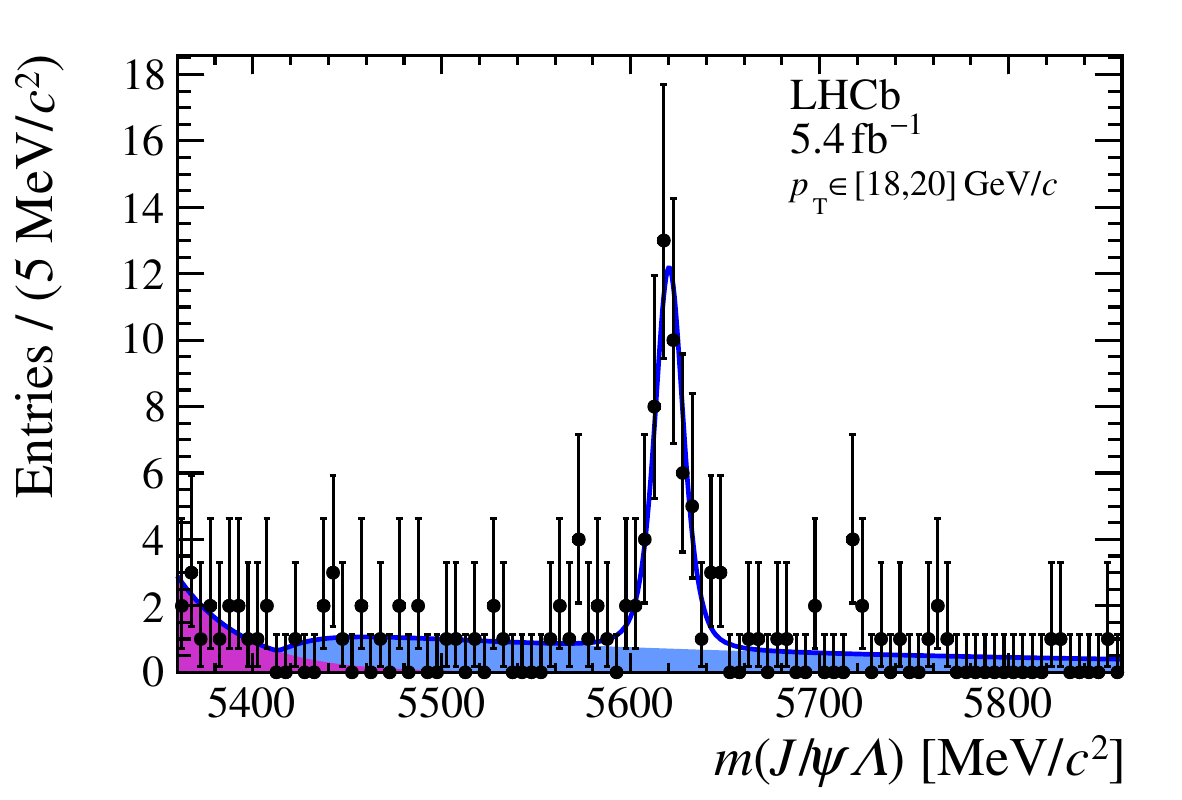} 
    \caption{
    Reconstructed $\jpsi\Lz$ mass distribution of selected \LbJpsiLambda candidates in bins of \Lz baryon \pt, reconstructed from long-category tracks. 
    The line and shaded regions show the result of the fit described in the text. 
    }
    \label{fig:Lambda:long}
\end{figure}

\begin{figure}[!htb] 
    \centering
    \includegraphics[width=0.32\linewidth]{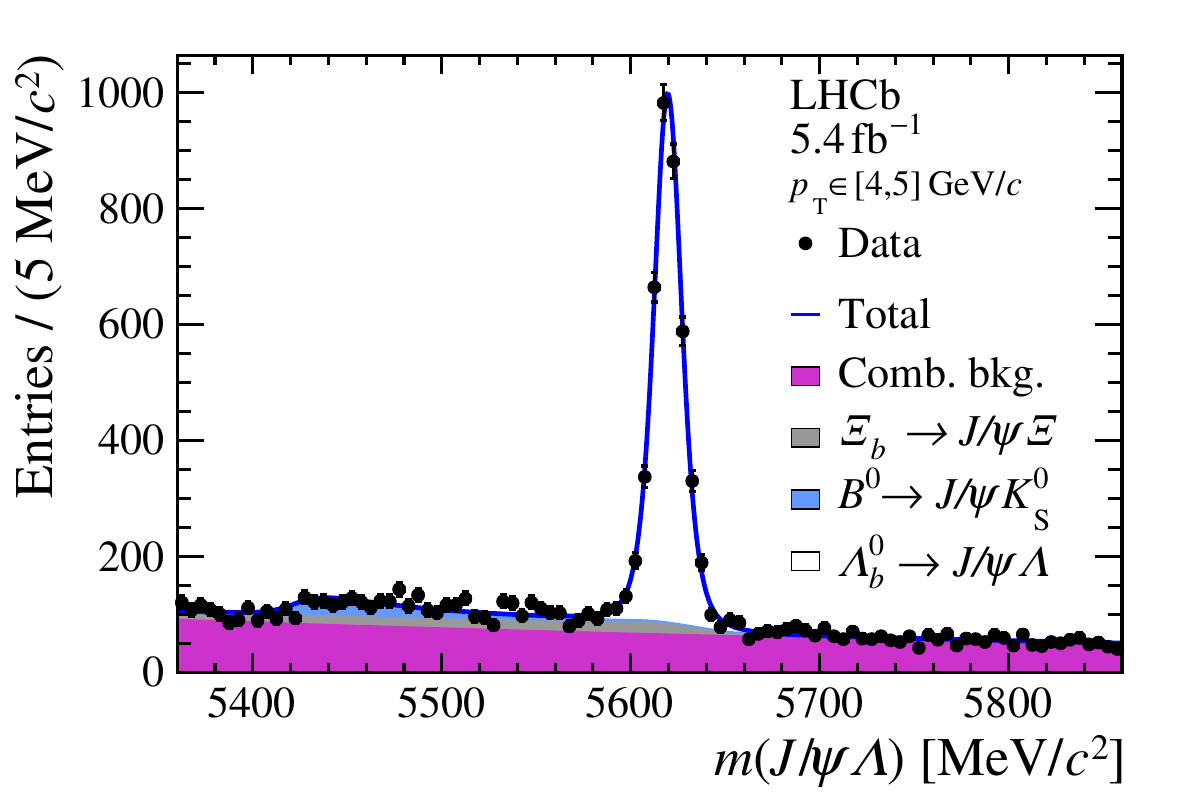} 
    \includegraphics[width=0.32\linewidth]{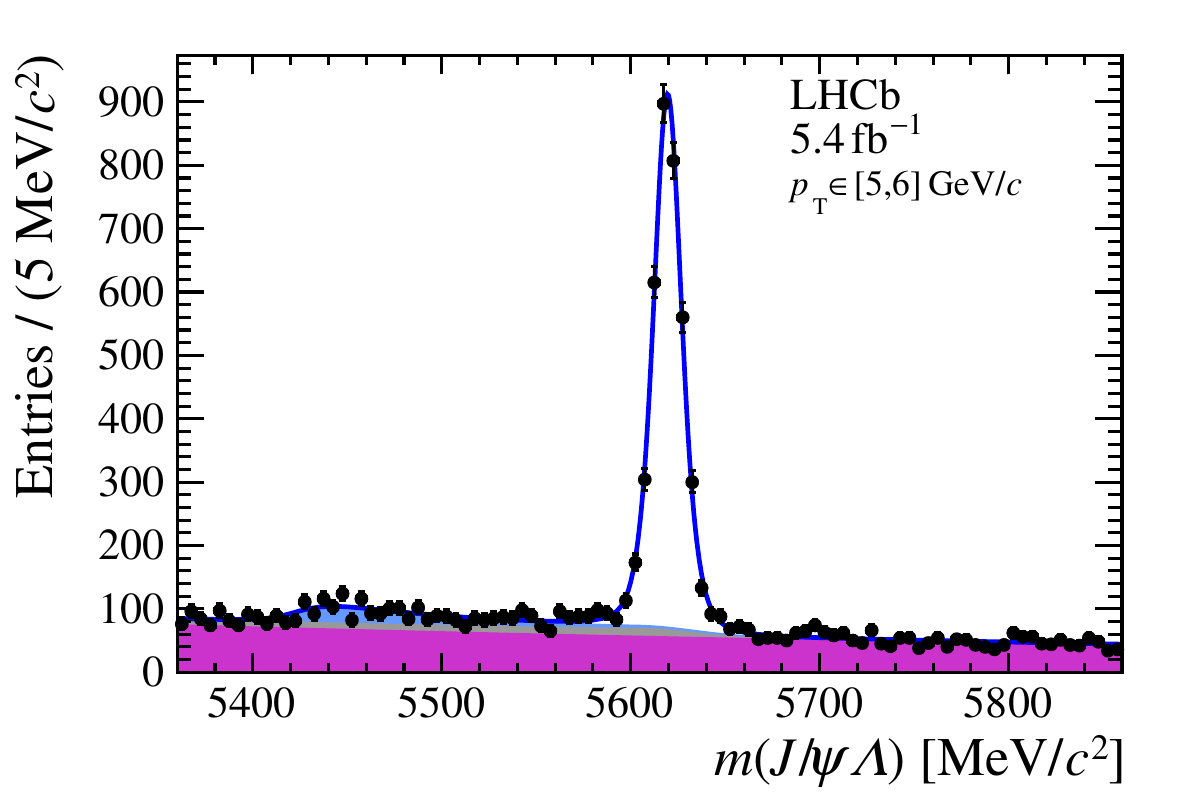} 
    \includegraphics[width=0.32\linewidth]{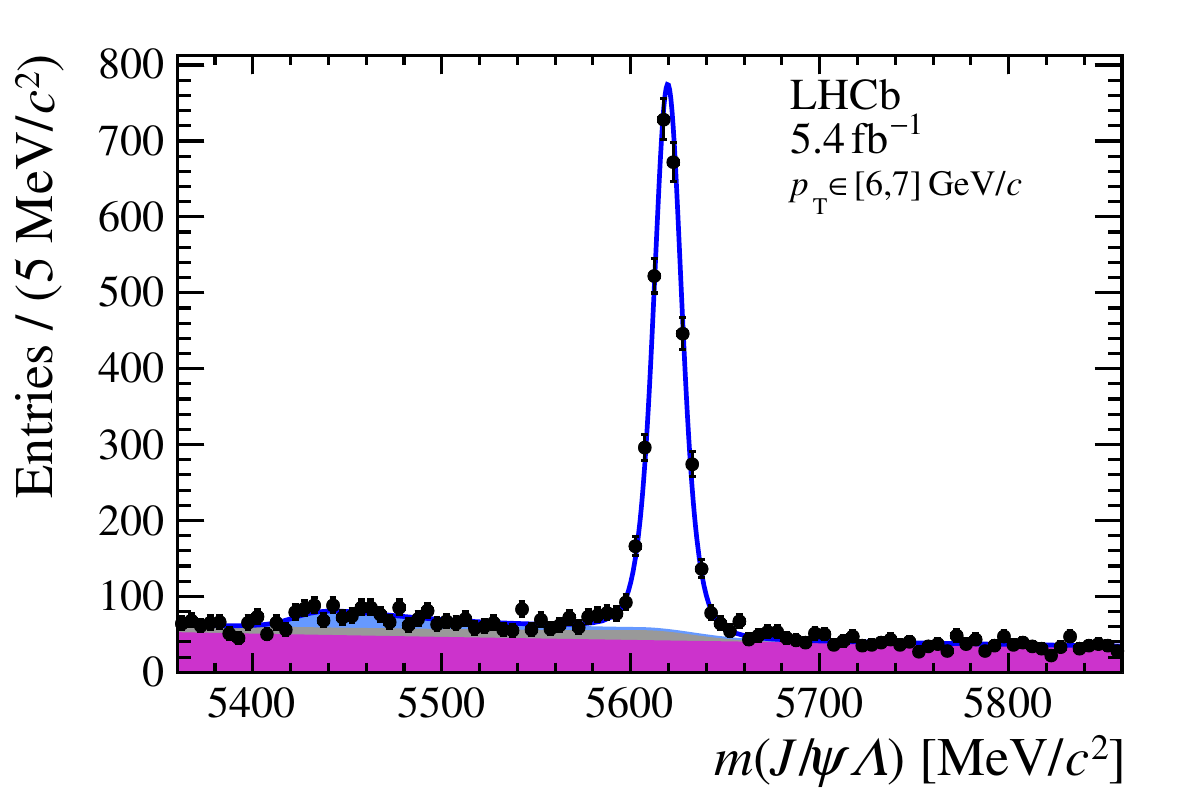} \\
    \includegraphics[width=0.32\linewidth]{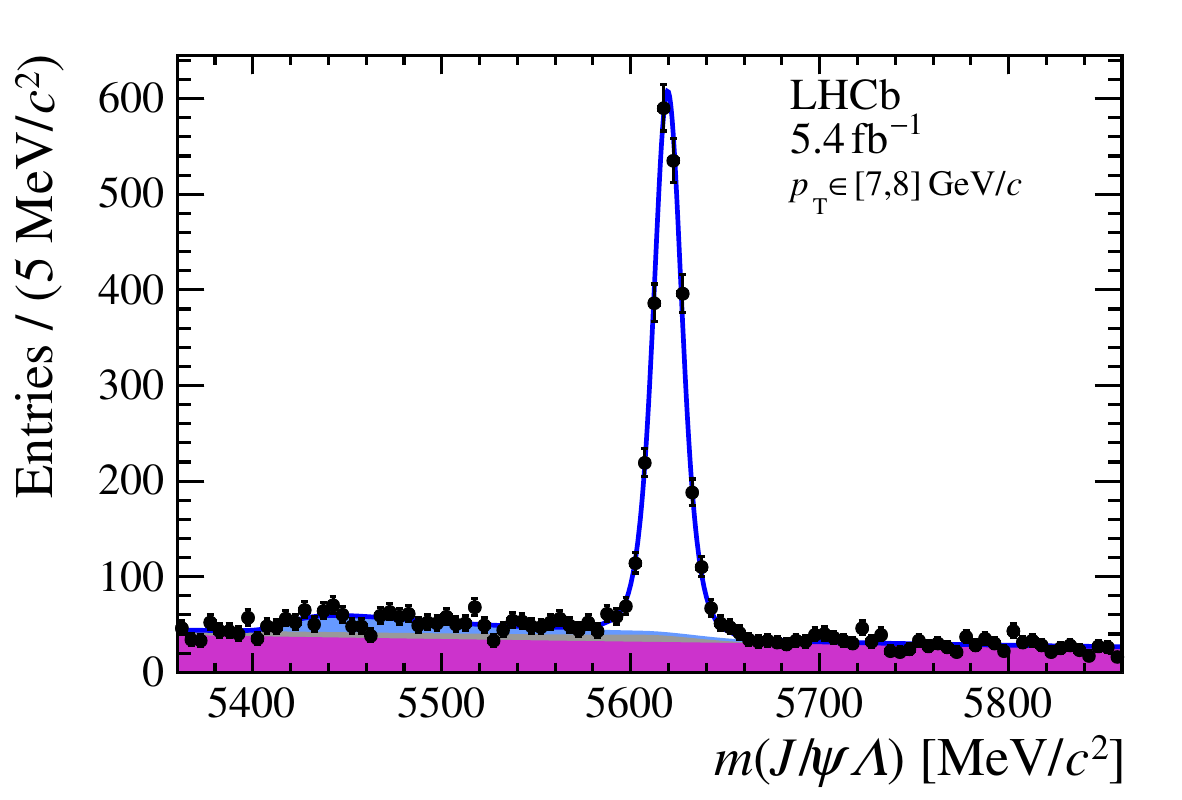} 
    \includegraphics[width=0.32\linewidth]{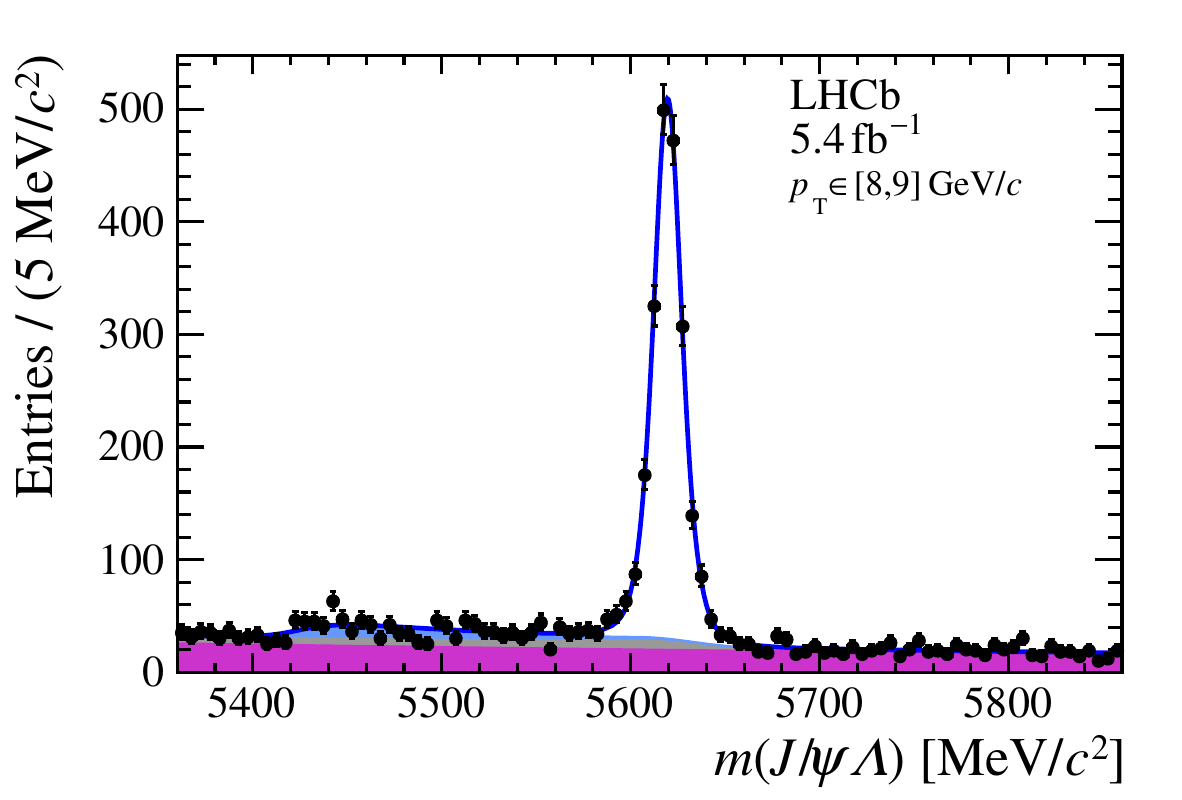} 
    \includegraphics[width=0.32\linewidth]{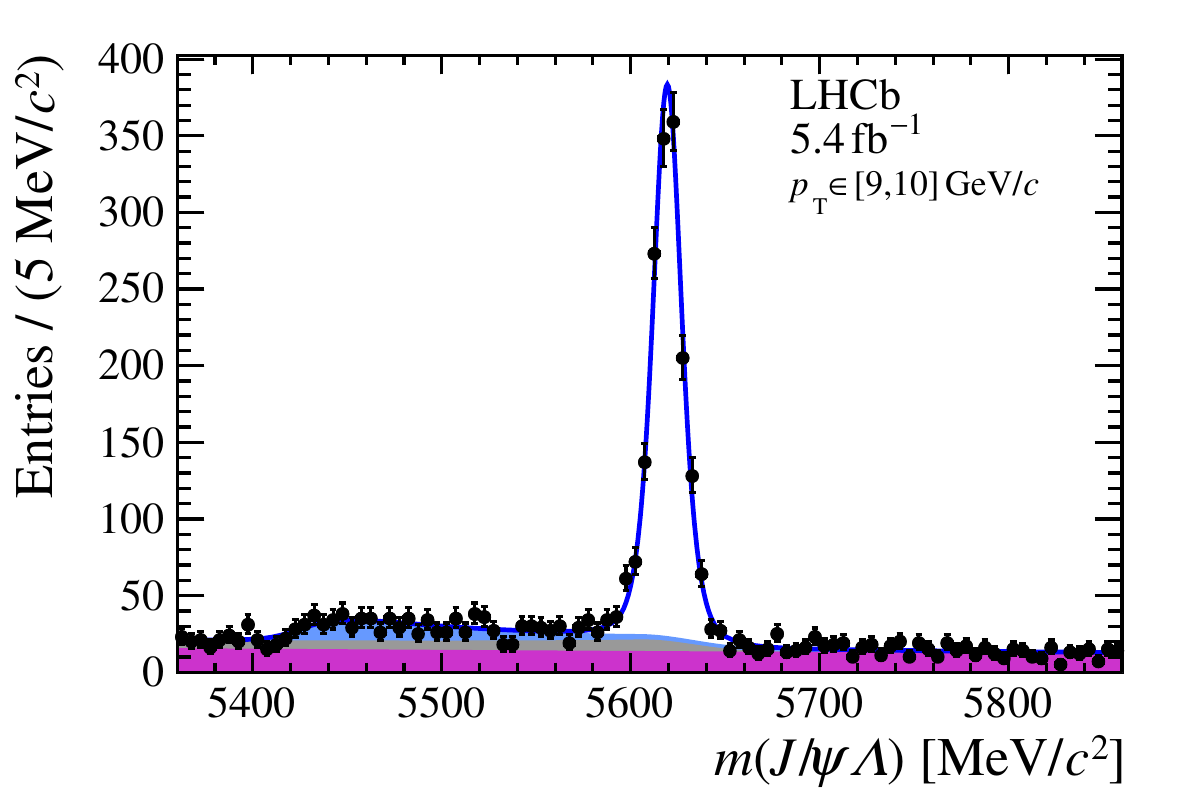} \\
    \includegraphics[width=0.32\linewidth]{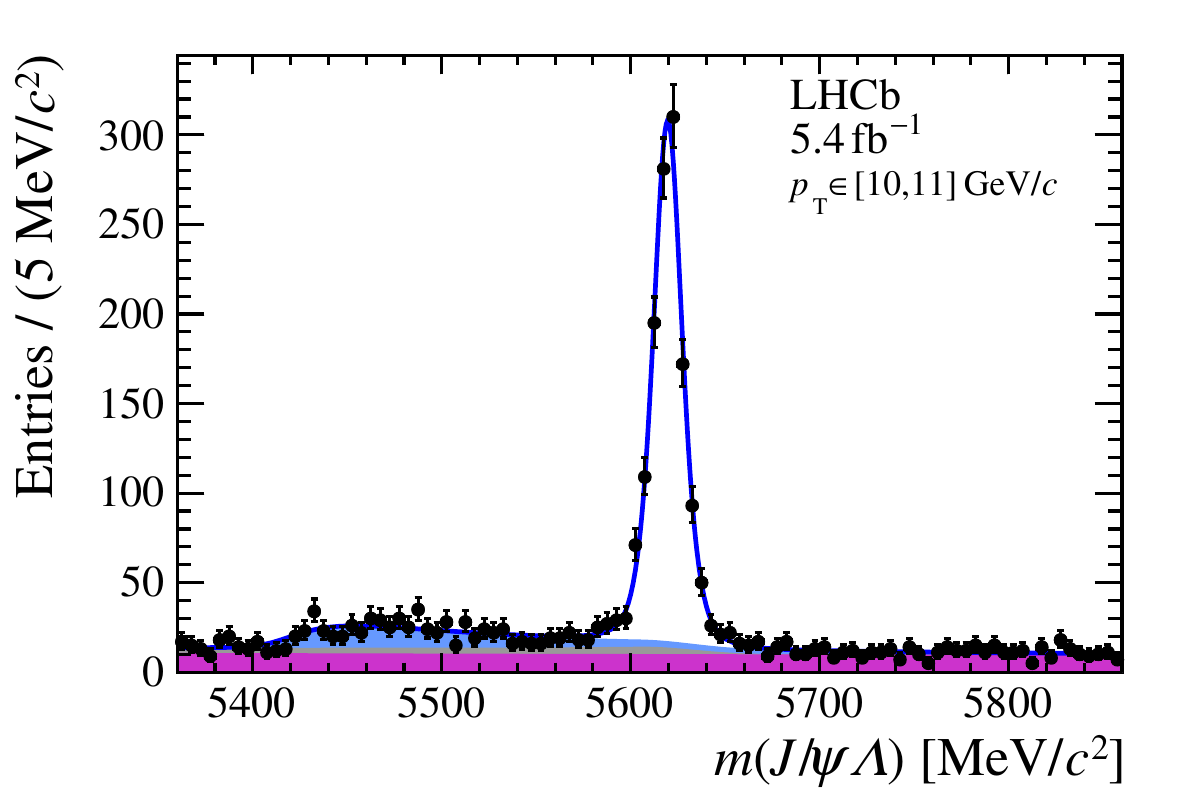} 
    \includegraphics[width=0.32\linewidth]{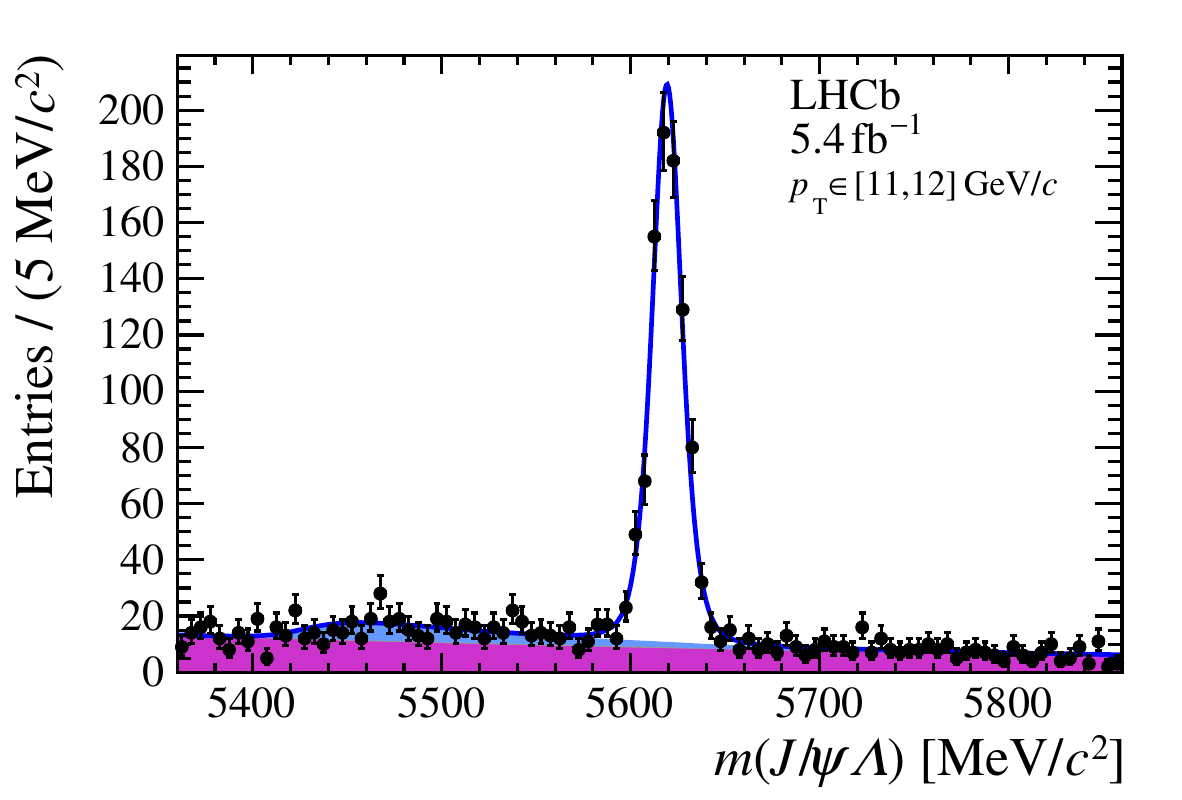} 
    \includegraphics[width=0.32\linewidth]{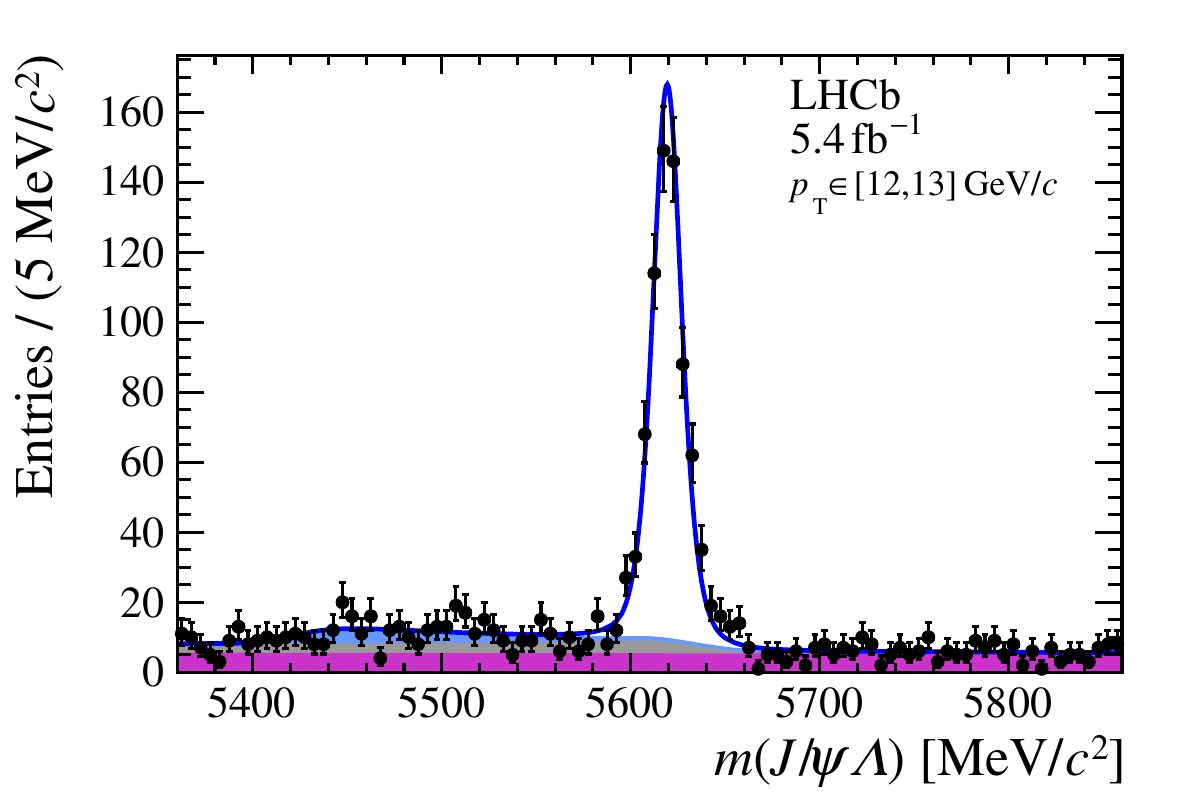} \\
    \includegraphics[width=0.32\linewidth]{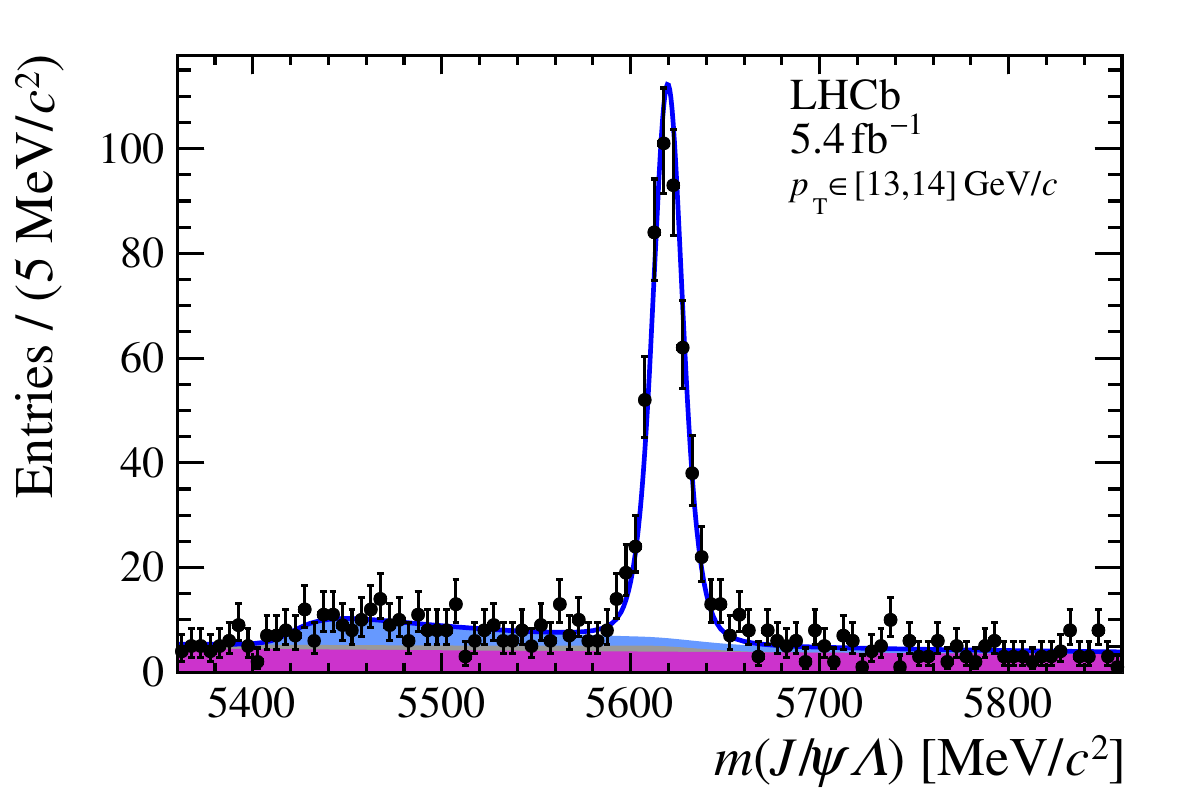} 
    \includegraphics[width=0.32\linewidth]{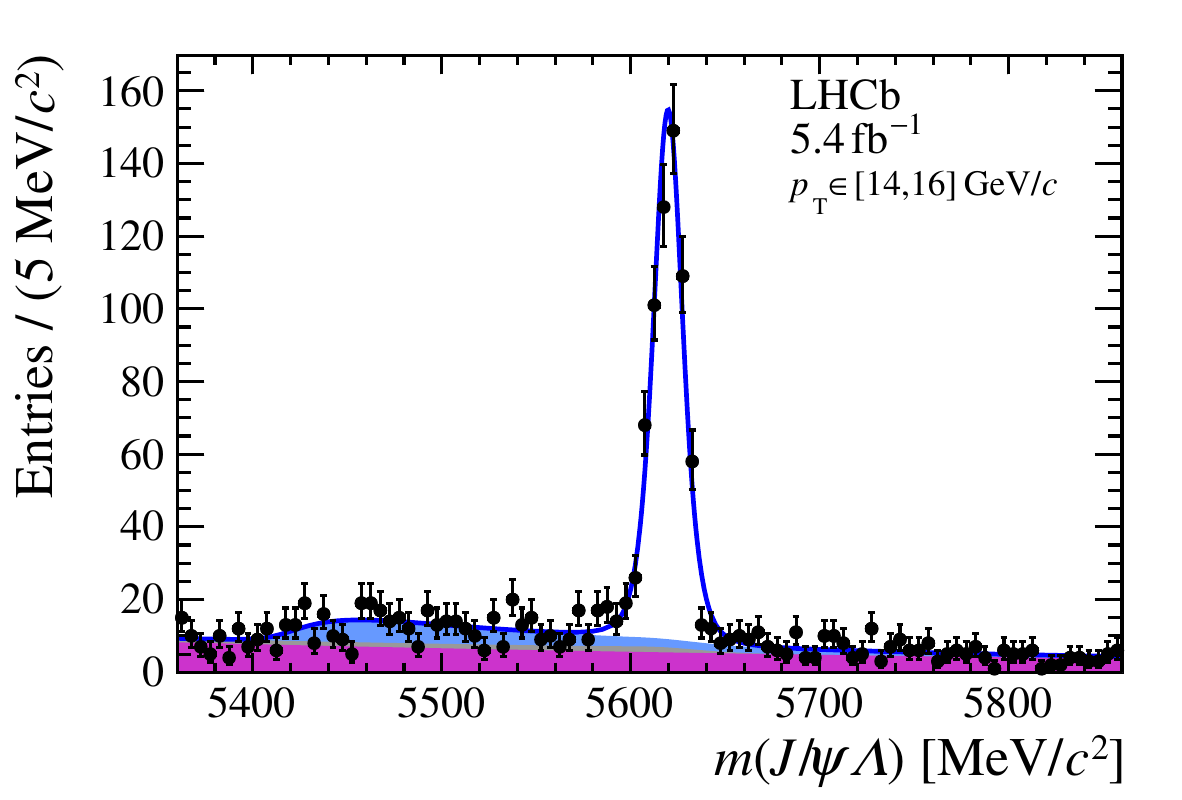} 
    \includegraphics[width=0.32\linewidth]{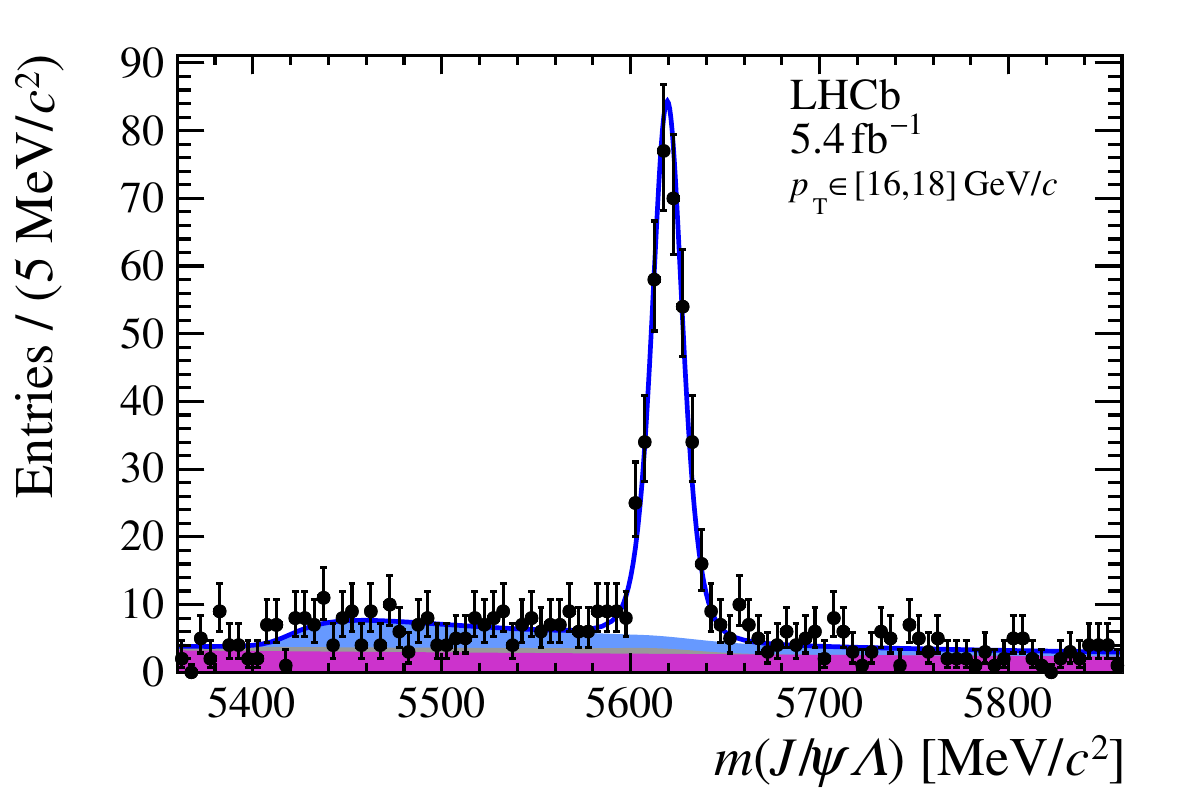} \\
    \includegraphics[width=0.32\linewidth]{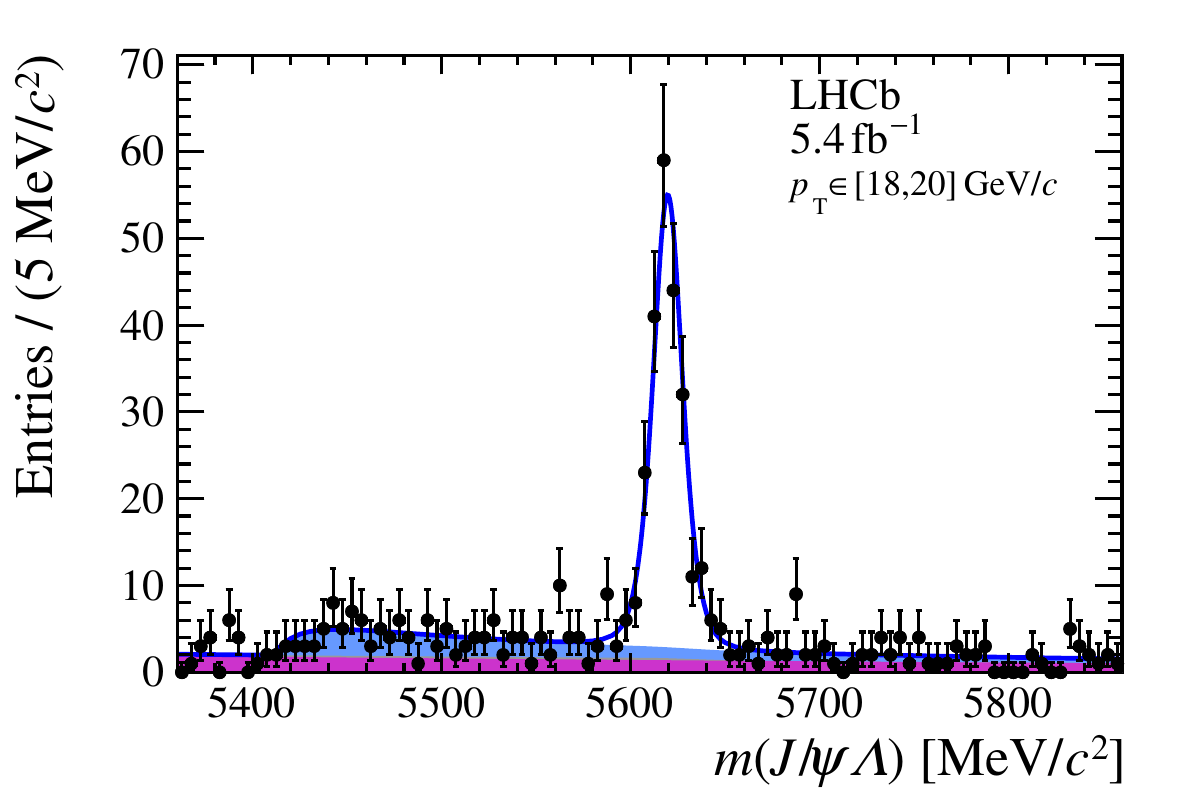} 
    \caption{
    Reconstructed $\jpsi\Lz$ mass distribution of selected \LbJpsiLambda candidates in bins of \Lb baryon \pt, reconstructed from downstream-category tracks. 
    The line and shaded regions show the result of the fit described in the text. 
    }
    \label{fig:Lambda:downstream}
\end{figure}

\clearpage

\addcontentsline{toc}{section}{References}
\bibliographystyle{LHCb}
\bibliography{main,standard,LHCb-PAPER,LHCb-CONF,LHCb-DP,LHCb-TDR}

\newpage
\centerline
{\large\bf LHCb collaboration}
\begin
{flushleft}
\small
R.~Aaij$^{38}$\lhcborcid{0000-0003-0533-1952},
A.S.W.~Abdelmotteleb$^{57}$\lhcborcid{0000-0001-7905-0542},
C.~Abellan~Beteta$^{51}$\lhcborcid{0009-0009-0869-6798},
F.~Abudin{\'e}n$^{57}$\lhcborcid{0000-0002-6737-3528},
T.~Ackernley$^{61}$\lhcborcid{0000-0002-5951-3498},
A. A. ~Adefisoye$^{69}$\lhcborcid{0000-0003-2448-1550},
B.~Adeva$^{47}$\lhcborcid{0000-0001-9756-3712},
M.~Adinolfi$^{55}$\lhcborcid{0000-0002-1326-1264},
P.~Adlarson$^{85}$\lhcborcid{0000-0001-6280-3851},
C.~Agapopoulou$^{14}$\lhcborcid{0000-0002-2368-0147}
% ...

{\footnotesize \it

% ...
$^{14}$Universit{\'e} Paris-Saclay, CNRS/IN2P3, IJCLab, Orsay, France\\
% ...
$^{38}$Nikhef National Institute for Subatomic Physics, Amsterdam, Netherlands\\
% ...
$^{47}$Instituto Galego de F{\'\i}sica de Altas Enerx{\'\i}as (IGFAE), Universidade de Santiago de Compostela, Santiago de Compostela, Spain\\
% ...
$^{51}$Physik-Institut, Universit{\"a}t Z{\"u}rich, Z{\"u}rich, Switzerland\\
% ...
$^{55}$H.H. Wills Physics Laboratory, University of Bristol, Bristol, United Kingdom\\
% ...
$^{57}$Department of Physics, University of Warwick, Coventry, United Kingdom\\
% ...
$^{60}$School of Physics and Astronomy, University of Glasgow, Glasgow, United Kingdom\\
$^{61}$Oliver Lodge Laboratory, University of Liverpool, Liverpool, United Kingdom\\
% ...
$^{69}$Syracuse University, Syracuse, NY, United States\\
% ...
$^{85}$Department of Physics and Astronomy, Uppsala University, Uppsala, Sweden, associated to $^{60}$\\
% ...
}
\end{flushleft}

\end{document}